\documentclass[aps,prd,reprint,showpacs,superscriptaddress,eqsecnum]{revtex4-1}

\usepackage{tikz}
\usepackage{amsmath}
\usepackage{amssymb}
\usepackage{graphicx}
\usepackage{bm}
\usepackage{multirow}
\usepackage{color}
\usepackage{combelow}
\usepackage{braket}

\def\feq{\ensuremath{f^{(\mathrm{eq})}}}

\def\erf{\ensuremath{\mathrm{erf}}}

\def\abs#1{\left|#1\right|}

\def\hata{{\hat{a}}}
\def\hatb{{\hat{b}}}
\def\hatc{{\hat{c}}}
\def\hatd{{\hat{d}}}

\def\hatR{{\hat{R}}}
\def\hatz{{\hat{z}}}
\def\hvarphi{{\hat{\varphi}}}
\def\heta{{\hat{\eta}}}
\def\wi{{\widetilde{\imath}}}
\def\wj{{\widetilde{\jmath}}}
\def\wk{{\widetilde{k}}}
\def\well{{\widetilde{\ell}}}
\def\vx{{\bm{x}}}
\def\vp{{\bm{p}}}
\def\vu{{\bm{u}}}

\def\vxi{{\bm{\xi}}}
\def\Rt{{\widetilde{R}}}
\def\hh{\mathfrak{h}}
\def\arctanh{\rm arctanh}
\def\wf{\widetilde{f}}

\begin{document}


\title{Lattice Boltzmann models based on the vielbein formalism \\
for the simulation of flows in curvilinear geometries}


\author{Sergiu \surname{Busuioc}}
\email[E-mail: ]{sergiu.busuioc@e-uvt.ro}

\author{Victor E. \surname{Ambru\cb{s}}}
\email[E-mail: ]{victor.ambrus@e-uvt.ro}
\thanks{Corresponding author.}                                                                                                                                                                                                                                                                                                                                                                                                                                                                                                                                                                                                                                                                 
\affiliation{Department of Physics, West University of Timi\cb{s}oara,\\
Vasile P\^arvan Avenue 4, 300223 Timi\cb{s}oara, Romania}


\graphicspath{{./}}

\date{\today}

\begin{abstract}
In this paper, we consider the Boltzmann equation with respect to orthonormal vielbein fields
in conservative form. This formalism allows the use of arbitrary 
coordinate systems to describe the space geometry, as well as of an adapted coordinate system in the 
momentum space, which is linked to the physical space through the use of vielbeins. Taking advantage 
of the conservative form, we derive the macroscopic equations in a covariant tensor notation, 
and show that the hydrodynamic limit can be obtained via the Chapman-Enskog expansion 
in the Bhatnaghar-Gross-Krook (BGK) approximation for the collision term.
We highlight that in this formalism, the component of the momentum which is perpendicular to some 
curved boundary can be isolated as a separate momentum coordinate, for which the half-range Gauss-Hermite
quadrature can be applied. We illustrate the capabilities of this formalism 
by considering two applications. The first one is the 
circular Couette flow between rotating coaxial cylinders, for which benchmarking data 
is available for all degrees of rarefaction, from the hydrodynamic to the ballistic regime. 
The second application concerns the flow in a gradually expanding channel.
We employ finite-difference lattice Boltzmann models based on 
half-range Gauss-Hermite quadratures for the implementation of diffuse reflection, 
together with the fifth order WENO and third-order TVD Runge-Kutta 
numerical methods for the advection and time-stepping, respectively.
\end{abstract}


\maketitle

\section{Introduction}

Rarefied gas flows, where non-equilibrium effects become important and the Navier-Stokes 
equations are no longer applicable, can be successfully described within the 
framework of the Boltzmann equation 
\cite{grad58,kogan69,cercignani88,cercignani00,liboff03,karniadakis05,struchtrup05,
shen05,gadelhaq06,sone07,sharipov16}.
Microfluidics specific effects (e.g. velocity slip, temperature jump) can be recovered 
by modelling the boundary conditions at the level of the Boltzmann distribution function 
$f \equiv f(\vx, \vp, t)$ (i.e. by imposing kinetic boundary conditions). 
According to the diffuse reflection concept, the 
particles reflected from the wall back into the fluid 
follow a Maxwellian distribution (all quantities are 
non-dimensionalized following the convention of 
Refs.~\cite{zhang1,zhang2,zhang3,zhangjfm,ambrus18pre}):
\begin{equation}
 f_w(p_n < 0) = \frac{n_w}{(2\pi m T_w)^{3/2}} \exp\left[-\frac{(\vp - m\vu_w)^2}{2m T_w}\right],
 \label{eq:diffuse_f}
\end{equation}
where $n_w$, $T_w$ and $\bm{u}_w$ are the particle number density, temperature and velocity of 
the wall. In the above, $p_n \equiv \vp \cdot \bm{n}$ represents the projection of the 
particle momentum vector on the outwards-directed normal $\bm{n}$ to the wall, such that 
particles for which $p_n <0$ travel from the wall back into the fluid domain.

Since the incident particle flux is a-priori essentially arbitrary, prescribing the 
distribution of emerging particles via Eq.~\eqref{eq:diffuse_f} induces a discontinuity
in the functional form of the distribution function \cite{gross57}. 
Furthermore, the impermeability of the wall is ensured by requiring that 
the mass flux through the boundary vanishes:
\begin{equation}
 \int_{p_n < 0} d^3p\, f_w\, p_n = -\int_{p_n > 0} d^3p\, f\, p_n,
 \label{eq:diffuse_flux}
\end{equation}

The correct numerical implementation of Eq.~\eqref{eq:diffuse_flux} requires 
the ability to recover half-range integrals of the distribution function.
This can be done by choosing the discrete set of momentum vectors and 
their associated quadrature weights following the prescription of 
half-range Gauss quadrature methods \cite{yang95,li03,li04,li09,lorenzani07,
frezzotti09,frezzotti11,gibelli12,guo13pre,ghiroldi14,ambrus14pre,ambrus14ipht,
ambrus14ijmpc,guo15pre,gibelli15,sader15,ambrus16jcp,ambrus16jocs,ambrus17arxiv,
ambrus18pre}. Since the Gauss quadratures are one-dimensional \cite{hildebrand87,shizgal15}, 
the integration over the momentum space must be split into a product of one-dimensional integrals.
The half-range integration can be performed using a half-range Gauss-Hermite quadrature 
only if the integration range along this direction is $[0, \infty)$ or 
$(-\infty, 0]$. This implies that, for the Cartesian split of the integration 
domain (i.e. when the integrals over $p_x$, $p_y$ and $p_z$ are performed separately),
the domain walls have to be orthogonal to the Cartesian axes. For example, 
for a wall perpendicular to the $z$ axis, the integration in Eq.~\eqref{eq:diffuse_flux}
is performed over the ranges $p_x, p_y \in (-\infty, \infty)$ and 
$p_z \in [0, \pm \infty)$. This results in a limitation of the applicability of the 
presently-available models based on half-range quadratures when curved or arbitrary 
boundaries are considered.

It is a common practice in the literature to exploit the symmetries of a non-Cartesian 
geometry by using curvilinear geometry-fitted coordinates 
\cite{yang95,mieussens00,li03,li04,li09,guo03,mendoza14ijmpc,lin14,watari16,debus16,hejranfar17pre,hejranfar17cf,velasco19}.
The coordinate system can be chosen such that the boundary is always orthogonal to the unit 
vector along one of the curvilinear coordinates. In order to apply the half-range quadrature 
along the direction perpendicular to the wall, one further step must be taken: the momentum space 
has to be adapted to the new coordinate system, such that the components of the momentum vector 
always point along the unit vectors corresponding to the curvilinear coordinates.

In Ref.~\cite{cardall13}, Cardall et al.~expressed the relativistic Boltzmann equation 
in conservative form with respect to a vielbein (i.e.~tetrad in $4D$ spacetime) 
field and a general choice 
for the parametrization of the momentum space.
In this paper, we present a formulation of the non-relativistic 
Boltzmann equation with respect to general coordinates.
In order to keep the momentum space tied to the new coordinate frame, we employ an orthonormal 
vielbein field (i.e.~a triad consisting of the non-commuting unit vectors of the coordinate frame) 
with respect to which the momentum space degrees of freedom are defined. The resulting Boltzmann 
equation contains inertial forces which ensure that freely-streaming particles travel along straight 
lines in the original Cartesian geometry. Key to this development is the use of the tools of 
differential geometry. It is worth mentioning that differential geometry 
and the vielbein formalism have been used previously in fluid dynamics,
in particular for the study of flows on curved surfaces 
\cite{nitschke12,reuther15,reuther16,reuther18}.

In order to demonstrate the robustness of our proposed formulation, we 
introduce the conservative form of the Boltzmann equation, with the help of which the Navier-Stokes 
equations with respect to general coordinates are derived via the Chapman-Enskog expansion.
This is the main result of this paper.

The applicability of our proposed scheme to rarefied flows enclosed inside curved boundaries 
is demonstrated by considering two applications, namely 
the circular Couette flow between coaxial cylinders and 
the flow in a gradually expanding channel, which are described in what follows.

In the first case, cylindrical coordinates are used to parametrize 
the flow domain, such that the boundaries are orthogonal to the radial ($R$) direction. After defining the 
momentum space with respect to the unit vectors along the radial, azimuthal and $z$ directions, 
the mixed-quadrature lattice Boltzmann (LB) models introduced in Ref.~\cite{ambrus16jcp} are employed.
These models allow the quadrature (half-range or full-range Gauss-Hermite) to be chosen 
on each axis separately. The implementation of the inertial forces requires the 
theory of distributions, as discussed in Ref.~\cite{ambrus17arxiv}. 

In Ref.~\cite{budinsky14}, a D2Q9 collide-and-stream LB
model was adapted to recover the Navier-Stokes equations with respect to the cylindrical 
coordinate system. In the resulting scheme, the velocity space parametrization is performed 
along the coordinate system unit vectors, however, its applicability is restricted to the 
hydrodynamic regime. Due to the collide-and-stream paradigm, the computational domain still 
required a two-dimensional discretization.

In Refs.~\cite{lin14,watari16,hejranfar17pre,hejranfar17cf}, the LB model was employed 
using a discretization with respect to cylindrical coordinates, but the momentum space degrees of 
freedom were the Cartesian ones. This discrepancy between the momentum space and the 
flow domain resulted in a broken symmetry which required a two-dimensional discretization 
of the flow domain. Furthermore, the aforementioned studies are limited to low Mach number flow 
regimes, where the flow is essentially incompressible. 
In our implementation of the circular Couette flow, the axial symmetry is preserved also in 
the momentum space,
such that the discretization of the flow domain can be performed in a one-dimensional 
fashion, along the radial coordinate (with only one point along the azimuthal and $z$ coordinates, 
where periodic boundary conditions apply), greatly reducing the total number of grid points
required to obtain accurate results. Also, the half-range quadratures employed in our models 
allow us to model highly compressible flows for which the profiles of the macroscopic velocity, 
number density, temperature and heat fluxes are correctly recovered.

The velocity sets employed in our models are prescribed via Gauss quadrature rules and 
are in general off-lattice (i.e. the velocity vectors cannot point simultaneously to 
neighbouring lattice sites). Therefore, the widely-used collide-and-stream paradigm 
is inapplicable with our models and we are forced to resort to finite-difference 
schemes \cite{meng11jcp,meng11pre1,meng11pre2,watari03,watari04,watari06,nanneli92,
succi95,mcnamara95,reider95,cao97,mei98,shi01,teng00,seta00,sofonea04,gan08,patil09,
piaud14,biciusca15,cristea04cejp,fede15,busuioc17aip,sofonea18}.
In order to ensure good accuracy of the spatial 
scheme, the fifth order Weighted Essentially Non-Oscillatory (WENO-5) scheme 
was employed \cite{jiang96,shu99,gan11,rezzolla13,ambrus18prc,hejranfar17pre,busuioc17arxiv}. 
For the time marching, the third-order total variation diminishing (TVD) Runge-Kutta method described in 
\cite{shu88,gottlieb98,henrick05,trangenstein07,ambrus18prc} was employed.
Furthermore, the resolution near the bounding cylinders is increased 
by performing a stretching of the radial grid points through a coordinate transformation which is 
compatible with our proposed numerical scheme, as described in Refs.~\cite{guo03,mei98}.

Our scheme is validated in the context of the circular Couette flow problem
in three flow regimes: the hydrodynamic (Navier-Stokes) regime, the transition regime 
and the ballistic (free-streaming) regime. In the hydrodynamic and ballistic regimes, 
our simulation results are compared with the analytic solution of the compressible Navier-Stokes
and collisionless Boltzmann equations, respectively. In the slip-flow and 
transition regimes, our results are compared with those reported in Ref.~\cite{aoki03} by Aoki et al.
In all cases, an excellent match is found and we conclude that our scheme can be successfully applied 
for the simulation of the circular Couette flow.

Since the aim of this paper is to demonstrate the applicability of the lattice Boltzmann models 
based on half-range Gauss-Hermite quadratures introduced in Refs.~\cite{ambrus16jcp,ambrus17arxiv} for the 
study of rarefied flows confined in non-rectangular geometries, our study of the 
circular Couette flow is limited to the case of pure diffuse reflection (unit accommodation 
coefficient). We therefore do not discuss other interesting aspects of the 
circular Couette flow, such as 
the Taylor-Couette instability appearing at large values of the Taylor number \cite{kong94,yoshida06}, 
or the inverted velocity profile due to sub-unitary accommodation coefficients 
\cite{tibbs97,aoki03,yuhong05,jung07,agrawal08,kim09,guo11,dongari12,kosuge15,akhlaghi15}.

The second application consists of the gradually expanding channel introduced by Roache in 
Ref.~\cite{roache82}. This configuration is interesting since the flow features exhibit 
scale invariance at sufficiently large values of the Reynolds number ${\rm Re}$.
In particular, the results for ${\rm Re} = 100$ already give a reasonable 
approximation of the flow features when ${\rm Re} \rightarrow \infty$. Subsequently, 
this problem was considered by $15$ participant groups who attended the fifth workshop of the 
International Association for Hydraulic Research (IAHR) Working Group on Refined 
Modelling of Flows, held in Rome on 24-25th May 1982 and was reported in 
Ref.~\cite{napolitano85} for benchmarking purposes.

Before ending the introduction, we note that our study is limited to the case when the 
quadrature method is based on a Cartesian split of the momentum space.
More efficient lattice Boltzmann algorithms may be developed 
by choosing a parametrization of the momentum space (after aligning the momentum 
space with respect to the triad) which shares the symmetries of the flow.
In particular, a cylindrical coordinate system in the momentum space, such as the 
shell-based models introduced in Ref.~\cite{watari03} and further employed 
in Refs.~\cite{watari16,sofonea06,sofonea09,gonnella09} may be more suitable 
for the simulation of flows with cylindrical symmetry. For flows with 
spherical symmetry, it may be convenient to parametrize the momentum space using 
spherical coordinates, as discussed in Refs.~\cite{romatschke11,ambrus12}.
However, to the best of our knowledge, none of the above mentioned models have been 
endowed with half-range 
capabilities. We thus postpone the study of flows in curvilinear geometries using 
non-Cartesian decompositions of the momentum space for future work.

The paper is structured as follows. In Sec.~\ref{sec:triad}, we lay the theoretical foundation 
for our scheme by introducing the non-relativistic Boltzmann equation in conservative form 
with respect to orthonormal 
vielbein fields (i.e. triads in $3{\rm D}$ space). In Subsec.~\ref{sec:triad:CE}, 
the Navier-Stokes equations are derived with respect to general coordinates via the Chapman-Enskog 
expansion. 
The numerical scheme and the implementation of the boundary conditions are discussed 
in Sec.~\ref{sec:num_sch}. 
The lattice Boltzmann algorithm is reviewed in Sec.~\ref{sec:LB}.
In Sec.~\ref{sec:couette}, the vielbein formalism is specialized to the case 
of the circular Couette flow and the numerical results are compared to 
analytic solutions in the Navier-Stokes (Subsec.~\ref{sec:couette:hydro}) and collisionless 
(Subsec.~\ref{sec:couette:bal}) regimes, as well as with the DVM results in Ref.~\cite{aoki03} 
in the transition regime. 
The flow through the gradually expanding channel is discussed in Sec.~\ref{sec:canalie}.
Our conclusions are presented in Sec.~\ref{sec:conc}. 
Appendices \ref{app:cov}--\ref{app:cons} contain supplementary mathematical details required 
in Sec.~\ref{sec:triad}, while Appendix~\ref{app:force} discusses the implementation 
of the momentum space derivative of the distribution function in the lattice Boltzmann
method employed in this paper.

\section{Boltzmann equation with respect to triads}\label{sec:triad}

\begin{figure}
\centering
\begin{tabular}{c}
%
%
%
%
%
%
\includegraphics[width=0.77\columnwidth]{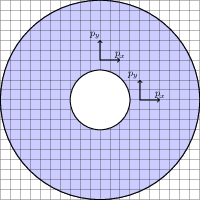}
\\
(a)
\\
%
%
%
%
%
\includegraphics[width=0.77\columnwidth]{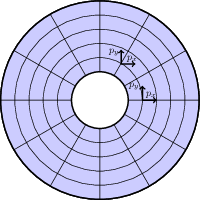}

\\
(b)
\\
\includegraphics[width=0.77\columnwidth]{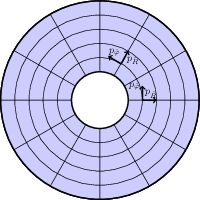}
\\
%
%
%
%
%
(c)
\end{tabular}
\caption{Circular Couette flow setup. 
(a) Cartesian grid and momentum space decomposition along the Cartesian axes;
(b) Cylindrical grid and momentum space decomposition along the Cartesian axes;
(c) Cylindrical grid and momentum space decomposition adapted to the curvilinear coordinates.}
\label{fig:decomp}
\end{figure}

To better illustrate the use of triads, we refer the reader to Fig.~\ref{fig:decomp},
where the space between two coaxial cylinders constitutes the flow domain. 
The spatial grid can be constructed in two ways: using Cartesian coordinates (a) or 
cylindrical/polar coordinates (b and c). Similarly, the momentum space degrees 
of freedom can be chosen along the Cartesian axes (a and b) or along the 
cylindrical axes (c). 

The grid in Fig.~\ref{fig:decomp}(a) requires a staircase (polygonal) approximation 
of the boundary and thus the results are dependent on the resolution of the grid around the 
boundary. The resulting grid is $2D$.

In Fig.~\ref{fig:decomp}(b), a cylindrical coordinate system ($R$, $\varphi$) is used to describe the 
flow domain. This ensures the exact representation of the boundary. However, the momentum space
degrees of freedom point along the Cartesian axes ($p_x$, $p_y$). The resulting setup is not invariant under 
rotations since a rotation about the symmetry axis also rotates the momentum space. Thus, a $2D$ grid is required.

The final step is to orient the momentum space along the cylindrical coordinates ($p_{\hat{R}}$, $p_\hvarphi$),
as shown in Fig.~\ref{fig:decomp}(c). This results in a representation of the flow domain and particle momenta
which is fully symmetric with respect to rotations about the symmetry axis. In order to achieve 
the alignment of the momentum space along the new coordinate system, an orthonormal triad must be employed, 
as described in the current Section.

The Boltzmann equation when non-Cartesian coordinates are used for the spatial domain and the
momentum space degrees of freedom are taken with respect to a triad is derived in 
Subsec.~\ref{sec:triad:boltz}. Using the conservative form of this equation derived 
in Subsec.~\ref{sec:triad:cons}, the application 
of the Chapman-Enskog procedure for the derivation of the conservation equations in the 
hydrodynamic limit is illustrated in Subsec.~\ref{sec:triad:CE}.

\subsection{Advective form}\label{sec:triad:boltz}

The Boltzmann equation with respect to the Cartesian coordinates 
$\{x, y, z\}$ can be written as:
\begin{equation}
 \frac{\partial f}{\partial t} + \frac{p^i}{m} \frac{\partial f}{\partial x^i} + 
 F^i \frac{\partial f}{\partial p^i} = J[f],\label{eq:boltz}
\end{equation}
where $f$ is the Boltzmann distribution function, $m$ is the mass of the fluid particles, 
while $p^i$ and $F^i$ represent the Cartesian components of the fluid particle momentum 
and of the external force, respectively.

In certain situations, it is convenient to introduce a set of arbitrary coordinates 
$\{x^{\widetilde{1}}, x^{\widetilde{2}}, x^{\widetilde{3}}\}$, 
where $x^\wi \equiv x^\wi(x, y, z)$ (in this paper, we restrict our analysis to 
time-independent coordinate transformations). This coordinate transformation
induces a metric $g_{\wi\wj}$, as follows:
\begin{align}
 ds^2 =& \delta_{ij} dx^i dx^j = dx^2 + dy^2 + dz^2 \nonumber\\
 =& g_{\wi\wj} dx^\wi dx^\wj,\label{eq:ds2}
\end{align}
such that
\begin{equation}
 g_{\wi\wj} = \delta_{ij} \frac{\partial x^i}{\partial x^\wi} \frac{\partial x^j}{\partial x^\wj}.
\end{equation}
The Boltzmann equation \eqref{eq:boltz} can be written in advective form
with respect to these new coordinates 
as follows:
\begin{equation}
 \frac{\partial f}{\partial t} + \frac{p^\wi}{m} \frac{\partial f}{\partial x^\wi} + 
 \left(F^\wi - \frac{1}{m} \Gamma^\wi{}_{\wj\wk} p^\wj p^\wk\right) 
 \frac{\partial f}{\partial p^\wi} = J[f],\label{eq:boltz_cov}
\end{equation}
where the components $p^\wi$ and $F^\wi$ with respect to the new coordinates are related 
to the components $p^i$ and $F^i$ with respect to the old coordinates through:
\begin{equation}
 p^\wi = \frac{\partial x^{\wi}}{\partial x^i} p^i,\qquad
 F^\wi = \frac{\partial x^{\wi}}{\partial x^i} F^i.
\end{equation}
The Christoffel symbols $\Gamma^\wi{}_{\wj\wk}$ appearing in Eq.~\eqref{eq:boltz_cov} 
are defined as:
\begin{align}
 \Gamma^\wi{}_{\wj\wk} =& \frac{\partial x^\wi}{\partial x^\ell} 
 \frac{\partial^2 x^\ell}{\partial x^\wj \partial x^\wk} \nonumber\\
 =& \frac{1}{2}g^{\wi\well} \left(\partial_\wk g_{\well\wj} + \partial_\wj g_{\well \wk} -
 \partial_\well g_{\wj\wk}\right).\label{eq:christoffel}
\end{align}
Further details regarding the connection between Eqs.~\eqref{eq:boltz} and \eqref{eq:boltz_cov}
can be found in Appendix~\ref{app:cov}.

The above formalism is sufficient to adapt the coordinate system to a curved boundary. 
However, the transition to an LB model is not straightforward, since the 
momentum space has an intrinsic dependence on the coordinates. Indeed, the 
Maxwellian distribution corresponding to a particle number density $n$, 
macroscopic velocity $\bm{u}$ and temperature $T$ has the expression:
\begin{equation}
 \feq = \frac{n}{(2\pi m T)^{\frac{3}{2}}} \exp\left[-\frac{g_{\wi\wj} (p^\wi - mu^\wi)(p^\wj - mu^\wj)}{2mT}\right],
 \label{eq:feq_wi}
\end{equation}
while its moments are calculated as:
\begin{equation}
 M^{\wi_1, \dots \wi_n}_{\rm eq} = 
 \sqrt{g} \int d^3\widetilde{p}\, \feq p^{\wi_1} \cdots p^{\wi_n},
 \label{eq:momg}
\end{equation}
where $g$ is the determinant of the metric tensor $g_{\wi\wj}$.

In order to eliminate the burden of this metric dependence in the expression for the Maxwellian,
it is convenient to introduce a triad (vielbein) with respect to which the metric is diagonal:
\begin{equation}
 g_{\wi\wj} dx^\wi \otimes dx^\wj = 
 \delta_{\hata\hatb} \omega^\hata \otimes \omega^\hatb,
\end{equation}
where the triad one-forms $\omega^\hata$ are defined as:
\begin{equation}
 \omega^\hata = \omega^\hata_\wj dx^\wj,
\end{equation}
such that:
\begin{equation}
 g_{\wi\wj} = \delta_{\hata\hatb} \omega^\hata_\wi \omega^\hatb_\wj.\label{eq:gij_omega}
\end{equation}
The above equation allows three degrees of freedom for the system $\{\omega^\hata_\wj\}$,
corresponding to the invariance of the right hand side of Eq.~\eqref{eq:gij_omega} under rotations 
with respect to the hatted indices. It is possible to define triad vectors dual to the above 
one-forms by introducing the following inner product:
\begin{equation}
 \braket{\omega^\hatb, e_\hata} \equiv \omega^\hatb_\wi e_\hata^\wi = \delta^\hatb{}_\hata,
\end{equation}
where 
\begin{equation}
 e_\hata = e_\hata^\wi \frac{\partial }{\partial x^\wi}.
\end{equation}

Using the above triad, the components of vectors can be expressed as follows:
\begin{equation}
 p^\hata = \omega^\hata_\wi p^\wi,
\end{equation}
such that 
\begin{equation}
 g_{\wi\wj} p^\wi p^\wj = \delta_{\hata\hatb} p^\hata p^\hatb.
\end{equation}
Thus, the metric dependence in the Maxwellian \eqref{eq:feq_wi} disappears:
\begin{equation}
 \feq = \frac{n}{(2\pi m T)^{\frac{3}{2}}} \exp\left[-\frac{\delta_{\hata\hatb} (p^\hata - mu^\hata)(p^\hatb - mu^\hatb)}{2mT}\right],
 \label{eq:feq_triad}
\end{equation}
allowing its moments to be written as:
\begin{equation}
 M^{\hata_1, \dots \hata_s}_{\rm eq} = 
 \int d^3\hat{p}\, \feq p^{\hata_1} \cdots p^{\hata_s}.
 \label{eq:momeq}
\end{equation}

The expressions for the lower order moments of $\feq$ are listed below:
\begin{align}
 &M_{\rm eq} = n, \quad 
 M_{\rm eq}^\hata = \rho u^\hata, \quad 
 M_{\rm eq}^{\hata\hatb} = m(P \delta^{\hata\hatb} + \rho u^\hata u^\hatb), \nonumber\\
 &M_{\rm eq}^{\hata\hatb\hatc} = m^2 P(u^\hata \delta^{\hatb\hatc} + 
 u^\hatb \delta^{\hata\hatc} + u^\hatc \delta^{\hata\hatb}) + 
 m^2 \rho u^\hata u^\hatb u^\hatc, \nonumber\\
 &M_{\rm eq}^{\hata\hatb\hatc\hatd} = m^2 P T(\delta^{\hata\hatb}\delta^{\hatc\hatd} +
 \delta^{\hata\hatc}\delta^{\hatb\hatd} + \delta^{\hata\hatd}\delta^{\hatb\hatc}) \nonumber\\
 &\phantom{M_{\rm eq}^{\hata\hatb\hatc\hatd} =}
 + m^3 P(u^\hata u^\hatb \delta^{\hatc\hatd} + 
 u^\hata u^\hatc \delta^{\hatb\hatd} + 
 u^\hata u^\hatd \delta^{\hatb\hatc} \nonumber\\
 &\phantom{M_{\rm eq}^{\hata\hatb\hatc\hatd} =+ m^3 P}+ 
 u^\hatb u^\hatc \delta^{\hata\hatd} + 
 u^\hatb u^\hatd \delta^{\hata\hatc} + 
 u^\hatc u^\hatd \delta^{\hata\hatb})\nonumber\\
 &\phantom{M_{\rm eq}^{\hata\hatb\hatc\hatd} =} \qquad\qquad\qquad\qquad+ 
 m^3 \rho u^\hata u^\hatb u^\hatc u^\hatd. 
 \label{eq:momeq_aux}
\end{align}
It will be useful to introduce at this point the notation 
for the moments of the distribution function $f$:
\begin{equation}
 M^{\hata_1, \dots \hata_s} = 
 \int d^3\hat{p}\, f\, p^{\hata_1} \cdots p^{\hata_s}.
 \label{eq:mom}
\end{equation}

The Boltzmann equation can now be written in advective form
in terms of the 
triad components of the momentum vectors, as follows:
\begin{equation}
 \frac{\partial f}{\partial t} + \frac{p^\hata}{m} e_\hata^\wi \frac{\partial f}{\partial x^\wi} + 
 \left(F^\hata - \frac{1}{m} \Gamma^\hata{}_{\hatb\hatc} p^\hatb p^\hatc\right) 
 \frac{\partial f}{\partial p^\hata} = J[f], 
 \label{eq:boltz_triad}
\end{equation}
where the connection coefficients $\Gamma^\hata{}_{\hatb\hatc}$ are defined by:
\begin{align}
 \Gamma^\hata{}_{\hatb\hatc} =& \omega^\hata_\wi \Gamma^\wi{}_{\wj\wk} e^\wj_\hatb e^\wk_\hatc -
 e^\wi_\hatb e^\wj_\hatc \frac{\partial \omega^\hata_\wi}{\partial x^\wj}\nonumber\\
 =& \frac{1}{2} \delta^{\hata \hatd} \left(c_{\hatd\hatb\hatc} + 
 c_{\hatd\hatc\hatb} - c_{\hatb\hatc\hatd}\right),
 \label{eq:Gamma_def}
\end{align}
while the Cartan coefficients 
$c_{\hatb\hatc}{}^\hata = \delta^{\hata\hatd} c_{\hatb\hatc\hatd}$ can be obtained using:
\begin{equation}
 c_{\hatb\hatc}{}^{\hata} = \braket{\omega^\hata, [e_\hatb, e_\hatc]},
\end{equation}
while $c_{\hatb\hatc\hatd} = \delta_{\hatd\hata} c_{\hatb\hatc}{}^{\hata}$.
The vector $[e_\hatb, e_\hatc]$ represents the commutator of the triad vectors 
$e_\hatb$ and $e_\hatc$, having the components:
\begin{equation}
 [e_\hatb, e_\hatc]^\wi = e_\hatb^{\wj} \partial_\wj e^{\wi}_\hatc - e^\wj_\hatc \partial_\wj e^\wi_\hatb.
\end{equation}
More details on the connection between Eqs.~\eqref{eq:boltz_cov} and \eqref{eq:boltz_triad} can be 
found in Appendix~\ref{app:triad}. Since the numerical implementations of hyperbolic
equations in advective form are in general non-conservative \cite{leveque02},
we will not consider the advective form \eqref{eq:boltz_triad} of the Boltzmann 
equation further in this paper.

\subsection{Conservative form}\label{sec:triad:cons}

The Boltzmann equation in advective form \eqref{eq:boltz_triad} 
hides the conservation 
laws both analytically and numerically. Following Ref.~\cite{cardall13}, 
Eq.~\eqref{eq:boltz_triad} can be written in conservative form as follows:
\begin{multline}
 \frac{\partial f}{\partial t} + \frac{1}{\sqrt{g}}
 \frac{\partial}{\partial x^{\wi}} \left(\frac{p^\hata}{m} e_\hata^\wi f \sqrt{g}\right)\\
 + \frac{\partial}{\partial p^\hata} \left[\left(F^\hata - 
 \frac{1}{m} \Gamma^\hata{}_{\hatb\hatc} p^\hatb p^\hatc\right) f\right]
 = J[f].
 \label{eq:boltz_noncons}
\end{multline}
The derivation of Eq.~\eqref{eq:boltz_noncons} is presented in Appendix~\ref{app:cons}. 

Multiplying Eq.~\eqref{eq:boltz_noncons} by $p^{\hata_1} p^{\hata_2} \cdots p^{\hata_s}$ and 
integrating over the momentum space, it can be shown that:
\begin{multline}
 \partial_t M^{\hata_1 \hata_2 \dots \hata_s} + 
 \frac{1}{m}\nabla_{\hatb} M^{\hatb \hata_1 \dots \hata_s}\\
 = \frac{1}{m} (F^{\hata_1} M^{\hata_2 \dots \hata_s} + \dots) + 
 \mathcal{S}^{\hata_1 \hata_2 \dots \hata_s},
 \label{eq:mom_eq}
\end{multline}
where the expression bewteen the parentheses on the right-hand side is symmetric with respect to the indices $\hata_1$, $\dots\hata_s$,
containing $s$ terms. The $s$'th order moment of $f$ is defined in Eq.~\eqref{eq:mom}, while
the covariant derivative $\nabla_\hata$ acts on the tensor $M^{\hata_1 \dots \hata_s}$ as follows:
\begin{multline}
 \nabla_\hata M^{\hata_1 \dots \hata_s} = e_\hata^\wi \partial_\wi M^{\hata_1 \dots \hata_s} + 
 \Gamma^{\hata_1}{}_{\hatb\hata} M^{\hatb \hata_2 \dots \hata_s} \\
 + \Gamma^{\hata_2}{}_{\hatb\hata} M^{\hata_1 \hatb \hata_3 \dots \hata_s} + \dots 
 + \Gamma^{\hata_s}{}_{\hatb\hata} M^{\hata_1 \hata_2 \dots \hata_{s-1} \hatb}.
\end{multline}
The source term $\mathcal{S}^{\hata_1 \hata_2 \dots \hata_s}$ is defined as:
\begin{equation}
 \mathcal{S}^{\hata_1 \hata_2 \dots \hata_s} = \int d^3\hat{p} \, J[f] \,
 p^{\hata_1} \cdots p^{\hata_s}.
\end{equation}

It is now easy to derive the macroscopic fluid equations:
\begin{subequations}\label{eq:macro}
\begin{gather}
 \frac{Dn}{Dt} + n (\nabla \cdot \vu) = 0,\label{eq:macro_n}\\
 \rho \frac{Du^\hata}{Dt} = nF^{\hata} - \nabla_{\hatb} T^{\hata\hatb},\label{eq:macro_u}\\
 n\frac{De}{Dt} + \nabla_\hata q^\hata + T^{\hata\hatb} \nabla_\hata u_\hatb = 0,\label{eq:macro_T}
\end{gather}
\end{subequations}
where $D/Dt = \partial_t + u^\hata \nabla_\hata$ is the material derivative, while 
$e = \frac{3}{2} T$ is the internal energy per constituent. 
The relations between the distribution function $f$ and the 
particle number density $n$, macroscopic velocity $u^\hata$, 
stress tensor $T^{\hata\hatb}$ and heat flux $q^\hata$ are listed below:
\begin{subequations}\label{eq:macro_def}
\begin{align}
 n =& \int d^3\hat{p} \, f, \label{eq:macro_def_n}\\
 u^\hata =& \frac{1}{\rho} \int d^3\hat{p} \, f\,p^\hata, \label{eq:macro_def_u}\\ 
 T^{\hata\hatb} =& \int d^3\hat{p} \, f\,\frac{\xi^\hata \xi^\hatb}{m}, \label{eq:macro_def_Tab}\\ 
 q^\hata =& \int d^3\hat{p} \, f\,\frac{\vxi^2}{2m} \frac{\xi^\hata}{m},\label{eq:macro_def_q}
\end{align}
\end{subequations}
where $\rho = mn$, $\xi^\hata = p^\hata - mu^\hata$, and 
$\vxi^2 = \delta_{\hata\hatb} \xi^\hata \xi^\hatb$.
The total number of particles $N_{\rm tot}$ inside the simulation domain can 
be computed using:
\begin{equation}
 N_{\rm tot} = \int d^3x \sqrt{g} n = \int d^3x \int d^3\hat{p} \wf,
 \label{eq:Ntot_def}
\end{equation}
where $\wf = f \sqrt{g}$.

\subsection{Chapman-Enskog Expansion}\label{sec:triad:CE}

In order to illustrate the application of the Chapman-Enskog procedure,
we consider the Bhatnaghar-Gross-Krook (BGK) single-time approximation for the collision term:
\begin{equation}
 J[f] = -\frac{1}{\tau} (f - \feq).\label{eq:BGK}
\end{equation}
We note that this simplified implementation of the collision term has 
several drawbacks, including the fact that the Prandtl number ${\rm Pr}$ is 
fixed at $1$, while its value for, e.g., hard sphere molecules is $2/3$.
This drawback (and others) can be corrected, i.e. by employing the 
Shakhov extension of the BGK collision term \cite{shakhov68,titarev07,graur09,graur13,ho15}.
In the interest of simplicity, in this paper we only consider the 
BGK implementation of the collision term, since the generalization 
of our proposed scheme to more complex formulations of $J[f]$ 
is straightforward.

The ``simplified version'' of the Chapman-Enskog expansion entails treating $\tau$ and the difference 
$\delta f = f - \feq$ as small quantities, such that $\delta f / \tau$ is of the same order as the 
left-hand side of Eq.~\eqref{eq:boltz_noncons} when $f \simeq \feq$.
Ignoring higher-order terms, the following expression is obtained for $\delta f$:
\begin{multline}
 \delta f = -\tau \left\{\frac{\partial \feq}{\partial t} + 
 \frac{1}{\sqrt{g}} \frac{\partial}{\partial x^{\wi}} \left(\frac{p^\hata}{m} e_\hata^\wi \feq \sqrt{g}\right)\right.\\
 \left.+ \frac{\partial}{\partial p^\hata} \left[\left(F^\hata - 
 \frac{1}{m} \Gamma^\hata{}_{\hatb\hatc} p^\hatb p^\hatc\right) \feq\right]\right\}.
 \label{eq:deltaf}
\end{multline}
The collision invariants $\psi \in \{1, p^{\hata}, \vp^2/2m\}$ are preserved only if:
\begin{equation}
 \int d^3\hat{p}\, \delta f = 
 \int d^3\hat{p}\, \delta f \,p^{\hata} = 
 \int d^3\hat{p}\, \delta f \, \frac{\vp^2}{2m} = 0,
\end{equation}
where $\vp^2 \equiv \delta_{\hata\hatb} p^\hata p^\hatb$ represents the 
squared norm of $\vp$ written in terms of its vielbein components.

The deviation from equilibrium $\delta f$ induces a deviation $\delta T^{\hata\hatb}$ from 
the equilibrium stress-tensor, as well as a heat flux:
\begin{equation}
 T^{\hata\hatb} = \delta^{\hata\hatb} P + \delta T^{\hata\hatb}, \qquad 
 q^\hata = \delta q^{\hata},\label{eq:noneq}
\end{equation}
where $P = n T$ is the ideal gas pressure.
The non-equilibrium quantities $\delta T^{\hata\hatb}$ and $\delta q^\hata$ can be obtained as follows:
\begin{subequations}
\begin{align}
 \delta T^{\hata\hatb} =& \int d^3\hat{p} \, \delta f \frac{\xi^\hata \xi^\hatb}{m}
  = \int d^3\hat{p} \, \delta f \frac{p^\hata p^\hatb}{m}, \label{eq:deltaT_def}\\
 \delta q^\hata =& \int d^3\hat{p} \, \delta f\, \frac{\vxi^2}{2m} \frac{\xi^\hata}{m} 
 = \int d^3\hat{p} \, \delta f\, \frac{\vp^2}{2m} \frac{p^\hata}{m} -u_\hatb \delta T^{\hata\hatb}.\label{eq:deltaq_def}
\end{align}
\end{subequations}
Substituting Eq.~\eqref{eq:deltaf} into Eq.~\eqref{eq:deltaT_def} yields:
\begin{subequations}
\begin{multline}
 \delta T^{\hata\hatb} = -\tau \\
 \times \left[\frac{1}{m} \partial_t M^{\hata\hatb}_{\rm eq} + 
 \frac{1}{m^2}\nabla_\hatc M^{\hata\hatb\hatc}_{\rm eq} - 
 n(u^\hata F^\hatb + u^\hatb F^\hata)\right],\label{eq:deltaT_aux}
\end{multline}
while the heat flux can be obtained as:
\begin{multline}
 \delta q^\hata + u_\hatb \delta T^{\hata\hatb} = -\tau \Big\{\delta_{\hatc\hatd}\left(
 \frac{1}{2m^2} \partial_t M^{\hata\hatc\hatd}_{\rm eq} + 
 \frac{1}{2m^3} \nabla_\hatb M^{\hata\hatb\hatc\hatd}_{\rm eq}\right)\\
 - \frac{5 n T}{2m} F^\hata - \frac{n}{2}[F^\hata \vu^2 + 2u^\hata (\vu \cdot \bm{F})]\Big\}.
 \label{eq:deltaq_aux}
\end{multline}
\end{subequations}
The time derivatives appearing in Eqs.~\eqref{eq:deltaT_aux} and \eqref{eq:deltaq_aux} 
can be eliminated since, at first order, $n$, $u^\hata$ and $T$ satisfy the Euler equations,
obtained by setting $T^{\hata\hatb} = \delta^{\hata\hatb} P$ and 
$q^\hata = 0$ in Eq.~\eqref{eq:macro}:
\begin{align}
 \frac{Dn}{Dt} + n \nabla_\hata u^\hata =& 0, \nonumber\\
 \rho \frac{Du^\hata}{Dt} + \nabla^{\hata} P=& nF^{\hata},\nonumber\\
 n\frac{De}{Dt} + P \nabla_\hata u^\hata =& 0.
\end{align}

Using the explicit expressions \eqref{eq:momeq_aux} for the 
moments of $\feq$, a straightforward but tedious calculation shows that 
$\delta T^{\hata \hatb}$ and $\delta q^\hata$ can be expressed as:
\begin{subequations}\label{eq:CE}
\begin{align}
 \delta T^{\hata\hatb} =& -\mu \left(\nabla^\hata u^\hatb + \nabla^\hatb u^\hata 
 - \frac{2}{3} \delta^{\hata\hatb} \nabla_\hatc u^\hatc\right),\label{eq:CE_T}\\
 \delta q^\hata =& -\kappa \nabla^\hata T, \label{eq:CE_q}
\end{align}
where the dynamic viscosity $\mu$ and the coefficient 
of thermal conductivity $\kappa$ are given by:
\begin{equation}
 \mu = \tau P, \qquad 
 \kappa = \frac{5}{2m} \tau P.\label{eq:CE_tcoeff}
\end{equation}
\end{subequations}


\section{Numerical scheme}\label{sec:num_sch}

The aim of this Section is to derive 
numerical implementations of Eq.~\eqref{eq:boltz_noncons} 
which are manifestly conservative. To this end, we 
also introduce the following form of the Boltzmann equation,
obtained by multiplying Eq.~\eqref{eq:boltz_noncons} with $\sqrt{g}$:
\begin{multline}
 \frac{\partial \wf}{\partial t} + 
 \frac{\partial}{\partial x^{\wi}} \left(\frac{p^\hata}{m} e_\hata^\wi \wf\right)
 + \frac{\partial}{\partial p^\hata} \left[\left(F^\hata - 
 \frac{1}{m} \Gamma^\hata{}_{\hatb\hatc} p^\hatb p^\hatc\right) \wf\right]\\
 = J[f] \sqrt{g},
 \label{eq:boltz_cons}
\end{multline}
where the following notation was introduced:
\begin{equation}
 \wf = f \sqrt{g}.\label{eq:wf_def}
\end{equation}
The advantage of the formulation \eqref{eq:boltz_cons} is that the
spatial derivatives corresponding to the advection term do not 
have any position-dependent prefactors, such that a conservative 
numerical implementation is straightforward. The disadvantage 
of this formulation is that performing the evolution and advection 
at the level of $\wf$ can introduce fluctuations in the numerical 
solution, which prevent, e.g., a solution of the form $f = {\rm const}$ 
to be exactly achieved \cite{leveque02}. For definiteness, we shall 
refer to the formulation starting from Eq.~\eqref{eq:boltz_cons} as 
the $\widetilde{f}$ formulation.

Our second (and preferred) implementation is inspired from the methodology 
proposed in Refs.~\cite{falle96,downes02} and starts again from the Boltzmann 
equation in the form presented in Eq.~\eqref{eq:boltz_noncons}.
For simplicity, we restrict the construction of the numerical scheme to 
the case when $\sqrt{g}$ is separable, i.e.:
\begin{equation}
 \sqrt{g} = \sqrt{g_{\widetilde{1}} g_{\widetilde{2}} g_{\widetilde{3}}},
 \label{eq:gi}
\end{equation}
where the factors $g_{\wi} \equiv g_{\wi}(x^\wi)$ each depend only on one 
coordinate ($x^\wi$).
The above assumption is valid for both examples considered in this paper 
(circular Couette flow and flow through the gradually expanding channel). 
An extension of the present methodology to a non-separable metric determinant 
is straightforward but for simplicity, we do not discuss this case here.
The main idea is to define a new set of coordinates, $\chi^{\wi}$, 
such that the $1/\sqrt{g}$ factor in front of the spatial 
derivatives in Eq.~\eqref{eq:boltz_noncons} is absorbed into the derivative.
This can be achieved when $\chi^\wi$ is introduced as follows:
\begin{equation}
 \chi^{\wi} = 
 \int^{x^{\wi}} dx^\wi \sqrt{g_\wi},
 \label{eq:chi_def}
\end{equation}
such that $\partial \chi^\wi / \partial x^\wi = \sqrt{g_\wi}$. 
The lower integration end is not relevant, since only differences 
of the form $\delta \chi^{\wi}$ appear in the numerical implementation 
and is thus left arbitrary. 
The above definition for $\chi^\wi$ is inspired from Refs.~\cite{falle96} and \cite{downes02},
where a similar definition was employed for the cylindrical and spherical coordinate systems, 
respectively
(more details will be given in Sec.~\ref{sec:couette}).
The advantage of performing the derivative with respect to $\chi^\wi$ is that 
the numerical procedure can be constructed to exactly preserve (up to machine 
precision) the conservation of the total number of particles, as will be shown 
in Subsec.~\ref{sec:num_sch:n}. For definiteness, we shall refer to the 
formulation based on the change of variables in the spatial derivative 
given by Eq.~\eqref{eq:chi_def} as the $\chi$ formulation.

For the flows considered in this paper, Eqs.~\eqref{eq:boltz_cons} and \eqref{eq:boltz_noncons} 
can be put in the form:
\begin{equation}
 \frac{\partial F}{\partial t} + \sum_{\wi} \frac{\partial (V^{\wi} F)}{\partial \chi^\wi} = S,
 \label{eq:numsch_eq}
\end{equation}
where the source term $S$ contains the external and inertial forces 
(involving the momentum derivatives of $f$) and the collision term. 
In the $\widetilde{f}$ formulation \eqref{eq:boltz_cons}, 
$F = \wf \equiv f \sqrt{g}$ 
and $\chi^\wi = x^\wi$ is the coordinate on direction $\wi$.
In the $\chi$ formulation \eqref{eq:boltz_noncons},
$F = f$ and $\chi^{\wi}$ is defined in 
Eq.~\eqref{eq:chi_def}.
The advection velocity $V^\wi$ is in general point dependent
and is given in the $\widetilde{f}$ formulation by 
$V^\wi = \frac{p^\hata}{m} e_\hata^\wi$, while in the $\chi$
formulation, it has the expression $V^\wi = \sqrt{g_\wi} \, \frac{p^\hata}{m} e_\hata^\wi$.

\subsection{Time-stepping}\label{sec:num_sch:time}

\begin{table}
\caption{Butcher tableau for the third-order Runge-Kutta 
time-stepping procedure described in Eq.~\eqref{eq:rk3}.\label{tab:rk3}}
\begin{ruledtabular}
\begin{tabular}{r|rrr}
0 & & & \\
1 & 1 & & \\
1/2 & 1/4 & 1/4 & \\\hline
& 1/6 & 1/6 & 2/3
\end{tabular}
\end{ruledtabular}
\end{table}

Equation \eqref{eq:numsch_eq} can be put in the following form
\begin{equation}
 \partial_t F = L[F],\label{eq:L_def}
\end{equation}
where $L[F]$ is an integro-differential operator with respect 
to the spatial and momentum space coordinates acting on $F$.
Let us consider an equidistant discretization of the time variable, 
such that at step $\ell$, the 
value of the time coordinate is $t_\ell = \ell \delta t$ (we assume that 
$t_0 = 0$ is the initial time).
If $F_\ell \equiv F(t_\ell)$ at time $t = t_\ell$ is known, its value at 
$t_{\ell+1} = t_\ell + \delta t$ can be obtained using the 
third-order total variation diminishing (TVD) Runge-Kutta method described in
Refs.~\cite{shu88,gottlieb98,henrick05,trangenstein07,rezzolla13,ambrus18prc}:

\begin{align}
 F_\ell^{(1)} =& F_\ell + \delta t \, L[F_\ell], \nonumber\\
 F_\ell^{(2)} =& \frac{3}{4} F_\ell + \frac{1}{4} F_\ell^{(1)} + 
 \frac{1}{4} \delta t\, L[F_\ell^{(1)}],\nonumber\\
 F_{\ell+1} =& \frac{1}{3} F_\ell + \frac{2}{3} F_\ell^{(2)} + 
 \frac{2}{3} \delta t\, L[F_\ell^{(2)}]. \label{eq:rk3}
\end{align}

The Butcher tableau \cite{butcher08} corresponding to this scheme is given in 
Table~\ref{tab:rk3}.

\subsection{Coordinate stretching}\label{sec:num_sch:stretch}

\begin{figure}
\begin{center}
\begin{tabular}{c}
\includegraphics[width=0.45\textwidth]{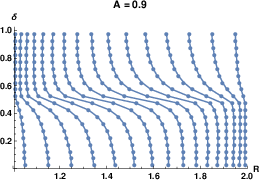}\\
(a)\\
\includegraphics[width=0.45\textwidth]{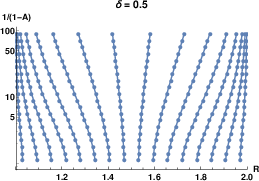}\\
(b)
\end{tabular}
\end{center}
\caption{Effect of grid stretching on 16 points between $R_{\rm in} = 1$ and $R_{\rm out} = 2$.
(a) The parameter $\delta$ controls the positioning of the stretching center (i.e. the 
point where the grid is the coarsest).
(b) The parameter $A$ contains the amplitude of the stretching, with $A = 0$ and 
$A = 1$ corresponding to equidistant and infinitely-stretched points, respectively.
}
\label{fig:stretch}
\end{figure}

As pointed out in Refs.~\cite{mei98,guo03}, the correct recovery of the Knudsen layer 
in wall-bounded flows requires a substantially finer mesh near the walls than in the 
bulk of the channel. This can be efficiently 
achieved by performing a coordinate stretching such that the resulting grid 
is finer near the boundaries and coarser in the interior of the
channel. Assuming that the walls are orthogonal to the $x^{\widetilde{1}}$ direction, 
we consider the following coordinate transformation:
\begin{equation}
 x^{\widetilde{1}}(\eta) = x^{\widetilde{1}}_{\rm left} + 
 (x^{\widetilde{1}}_{\rm right} - x^{\widetilde{1}}_{\rm left})
 \left(\delta + \frac{A_0}{A} \tanh\eta\right),
 \label{eq:eta_def}
\end{equation}
where $x^{\widetilde{1}}_{\rm left}$ and $x^{\widetilde{1}}_{\rm right}$ 
are the coordinates of the left and right domain boundaries, respectively.
The constants $\delta$ and $A$ are free parameters, while $A_0$ is chosen as:
\begin{equation}
 A_0 = {\rm max}(\delta, 1 - \delta).
\end{equation}
The above definition of $A_0$ allows the range of $\delta$ to be 
$\delta \in [0,1]$, while $A \in (0,1)$.  As illustrated in 
Fig.~\ref{fig:stretch}(a), the parameter $\delta$ controls the 
position of the stretching center (i.e. when $\eta = 0$), such 
that when $\delta = 0$ and $1$, the coarsest region 
is near the left and right boundary, respectively. 

The parameter $A$ controls the grid stretching, such that as $A \rightarrow 0$, the 
grid becomes equidistant, while when $A \rightarrow 1$, the grid becomes 
infinitely stretched near the stretching center at 
$x^{\widetilde{1}} = x^{\widetilde{1}}_{\rm left} (1 - \delta) 
+ x^{\widetilde{1}}_{\rm right} \delta$.
This is illustrated in Fig.~\ref{fig:stretch}(b).

The range of $\eta$ is $\eta \in [\eta_{\rm left}, \eta_{\rm right}]$, 
where $\eta_{\rm left}$ and $\eta_{\rm right}$ can be found by setting 
$x^{\widetilde{1}} = x^{\widetilde{1}}_{\rm left}$ and 
$x^{\widetilde{1}} = x^{\widetilde{1}}_{\rm right}$ in Eq.~\eqref{eq:eta_def}:
\begin{equation}
 \eta_{\rm left} = -\arctanh\frac{A \delta}{A_0}, \qquad 
 \eta_{\rm right} = \arctanh\frac{A(1 -\delta)}{A_0}.
 \label{eq:eta_inout}
\end{equation}
In the special case when $\delta = 0.5$, the range of $\eta$ is 
$\eta \in [-\arctanh A, \arctanh A]$, since $A_0 = 0.5$.

In the current formulation, the grid stretching is a coordinate transformation 
which changes the line element \eqref{eq:ds2}. In particular, the Boltzmann 
equation can be re-derived with respect to the stretched coordinate $\eta$ and 
its associated momentum $p^\heta$ and a different conservative formulation 
is obtained compared to the case when the grid is not stretched. This 
will be further discussed in the context of the circular Couette flow in 
Sec.~\ref{sec:couette}.

\subsection{Implementation of advection}\label{sec:num_sch:weno}

\begin{table}
\caption{Limiting values for the weighting factors $\overline{\omega}_q$ \eqref{eq:weno5_omega} 
employed in the computation of the WENO-5 flux \eqref{eq:weno5_flux}.\label{tab:weno}}
\begin{ruledtabular}
\begin{tabular}{r|rrr}
 & $\overline{\omega}_1$ & $\overline{\omega}_2$ & $\overline{\omega}_3$ \\\hline
$\sigma_1 = \sigma_2 = \sigma_3$ & $0.1$ & $0.6$ & $0.3$ \\\hline
$\sigma_2 = \sigma_3 = 0$ & $0$ & $2/3$ & $1/3$ \\
$\sigma_3 = \sigma_1 = 0$ & $1/4$ & $0$ & $3/4$ \\
$\sigma_1 = \sigma_2 = 0$ & $1/7$ & $6/7$ & $0$ \\\hline
$\sigma_1 = 0$ & $1$ & $0$ & $0$\\
$\sigma_2 = 0$ & $0$ & $1$ & $0$\\
$\sigma_3 = 0$ & $0$ & $0$ & $1$
\end{tabular}
\end{ruledtabular}
\end{table}

The examples considered in this paper are either one-dimensional (the circular 
Couette flow discussed in Sec.~\ref{sec:couette}) or two-dimensional (the 
gradually expanding channel discussed in Sec.~\ref{sec:canalie}), hence the 
flow can always be assumed to be homogeneous with respect to the $z$ axis 
(we will take advantage of this simplification in Sec.~\ref{sec:LB:red}, where 
the $z$ degree of freedom of the momentum space will be eliminated by 
introducing reduced distribution functions).
The simulation domain is thus divided into 
$N_{\widetilde{1}} \times N_{\widetilde{2}}$ cells centered on 
$\bm{x}_{s,p} = (x^{\widetilde{1}}_s, x^{\widetilde{2}}_p)$ 
($1 \le s \le N_{\widetilde{1}}$, $1 \le p \le N_{\widetilde{2}}$).
Each cell $(s,p)$ has four interfaces, located at 
$\bm{x}_{s+1/2,p}$, $\bm{x}_{s-1/2,p}$, 
$\bm{x}_{s,p+1/2}$ and $\bm{x}_{s,p-1/2}$.
The domain boundary consists of the outer interfaces of the outer cells, 
having coordinates $\bm{x}_{{\rm left};p} = \bm{x}_{1/2,p}$, 
$\bm{x}_{{\rm right};p} = \bm{x}_{N_{\widetilde{1}}+1/2,p}$, 
$\bm{x}_{{\rm bottom};s} = \bm{x}_{s,1/2}$ and
$\bm{x}_{{\rm top};s} = \bm{x}_{s,N_{\widetilde{2}}+1/2}$.
With this notation, the advection part of Eq.~\eqref{eq:numsch_eq} 
can be written as follows:
\begin{multline}
 \sum_{\wi} \left(\frac{\partial (V^\wi F)}{\partial \chi^\wi}\right)_{s,p} \simeq \\
 \frac{V^{\widetilde{1}}_{s+1/2,p} \mathcal{F}_{\,\widetilde{1};s+1/2,p} - 
 V^{\widetilde{1}}_{s-1/2,p} \mathcal{F}_{\,\widetilde{1};s-1/2,p}}
 {\chi^{\widetilde{1}}_{s+1/2} - \chi^{\widetilde{1}}_{s-1/2}} \\+ 
 \frac{V^{\widetilde{2}}_{s,p+1/2} \mathcal{F}_{\,\widetilde{2};s+1/2,p} -
 V^{\widetilde{2}}_{s,p-1/2} \mathcal{F}_{\,\widetilde{2};s,p-1/2}}
 {\chi^{\widetilde{2}}_{p+1/2} - \chi^{\widetilde{2}}_{p-1/2}},
 \label{eq:flux_def}
\end{multline}
where directional splitting was applied, i.e. 
the advection along each direction $x^\wi$ is performed independently.
The quantities bearing the indices $s+1/2,p$ are evaluated at 
the interfaces between cells $(s+1,p)$ and $(s,p)$, etc. The fluxes 
$\mathcal{F}_{\,\widetilde{1};s\pm 1/2,p}$ correspond to the advection of 
$F$ along $V^{\widetilde{1}}_{s \pm 1/2,p}$ with respect to 
the coordinate $\chi^{\widetilde{1}}$, while 
the fluxes $\mathcal{F}_{\,\widetilde{2};s, p\pm 1/2}$ correspond to the advection of 
$F$ along $V^{\widetilde{2}}_{s; p \pm 1/2}$ with respect to the coordinate $\chi^{\widetilde{2}}$.
These fluxes are calculated using the fifth-order weighted essentially 
non-oscillatory (WENO-5) scheme 
\cite{jiang96,shu99,gan11,rezzolla13,ambrus18prc,hejranfar17pre}. We employ the WENO-5 
scheme as described in Ref.~\cite{ambrus18pre,ambrus18prc}, where the addition of a small 
quantity $\varepsilon$ in order to avoid division by $0$ operations is not required. 
For definiteness, we give below the procedure for constructing 
the flux $\mathcal{F}_{\widetilde{1}; s+1/2, p}$ for the case 
when $V^{\widetilde{1}}_{s+1/2, p} > 0$:
\begin{multline}
\mathcal{F}_{\,\widetilde{1};s+1/2, p} = 
\overline{\omega}_1\mathcal{F}^1_{\,\widetilde{1};s+1/2,p} \\
+\overline{\omega}_2\mathcal{F}^2_{\,\widetilde{1};s+1/2,p} + 
\overline{\omega}_3\mathcal{F}^3_{\,\widetilde{1};s+1/2,p},
\label{eq:weno5_flux}
\end{multline}
where the interpolating functions $\mathcal{F}^q_{\widetilde{1};s + 1/2,p}$ ($q = 1,2,3$) 
are given by:
\begin{align}
\mathcal{F}^1_{\,\widetilde{1};s+1/2,p} =& 
\frac{1}{3}F_{s-2,p} - \frac{7}{6} F_{s-1,p} + \frac{11}{6} F_{s,p}, \nonumber \\
\mathcal{F}^2_{\,\widetilde{1};s+1/2,p} =& 
-\frac{1}{6}F_{s-1,p} + \frac{5}{6} F_{s,p} + \frac{1}{3} F_{s+1,p}, \nonumber \\
\mathcal{F}^3_{\,\widetilde{1};s+1/2,p} =& 
\frac{1}{3}F_{s,p} + \frac{5}{6} F_{s+1,p} - \frac{1}{6} F_{s+2,p},
\end{align}
while the weighting factors $\overline{\omega}_q$ are defined as:
\begin{equation}
\overline{\omega}_q = \frac{\widetilde{\omega}_q}
{\widetilde{\omega}_1+\widetilde{\omega}_2+\widetilde{\omega}_3}, \qquad 
\widetilde{\omega}_q = \frac{\delta_q}{\sigma^2_q}.\label{eq:weno5_omega}
\end{equation}
The ideal weights $\delta_q$ are:
\begin{equation}
 \delta_1 = 1/10, \qquad \delta_2 = 6/10,\qquad \delta_3 = 3/10, 
\end{equation}
while the smoothness indicator $\sigma_q$ are given by:
\begin{align}
\sigma_1 =& \frac{13}{12} \left(F_{s-2,p} -2F_{s-1,p} + F_{s,p} \right)^2 \nonumber\\
& + \frac{1}{4} \left( F_{s-2,p} - 4F_{s-1,p} + 3F_{s,p} \right)^2, 
\nonumber \\
\sigma_2 =& \frac{13}{12} \left(F_{s-1,p} -2F_{s,p} + F_{s+1,p} \right)^2 \nonumber\\
& + \frac{1}{4} \left( F_{s-1,p} - F_{s+1,p} \right)^2,
\nonumber \\
\sigma_3 =& \frac{13}{12} \left(F_{s,p} -2F_{s+1,p} + F_{s+2,p} \right)^2 \nonumber\\
&+ \frac{1}{4} \left( 3F_{s,p} -4 F_{s+1,p} + F_{s+2,p} \right)^2.
\label{eq:weno_sigma}
\end{align}
It is customary to add in the denominators of $\widetilde{\omega}_q$ a small quantity 
$\varepsilon$ (usually taken 
as $10^{-6}$) to avoid division by $0$ operations. However, as pointed out in 
Ref.~\cite{henrick05}, the effect of 
this alteration on the smoothness indicators is strongly dependent on the given problem, since 
$\varepsilon$ becomes a dimensional quantity. Furthermore, the accuracy of the resulting 
scheme depends on the value 
of $\varepsilon$. Since at higher orders, the distribution functions corresponding to 
large velocities can 
have values which are significantly smaller than those for smaller velocities, we cannot 
predict the effect 
of employing a unitary value for $\varepsilon$ for the advection of all distribution functions. 
Therefore, we prefer to follow 
Refs.~\cite{ambrus18pre,ambrus18prc} and compute the limiting values of 
$\overline{\omega}_q$ when one, two or all three of the smoothness indicators 
vanish as indicated in Table~\ref{tab:weno}.

\subsection{Particle number conservation}\label{sec:num_sch:n}

The Boltzmann equation implies the fluid equations \eqref{eq:macro}, 
which ensure that the total number of particles $N_{\rm tot}$ \eqref{eq:Ntot_def}
per unit length, the total momentum $\mathcal{P}$ and 
the total energy $\mathcal{E}$ are conserved within the fluid.
However, the gas-wall interaction can induce changes in these parameters. 
In this paper, we will consider diffuse-reflection boundary conditions for 
impermeable walls, such that $N_{\rm tot}$ is preserved at all times, while 
$\mathcal{P}$ and $\mathcal{E}$ are allowed to vary. Thus, in this Subsection, we 
will only consider the conservation of $N_{\rm tot}$. 

After the discretization of space and time, 
the only changes that can be induced in $N_{\rm tot}(t)$ 
are due to the operator $L[F]$. In the following, the $\widetilde{f}$ and 
$\chi$ formulations will be treated separately. 

In the $\widetilde{f}$ formulation, $F = \wf = f\sqrt{g}$ and the time evolution 
of $N_{\rm tot}$ \eqref{eq:Ntot_def} can be obtained by integrating 
Eq.~\eqref{eq:L_def} with respect to the momentum space and 
over the entire fluid domain:
\begin{equation}
 \partial_t N_{\rm tot}(t) = \int d^3\widetilde{x} 
 \int d^3\hat{p}\,L[\wf].\label{eq:dtotalNcons_aux} 
\end{equation}
The momentum space integral of the source term in Eq.~\eqref{eq:numsch_eq} 
vanishes, since $1$ is a collision invariant, while the zeroth-order moment 
of the force term is zero. For simplicity, an equidistant grid is considered,
such that Eq.~\eqref{eq:dtotalNcons_aux} reduces to:
\begin{equation}
 \partial_t N_{\rm tot}(t) = -\int d^3\hat{p} \int d^3\widetilde{x} \,
 \sum_{\wi} \frac{\partial (V^\wi f \sqrt{g})}{\partial x^\wi},\label{eq:dtotalNcons}
\end{equation}
where we took into account that $\chi^\wi = x^\wi$ in the $\widetilde{f}$ formulation.
The integration domain can be split into cells and the advection term, 
replaced via Eq.~\eqref{eq:flux_def}, can be considered constant 
within each cell, such that Eq.~\eqref{eq:dtotalNcons} becomes simply:
\begin{multline}
 \partial_t N_{\rm tot}(t) = -\delta z \int d^3\hat{p} \sum_{s = 1}^{N_{\widetilde{2}}}
 \sum_{p = 1}^{N_{\widetilde{2}}}
 \left[\delta x^{\widetilde{2}} (\widetilde{\mathcal{F}}_{\,\widetilde{1}; s + 1/2,p} \right.\\
 \left. - \widetilde{\mathcal{F}}_{\,\widetilde{1}; s - 1/2,p})
 + \delta x^{\widetilde{1}} (\widetilde{\mathcal{F}}_{\,\widetilde{2}; s,p+1/2} - 
 \widetilde{\mathcal{F}}_{\,\widetilde{2}; s,p-1/2})\right],
 \label{eq:dtotalN_cons}
\end{multline}
where $\delta z$ represents the height of the fluid domain and
the notations $\widetilde{\mathcal{F}}_{\widetilde{1};s+1/2,p}$ 
and $\widetilde{\mathcal{F}}_{\widetilde{2};s,p+1/2}$ indicate 
that the fluxes 
are computed by replacing $F_{s,p}$ with $\wf_{s,p} = f_{s,p} \sqrt{g_{s,p}}$ in 
Eq.~\eqref{eq:weno5_flux}.
The bulk terms cancel out and $\partial_t N_{\rm tot}(t)$ reduces to:
\begin{multline}
 \partial_t N_{\rm tot}(t) = -\delta z \int d^3\hat{p} \left[ 
 \sum_{p = 1}^{N_{\widetilde{2}}} 
 \delta x^{\widetilde{2}}
 (\widetilde{\mathcal{F}}_{\widetilde{1}; N_{\widetilde{1}} + 1/2,p} -
 \widetilde{\mathcal{F}}_{\widetilde{1};1/2,p})\right.\\ 
 \left.+\sum_{s = 1}^{N_{\widetilde{1}}} 
 \delta x^{\widetilde{1}}(\widetilde{\mathcal{F}}_{\widetilde{2}; s,N_{\widetilde{2}} + 1/2} -
 \widetilde{\mathcal{F}}_{\widetilde{2}; s,1/2})\right].
\end{multline}
Thus, the conservation of the total number of particles is conditioned by the requirement that 
the momentum-space integrals of the fluxes at the outer interfaces of the outer cells 
cancel. Ensuring that these momentum space integrals vanish is the subject of 
Subsec.~\ref{sec:num_sch:bound}, which is dedicated to the discussion of the implementation 
of the boundary conditions.

In the case of the $\chi$ approach, $F = f$ 
while $\sqrt{g}$ appears explicitly in \eqref{eq:dtotalNcons_aux}:
\begin{equation}
 \partial_t N_{\rm tot}(t) = \int d^3\widetilde{x}\, \sqrt{g} 
 \int d^3\hat{p}\,L[f].\label{eq:dtotalN_aux} 
\end{equation}
As before, the momentum space integral of the source term vanishes and 
the only contributions to $\partial_t N_{\rm tot}(t)$ come from the 
advection part of $L[f]$. Treating again the advection terms as constants 
over the domain cells, the integral of $\sqrt{g}$ can be performed over 
each cell by keeping in mind the definition of $\chi^\wi$ \eqref{eq:chi_def}, 
such that:
\begin{align}
 &\int_{(s,p)} d^3\widetilde{x} \, \sqrt{g} 
 \frac{V^{\widetilde{1}}_{s+1/2,p} \mathcal{F}_{\,\widetilde{1}; s+1/2,p} -
 V^{\widetilde{1}}_{s-1/2,p} \mathcal{F}_{\,\widetilde{1}; s-1/2,p}}
 {\delta \chi^{\widetilde{1}}_{s}}\nonumber\\
 &\quad = \delta z \delta\chi^{\widetilde{2}}_{p}
 (V^{\widetilde{1}}_{s+1/2,p} \mathcal{F}_{\,\widetilde{1}; s+1/2,p} -
 V^{\widetilde{1}}_{s-1/2,p} \mathcal{F}_{\,\widetilde{1}; s-1/2,p}), \nonumber\\
 &\int_{(s,p)} d^3\widetilde{x}\, \sqrt{g} 
 \frac{V^{\widetilde{2}}_{s,p+1/2} \mathcal{F}_{\,\widetilde{2}; s,p+1/2} -
 V^{\widetilde{2}}_{s,p-1/2} \mathcal{F}_{\,\widetilde{2}; s,p-1/2}}
 {\delta\chi^{\widetilde{2}}_{p}}\nonumber\\
 &\quad = \delta z \delta \chi^{\widetilde{1}}_{s} 
 (V^{\widetilde{2}}_{s,p+1/2} \mathcal{F}_{\,\widetilde{2}; s,p+1/2} -
 V^{\widetilde{2}}_{s,p-1/2} \mathcal{F}_{\,\widetilde{2}; s,p-1/2}),
\end{align}
where $\delta \chi^{\widetilde{1}}_{s} = 
\chi^{\widetilde{1}}_{s+1/2} - \chi^{\widetilde{1}}_{s-1/2}$ and 
$\delta \chi^{\widetilde{2}}_{p} = 
\chi^{\widetilde{2}}_{p+1/2} - \chi^{\widetilde{2}}_{p-1/2}$.
The bulk terms again cancel and Eq.~\eqref{eq:dtotalN_aux} becomes:
\begin{multline}
 \partial_t N_{\rm tot}(t) = -\delta z \int d^3\hat{p} \\
 \left[ 
 \sum_{p = 1}^{N_{\widetilde{2}}} \delta \chi^{\widetilde{2}}_{p}
 (V^{\widetilde{1}}_{N_{\widetilde{1}} + 1/2,p}
 \mathcal{F}_{\widetilde{1}; N_{\widetilde{1}} + 1/2,p} -
 V^{\widetilde{1}}_{1/2,p}
 \mathcal{F}_{\widetilde{1};1/2,p})\right.\\ 
 \left.+\sum_{s = 1}^{N_{\widetilde{1}}} \delta \chi^{\widetilde{1}}_{s}
 (V^{\widetilde{2}}_{s,N_{\widetilde{2}} + 1/2}
 \mathcal{F}_{\widetilde{2}; s,N_{\widetilde{2}} + 1/2} -
 V^{\widetilde{2}}_{s,1/2}
 \mathcal{F}_{\widetilde{2}; s,1/2})\right].
\end{multline}
As in the $\widetilde{f}$ formulation, the conservation of the total number of 
particles relies on the exact cancellation of the numerical fluxes 
through the outer interfaces of the outer cells of the fluid domain.

\subsection{Order of advection scheme} \label{sec:num_sch:order}

Let us now discuss the order of our proposed scheme. For definiteness, 
the advection along the $x^{\widetilde{1}}$ direction is considered 
and for brevity, only the coordinate index along this direction is 
displayed. In particular, we are interested in deriving the accuracy 
of the approximation of the quantity:
\begin{equation}
 \frac{\partial (V F)}{\partial x} = \sqrt{g} \frac{\partial (V F)}{\partial \chi}.
 \label{eq:numsch_err_1}
\end{equation}
In our implementation, $\sqrt{g}$ is replaced by its cell average
\begin{equation}
 \sqrt{g}_s \simeq \frac{\chi_{s+1/2} - \chi_{s-1/2}}{\delta s},
\end{equation}
where $\delta s$ is the equidistant spacing on the $x$ direction 
(in the case of the equidistant grid, $\delta s = \delta x$, 
while for the stretched grid, $\delta s = \delta \eta$).
The derivative with respect to $\chi$ is approximated according to 
\eqref{eq:flux_def}, such that Eq.~\eqref{eq:numsch_err_1}
becomes:
\begin{equation}
 \left(\frac{\partial (V F)}{\partial x}\right)_s \simeq 
 \frac{V_{s+1/2} \mathcal{F}_{s+1/2} - V_{s-1/2} \mathcal{F}_{s-1/2}}{\delta s}.
 \label{seC:numsch_err_2}
\end{equation}
The right hand side of the above relation can be expanded with 
respect to $x = x_s$ as follows:
\begin{multline}
 \frac{V_{s+1/2} \mathcal{F}_{s+1/2} - V_{s-1/2} \mathcal{F}_{s-1/2}}{\delta s} \\
 \simeq \frac{\mathcal{F}_{s+1/2} - \mathcal{F}_{s-1/2}}{\delta s} 
 \left[V_{s} + \frac{(\delta s)^2}{8} 
 \left(\frac{\partial^2 V}{\partial x^2}\right)_{s} 
 + \dots \right] \\
 + \frac{\mathcal{F}_{s+1/2} + \mathcal{F}_{s-1/2}}{2}
 \left[\left(\frac{\partial V}{\partial x}\right)_{s} + 
 \frac{(\delta s)^2}{24} \left(\frac{\partial^3 V}{\partial x^3}\right)_{s} + \dots\right],
 \label{eq:numsch_err_aux}
\end{multline}
When $V$ is a constant, the error term is that of the scheme used to compute the 
fluxes, which ensures that 
$\frac{1}{\delta s}(\mathcal{F}_{s+1/2} - \mathcal{F}_{s-1/2}) = (\partial_x F)_s + 
O[(\delta s)^n]$, where $n$ is the order of accuracy of the scheme for Cartesian 
coordinates. In the case when $V$ depends on the coordinate, there are second order 
errors which are unavoidable in this construction. In the case when the WENO-5 procedure 
is employed to compute the fluxes $\mathcal{F}_{s+1/2}$, Eq.~\eqref{eq:numsch_err_aux}
reduces to:
\begin{multline}
 \frac{V_{s+1/2} \mathcal{F}_{s+1/2} - V_{s-1/2} \mathcal{F}_{s-1/2}}{\delta s}
 \simeq \left(\frac{\partial(V F)}{\partial x}\right)_s \\ 
 + \frac{(\delta s)^2}{24} \left\{\frac{\partial}{\partial x} \left[ 
 2\frac{\partial V}{\partial x} \frac{\partial f}{\partial x} + 
 f \frac{\partial^2 V}{\partial x^2}\right]\right\}_s + O[(\delta s)^4].\nonumber
\end{multline}
Even though the resulting implementation presents errors which are second 
order with respect to $\delta s$, we find the implementation of the numerical 
fluxes using the WENO-5 algorithm to be more accurate than when using second order 
schemes, such as the flux limiters scheme \cite{leveque02,sofonea04,trangenstein07}.

\subsection{Diffuse reflection boundary conditions}\label{sec:num_sch:bound}

In the case of diffuse reflection, the flux of particles returning 
into the fluid domain through the cell interfaces between the fluid 
and the walls follow Maxwellian distributions. In the flows considered in this paper, 
the walls are always perpendicular to the direction corresponding to the first coordinate
$x^{\widetilde{1}}$.
For definiteness, let us consider the case of the 
left boundary, for which the above condition reads:
\begin{equation}
 \mathcal{F}_{\,\widetilde{1};1/2,p} = \feq(n_{\rm left}, \vu_{\rm left}, T_{\rm left}) 
 \hfill (V^{\widetilde{1}}_{1/2,p} > 0).
 \label{eq:weno_flux_eq}
\end{equation}
We note that Eq.~\eqref{eq:weno_flux_eq} holds in both the $\widetilde{f}$
and in the $\chi$ formulations, since the $\sqrt{g}$ factor which multiplies the 
distribution function in the $\widetilde{f}$ approach ($\wf = f\sqrt{g}$)
can easily be absorbed into the unknown wall particle number density $n_{\rm left}$. 

The flux in Eq.~\eqref{eq:weno_flux_eq} can be easily achieved analytically by populating the 
ghost nodes at $s = -2$, $-1$ and $0$ according to
($V^{\widetilde{1}}_{1/2,p} > 0$):
\begin{equation}
 F_{-2,p} = F_{-1,p} = F_{0,p} = \feq(n_{\rm left}, \vu_{\rm left}, T_{\rm left}).
 \label{eq:weno_populate} 
\end{equation}
With the above definitions, Eq.~\eqref{eq:weno_sigma} shows that $\sigma_1 = 0$ for $s = 0$.
According to Table~\ref{tab:weno}, $\overline{\omega}_1 = 1$ and 
$\overline{\omega}_2 = \overline{\omega}_3 = 0$
when $\sigma_1 = 0$. Thus, Eq.~\eqref{eq:weno5_flux} implies that 
($V^{\widetilde{1}}_{1/2,p} > 0$):
\begin{equation}
 \mathcal{F}_{\,\widetilde{1};1/2,p} = \mathcal{F}_{\,\widetilde{1};1/2,p}^1 = 
 F_{0,p}.
\end{equation}

In order to calculate the fluxes at $s = 1/2$ and $s = 3/2$ for particles traveling 
towards the wall ($V^{\widetilde{1}}_{1/2,p} < 0$), the populations in the ghost nodes 
at $s = 0$ and $s = -1$ are obtained using a quadratic extrapolation:
\begin{align}
 F_{0,p} =& 3 F_{1,p} - 3 F_{2,p} + F_{3,p}, \nonumber\\
 F_{-1,p} =& 6 F_{1,p} - 8 F_{2,p} + 3F_{3,p}.\label{eq:extrapol}
\end{align}

Finally, mass conservation is ensured by requiring that:
\begin{equation}
 \int d^3\hat{p} \mathcal{F}_{\,\widetilde{1};1/2,p} V^{\widetilde{1}}_{1/2,p} = 0.
\end{equation}
This translates into the following equation for $n_w$:
\begin{equation}
 n_w = - \frac{\int_{V^{\widetilde{1}}_{1/2,p} < 0} d^3\hat{p} 
 \mathcal{F}_{\,\widetilde{1};1/2,p} V^{\widetilde{1}}_{1/2,p} } 
 {\int_{V^{\widetilde{1}}_{1/2,p} > 0} d^3\hat{p} 
 \feq(n = 1, \vu_{\rm left}, T_{\rm left}) V^{\widetilde{1}}_{1/2,p}}.
 \label{eq:nw}
\end{equation}

\section{Mixed quadrature LB models}\label{sec:LB}

In this Section, the construction of mixed quadrature LB models for flows in curvilinear 
geometries will be discussed. Since the flows considered in this paper are homogeneous 
with respect to the $z$ axis, the momentum degree of freedom along this axis can be integrated
out, giving rise to the reduced Boltzmann equations which will be discussed in 
Subsec.~\ref{sec:LB:red}.
The choice of quadrature for the two remaining directions 
is discussed in Subsec.~\ref{sec:LB:quad}. The implementation of the 
inertial forces arising due to the formulation of the Boltzmann equation 
with respect to triads is discussed in Subsec.~\ref{sec:LB:force}.

\subsection{Reduced Boltzmann equation}\label{sec:LB:red}

The flows considered in this paper are homogeneous with respect to the $z$ axis. Hence, 
it is convenient to define the following reduced distribution functions:
\begin{align}
 f' =& \int_{-\infty}^\infty dp^\hatz\, f, \nonumber\\
 f'' =& \int_{-\infty}^\infty dp^\hatz\, \frac{(p^\hatz)^2}{m} f.
 \label{eq:fred}
\end{align}
With the aid of these two reduced distributions, the macroscopic 
fields \eqref{eq:macro_def} can be written as:
\begin{subequations}\label{eq:macro_red}
\begin{align}
 n =& \int d^2\hat{p} \, f', \label{eq:macro_red_n}\\
 u^\hata =& \frac{1}{\rho} \int d^2\hat{p} \,p^\hata\, f', \label{eq:macro_red_u}\\ 
 T^{\hata\hatb} =& \int d^2\hat{p} \, \frac{\xi^\hata \xi^\hatb}{m} f', \label{eq:macro_red_Tab}\\ 
 q^\hata =& \int d^2\hat{p} \, \left(\frac{\vxi^2}{2m} f' + 
 \frac{1}{2}f''\right)\frac{\xi^\hata}{m},\label{eq:macro_red_q}
\end{align}
\end{subequations}
where the indices $\hata$, $\hatb \in \{\hat{1}, \hat{2}\}$. Moreover, the temperature 
is defined as:
\begin{equation}
 \frac{3}{2}n T = \int d^2\hat{p} \left(\frac{\vxi^2}{2m} f' + \frac{1}{2} f''\right).
\end{equation}
Thus, the function $f''$ appears only in the definitions of the temperature 
$T$ and heat flux $q^\hata$. 

\subsection{Choice of quadrature}\label{sec:LB:quad}

We perform the numerical simulations presented in this paper 
using the mixed quadrature lattice Boltzmann models introduced in 
Refs.~\cite{ambrus16jcp,ambrus16jocs,ambrus17arxiv}.
Depending on the flow regime under consideration, 
a mixture of the full-range Gauss-Hermite and half-range 
Gauss-Hermite quadratures can be employed. 

For definiteness, let us consider the case when the half-range Gauss-Hermite quadrature
of order $Q_1$ is employed along the first coordinate direction, while the full-range 
Gauss-Hermite quadrature of order $Q_2$ is employed along the second coordinate 
direction. Following the notation introduced in Refs.~\cite{ambrus16jcp,ambrus17arxiv},
this model can be denoted using:
\begin{equation}
 {\rm HH}(\mathcal{N}_1; Q_1) \times 
 {\rm H(\mathcal{N}_2; Q_2)} \label{eq:model}
\end{equation}
where $\mathcal{N}_a$ represents the order of the expansion 
of the equilibrium distribution $\feq$ with 
respect to axis $a$, as will be discussed in Sec.~\ref{sec:LB:feq}.

The choice of quadrature controls the discretization of the momentum space, as well as the 
momentum space integration. In particular, the moments \eqref{eq:mom} are evaluated as:
\begin{equation}
 M^{\hata_1, \dots \hata_s} = \sum_{i = 1}^{\mathcal{Q}_1} \sum_{j = 1}^{\mathcal{Q}_2} 
 f'_{ij} \prod_{\ell = 1}^s p^{\hata_\ell}_{ij}.
 \label{eq:mom_quad}
\end{equation}
A similar prescription holds for the macroscopic quantities appearing in 
Eq.~\eqref{eq:macro_red}.
The total number of quadrature points on axis $a$ is $\mathcal{Q}_a = Q_a$ 
for the full-range Gauss-Hermite quadrature and $\mathcal{Q}_a = 2Q_a$ for the 
half-range Gauss-Hermite quadrature.
In particular, $\mathcal{Q}_1 = 2Q_1$ and $\mathcal{Q}_2 = Q_2$ 
for the example considered in Eq.~\eqref{eq:model}.

The components of $\vp_{ij} = \{p^{\hat{1}}_i, p^{\hat{2}}_j\}$ are 
indexed on each direction separately, where $1 \le i \le \mathcal{Q}_1$ and 
$1 \le j \le \mathcal{Q}_2$.
For the half-range Gauss-Hermite quadrature, we use the convention that 
the points with $1\le i \le Q_1$ lie on the positive semi-axis of the radial 
direction, being given as the roots of the half-range Hermite polynomial 
$\hh_{Q_1}(x)$ of order $Q_1$:
\begin{equation}
 \hh_{Q_1}(p^{\hat{1}}_i) = 0, \qquad (1 \le i \le Q_1),\label{eq:hh_roots}
\end{equation}
while $p_{Q_1 + i}^{\hat{1}} = -p_i^{\hat{1}}$ ($1 \le i \le Q_1$).
On the direction where the full-range Gauss-Hermite quadrature is applied, 
the quadrature points are chosen as the roots of the Hermite polynomial 
$H_{Q_2}(x)$ of order $Q_2$:
\begin{equation}
 H_{Q_2}(p^{\hat{2}}_j) = 0.
\end{equation}

The link between $f'_{ij}$ and $f''_{ij}$ and the 
reduced Boltzmann distribution functions $f'(p^{\hat{1}}, p^{\hat{2}})$  and
$f''(p^{\hat{1}}, p^{\hat{2}})$ is given through:
\begin{equation}
 \begin{pmatrix}
  f'_{ij} \\ f''_{ij}
 \end{pmatrix}
 = \frac{w_i^{\hh}(Q_1) w_j^H(Q_2)}
 {\omega(p^{\hat{1}}_i) \omega(p^{\hat{2}}_j)}
 \begin{pmatrix}
  f'(p^{\hat{1}}_i, p^{\hat{2}}_j) \\ 
  f''(p^{\hat{1}}_i, p^{\hat{2}}_j)
 \end{pmatrix},
\end{equation}
where the weight function $\omega(x)$ for the half-range and full-range Hermite 
polynomials is:
\begin{equation}
 \omega(x) = \frac{1}{\sqrt{2\pi}} e^{-x^2/2}.
\end{equation}

The quadrature weights $w_j^H(Q_2)$ for the full-range Gauss-Hermite quadrature 
of order $Q_2$ are \cite{ambrus16jocs}:
\begin{equation}
 w_j^H(Q_2) = \frac{Q_2!}{H_{Q_2 + 1}^2(p^{\hat{2}}_j)}.
 \label{eq:w_H}
\end{equation}

The quadrature weights $w_i^\hh(Q_1)$ for the half-range Gauss-Hermite quadrature of 
order $Q_1$ are \cite{ambrus16jcp,ambrus16jocs}:
\begin{equation}
 w_i^\hh(Q_1) = \frac{p^{\hat{1}}_i a_{Q_1 - 1}^2}
 {\hh_{Q_1 - 1}^2(p^{\hat{1}}_i) [p^{\hat{1}}_i + \hh_{Q_1}^2(0) / \sqrt{2\pi}]},
 \label{eq:w_hh}
\end{equation}
where 
\begin{equation}
 a_\ell = \frac{\hh_{\ell+1,\ell+1}}{\hh_{\ell,\ell}}
 \label{eq:hh_a}
\end{equation}
is written in terms of the coefficients $\hh_{\ell,s}$ of $x^s$ 
in the polynomial expansion of $\hh_\ell(x)$:
\begin{equation}
 \hh_\ell(x) = \sum_{s = 0}^\ell \hh_{\ell,s} x^s. \label{eq:hh_exp}
\end{equation}

\subsection{Force terms}\label{sec:LB:force}

Since the functional dependence of the distribution function on the components of the momentum 
is removed through the discretization of the momentum space, an appropriate method for the 
computation of the momentum derivative of the distribution function must be employed.
Discrete velocity models (DVMs) usually rely on finite difference
techniques to perform 
the momentum space derivatives \cite{mieussens00,aoki03}. 
In this paper, we take the lattice Boltzmann 
approach introduced in Ref.~\cite{shan06}, according to which the momentum 
space derivative is projected on the space of orthogonal Hermite polynomials. 
More precisely, we follow Ref.~\cite{ambrus17arxiv} and write the terms involving 
the momentum derivatives of $f'$ and $f''$ as follows:
\begin{align}
 \left[\frac{\partial}{\partial p^{\hat{1}}}
 \begin{pmatrix}
  f' \\ f''
 \end{pmatrix} \right]_{ij} =& 
 \sum_{i' = 1}^{\mathcal{Q}_1} \mathcal{K}^{\hat{1}}_{i,i'} 
 \begin{pmatrix}
  f'_{i',j} \\ f''_{i',j}
 \end{pmatrix}, \nonumber\\
 \left[\frac{\partial}{\partial p^{\hat{1}}}
 \begin{pmatrix}
  p^{\hat{1}} f' \\ p^{\hat{1}} f''
 \end{pmatrix} \right]_{ij} =& 
 \sum_{i' = 1}^{\mathcal{Q}_1} \widetilde{\mathcal{K}}^{\hat{1}}_{i,i'} 
 \begin{pmatrix}
  f'_{i',j} \\ f''_{i',j}
 \end{pmatrix},
\end{align}
and similarly for the derivatives with respect to $p^{\hat{2}}$.

In the case of the full-range Gauss-Hermite quadrature, the matrix 
$\mathcal{K}^{\hata}_{k,k'}$ has the following form \cite{ambrus17arxiv}:
\begin{equation}
 \mathcal{K}^{\hata,H}_{k,k'} = -w_k^H(Q_a) \sum_{\ell = 0}^{Q_a - 1} \frac{1}{\ell!} 
 H_{\ell + 1}(p^{\hata}_k) H_\ell(p^{\hata}_{k'}),
 \label{eq:df_K_H}
\end{equation}
while in the case of the half-range Gauss-Hermite quadrature, it is given by 
\cite{ambrus17arxiv}:
\begin{multline}
 \mathcal{K}^{\hata,\hh}_{k,k'} =  w_k^\hh(Q_a) \sigma_k^\hata \Bigg\{
 \frac{1 + \sigma_k^\hata \sigma_{k'}^\hata}{2} \sum_{\ell = 0}^{Q_a - 2} 
 \hh_\ell(\abs{p_{k'}^\hata}) \\ \times \left[
 \frac{\hh_{\ell,0}}{\sqrt{2\pi}} \sum_{s = \ell + 1}^{Q_a - 1} \hh_{s,0} \hh_s(\abs{p_k^\hata})
 - \frac{\hh_{\ell, \ell}}{\hh_{\ell+1,\ell+1}} \hh_{\ell + 1}(\abs{p_k^\hata})\right]\\
 - \frac{1}{2\sqrt{2\pi}} \Phi^{Q_a}_0(\abs{p_k^\hata}) \Phi^{Q_a}_0(\abs{p_{k'}^\hata})\Bigg\}.
 \label{eq:df_K_hh}
\end{multline}
In the above, $\sigma_k^\hata$ and $\sigma_{k'}^\hata$ are the signs of $p_k^\hata$ and 
$p_{k'}^\hata$, respectively, having values 
$\sigma_k^\hata = 1$ for $1 \le k \le Q_a$ and $\sigma_k^\hata = -1$ when $Q_a < k \le 2Q_a$.
The function $\Phi^n_s(x)$ is defined as follows \cite{ambrus17arxiv}:
\begin{equation}
 \Phi^n_s(x) = \sum_{\ell = s}^{n} \hh_{\ell,s} \hh_\ell(x).
 \label{eq:Phi_def}
\end{equation}

The details regarding the expansions of $\partial (p^\hata f')/\partial p^\hata$
and $\partial (p^\hata f'')/\partial p^\hata$ with respect to the 
full-range and half-range Hermite polynomials are presented in 
Appendix~\ref{app:force}. Below we only quote the results. 
In the case when the full-range Gauss-Hermite quadrature is employed,
the matrix $\widetilde{\mathcal{K}}^{\hata}_{k,k'}$ reduces to:
\begin{equation}
 \widetilde{\mathcal{K}}^{\hata,H}_{k,k'} = -w_k^H(Q_a) \sum_{\ell = 0}^{Q_a - 2} \frac{1}{\ell!} 
 H_{\ell + 1}(p_k^\hata) [H_{\ell+1}(p_{k'}^\hata) + \ell H_{\ell - 1}(p_{k'}^\hata)].
 \label{eq:dfp_K_H}
\end{equation}
In the case of the half-range Gauss-Hermite quadrature, 
the kernel $\widetilde{\mathcal{K}}^{\hata,\hh}_{k,k'}$ is given by:
\begin{multline}
 \widetilde{\mathcal{K}}^{\hata,\hh}_{k,k'} = -w_k^\hh(Q_a) \frac{1 + \sigma_k^\hata
 \sigma_{k'}^\hata}{2}\sum_{\ell = 0}^{Q_a - 1} \hh_\ell(\abs{p_k^\hata})
 \Bigg[\ell\,\hh_\ell(\abs{p^\hata_{k'}}) \\
 + \frac{\hh_{\ell,0}^2 + \hh_{\ell-1,0}^2}{a_{\ell-1}\sqrt{2\pi}}
 \hh_{\ell-1}(\abs{p^\hata_{k'}}) + \frac{1}{a_{\ell-1} a_{\ell-2}} 
 \hh_{\ell-2}(\abs{p^\hata_{k'}})\Bigg],
 \label{eq:dfp_K_hh}
\end{multline}
where we use the convention that 
$\mathfrak{h}_{-1}(z) = \mathfrak{h}_{-2}(z) = 0$.

\subsection{Equilibrium distribution function}\label{sec:LB:feq}

We now present the construction of the equilibrium distribution function \eqref{eq:feq_triad}
appearing on the right hand side of Eq.~\eqref{eq:BGK}, as well 
as in the boundary conditions and in the initial state. 
After eliminating the $p^\hatz$ degree of freedom, $\feq$ is replaced by
\begin{align}
 f'_{\rm (eq)} =& \int_{-\infty}^\infty dp^\hatz\, \feq, \nonumber\\
 f''_{\rm (eq)} =& \int_{-\infty}^\infty dp^\hatz\, \frac{(p^\hatz)^2}{m} \feq 
 = T f'_{\rm (eq)}.
\end{align}

In discrete velocity models (DVMs), it is customary to evaluate the 
equilibrium distributions $f'_{\rm (eq)}$ and $f''_{\rm (eq)}$ directly, 
i.e. by computing the value of the Maxwellian for each given discrete 
momentum vector $\bm{p}_{ij}$ \cite{mieussens00,aoki03}.
On the other hand, the lattice Boltzmann (LB) approach is to replace the 
Maxwell-Boltzmann distribution with a polynomial approximation which 
ensures the exact recovery of its first few moments with a 
relatively small quadrature order.
Thus, in this paper, we take 
the LB approach and replace $f'_{\rm eq}$ and $f''_{\rm eq}$ with their 
polynomial approximations.

As discussed in Refs.~\cite{ambrus16jcp,ambrus16jocs}, $f'_{\rm (eq)}$ can be
factorized with respect to $p^{\hat{1}}$ and $p^{\hat{2}}$ as follows:
\begin{gather}
 f'_{\rm (eq)} = n \,g_1(p^{\hat{1}}) g_2(p^{\hat{2}}), \nonumber\\
 g_a (p^\hata) = \frac{1}{\sqrt{2\pi m T}} \exp\left[-\frac{(p^\hata - mu^\hata)^2}{2mT}\right].
 \label{eq:feq_red}
\end{gather}
Following the discretization of the momentum space, $f'_{\rm (eq)}$ is replaced by 
$f'_{{\rm (eq);} ij} = n\, g_{1,i} g_{2,j}$, while $f''_{{\rm (eq);} ij} = T f'_{{\rm (eq);} ij}$. 
For the case of the full-range Gauss-Hermite quadrature, the polynomial 
approximation of $g_{a,k}$ is \cite{ambrus16jcp,ambrus16jocs}:
\begin{equation}
 g_{a,k}^H = w_k^H(Q_a) \sum_{\ell = 0}^{\mathcal{N}_a} H_\ell(p^\hata_k) 
 \sum_{s = 0}^{\lfloor \ell / 2\rfloor} 
 \frac{(mT - 1)^s (mu^\hata)^{\ell - 2s}}{2^s s! (\ell -2s)!},
 \label{eq:eq_gak_H}
\end{equation}
where the expansion order $\mathcal{N}_a$ is a free parameter satisfying 
\begin{equation}
 0 \le \mathcal{N}_a < Q_a. \label{eq:norder_limit}
\end{equation}
An expansion of $g_{a,k}$ up to order $\mathcal{N}_a$ ensures the 
{\it exact} recovery of the moments 
\eqref{eq:momeq} for polynomials in $p^\hata$ of order less than or equal to $\mathcal{N}_a$.
In the case of the half-range Gauss-Hermite quadrature, the polynomial 
approximation of $g_{a,k}$ can be put in the following form \cite{ambrus16jcp,ambrus16jocs}:
\begin{multline}
 g_{a,k}^\hh = \frac{w_k^\hh(Q_a)}{2} \sum_{s = 0}^{\mathcal{N}_a} \left(\frac{mT}{2}\right)^{s/2} 
 \Phi_s^{\mathcal{N}_a}(\abs{p^\hata_k}) \\
 \times \left[(1 + \erf\zeta^\hata) P_s^+(\zeta^\hata) + 
 \frac{2}{\sqrt{\pi}} e^{-\zeta_{\hata}^2} P_s^*(\zeta^\hata)\right],
 \label{eq:eq_gak_hh}
\end{multline}
where $\Phi_s^{\mathcal{N}_a}$ is given in Eq.~\eqref{eq:Phi_def}, while
$\zeta^\hata = u^\hata \sqrt{m / 2T}$ when $p^\hata_k > 0$ and 
$\zeta^\hata = -u^\hata \sqrt{m / 2T}$ when $p^\hata_k < 0$. The polynomials 
$P_s^+(x)$ and $P_s^*(x)$ are defined as:
\begin{gather}
 P_s^\pm(x) = e^{\mp x^2} \frac{d^s}{dx^s} e^{\pm x^2}, \nonumber\\
 P_s^*(x) = \sum_{j = 0}^{s -1} \binom{s}{j} P_j^+(x) P_{s-j-1}^-(x).
\end{gather}


\section{Circular Couette flow}\label{sec:couette}

\begin{figure}
\centering
%
%
%
%
%
%
\includegraphics[width=0.9\columnwidth]{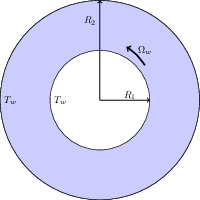}
\caption{Circular Couette flow setup.}
\label{fig:tcsetup}
\end{figure}

In this Section, the vielbein approach introduced in Sec.~\ref{sec:triad} 
is validated in the case of the circular Couette flow. 
The flow domain is bounded by two coaxial cylinders of radii 
$R_{\rm in} < R_{\rm out}$ which are kept at equal temperatures 
$T_w$, as shown in Fig.~\ref{fig:tcsetup}. 
The cylinders are free to rotate around their vertical axis 
(the $z$ axis). We are interested only in the stationary state 
and consider that the flow is homogeneous with respect to the 
$z$ and $\varphi$ directions. In order to take 
advantage of the $\varphi$ homogeneity, we employ
the vielbein approach.

This Section is structured as follows. 
In Subsec.~\ref{sec:couette:boltz}, the Boltzmann equation is
written with respect to the cylindrical coordinate system, in both 
the $\widetilde{f}$ and $\chi$ formulations, with or without 
grid stretching, while the ensuing macroscopic equations are 
discussed in Subsec.~\ref{sec:couette:macro}.
These formulations are discussed in Subsec.~\ref{sec:couette:cons},
where we demonstrate the failure of the $\widetilde{f}$ formulations to capture the constant solution when the 
two cylinders are at rest, as well as the solution corresponding to rigid rotation.
Subsections~\ref{sec:couette:hydro} and \ref{sec:couette:bal}
validate the $\chi$ implementation against analytic solutions in
the hydrodynamic and ballistic regimes.
In the transition regime, our scheme is validated 
against the DVM results presented in Ref.~\cite{aoki03} in 
Subsec.~\ref{sec:couette:trans}. 
A performance analysis of our vielbein-based implementation 
is presented in Subsec.~\ref{sec:couette:msites}.
Finally, conclusions are 
presented in Subsec.~\ref{sec:couette:conc}. The details regarding the 
mixed quadrature LB models employed for the simulations discussed 
in Subsecs.~\ref{sec:couette:hydro}, \ref{sec:couette:bal} 
and \ref{sec:couette:trans} are summarized in Table~\ref{tab:couette_q}.

The initial state for all the numerical simulations presented in this Section 
consists of a gas in thermal equilibrium having constant
density $n_0 = 1$, vanishing velocity $u^\hatR = u^\hvarphi = u^\hatz = 0$
and $T_0 = T_w = 1$.

\subsection{Boltzmann equation}\label{sec:couette:boltz}

Let us specialize the formalism of Section~\ref{sec:triad} to the case of the Couette flow between 
coaxial cylinders, described in Fig.~\ref{fig:tcsetup}. To describe the geometry of this flow, 
it is convenient to employ
cylindrical coordinates $\{x^\wi\} = \{R, \varphi, z\}$ through $x = R\cos\varphi$ and 
$y = R\sin\varphi$. The line element \eqref{eq:ds2} with respect to cylindrical coordinates is:
\begin{equation}
 ds^2 = dR^2 + R^2 d\varphi^2 + dz^2,\label{eq:ds2_cyl}
\end{equation}
while the triad vectors and the one-forms can be chosen as:
\begin{align}
 e_\hatR =& \partial_R, &
 e_\hvarphi =& R^{-1} \partial_\varphi, &
 e_\hatz =& \partial_z,\nonumber\\
 \omega^\hatR =& dR, &
 \omega^\hvarphi =& R \, d\varphi, &
 \omega^\hatz =& dz.
 \label{eq:triad_cyl}
\end{align}
The square root of the determinant of the metric in Eq.~\eqref{eq:ds2_cyl} is equal to
\begin{equation}
 \sqrt{g} = \sqrt{g_R} = R,
\end{equation}
while $\sqrt{g_\varphi} = \sqrt{g_z} = 1$ since the metric components 
do not depend on the $\varphi$ and $z$ coordinates.

The non-vanishing connection coefficients for the triad \eqref{eq:triad_cyl} are:
\begin{equation}
 \Gamma^\hatR{}_{\hvarphi\hvarphi} = -\frac{1}{R}, \qquad 
 \Gamma^\hvarphi{}_{\hatR\hvarphi} = \frac{1}{R},
\end{equation}
such that the Boltzmann equation in the $\widetilde{f}$ formulation \eqref{eq:boltz_cons} reads:
\begin{multline}
 \frac{\partial}{\partial t} 
 \begin{pmatrix}
  \wf' \\ \wf''
 \end{pmatrix}
 + \frac{p^\hatR}{m} \frac{\partial}{\partial R}
 \begin{pmatrix}
  \wf' \\ \wf''
 \end{pmatrix}
 + \frac{1}{mR} \left[(p^\hvarphi)^2 \frac{\partial}{\partial p^\hatR} 
 \begin{pmatrix}
  \wf' \\ \wf''
 \end{pmatrix}\right.\\
 \left.- p^\hatR \frac{\partial }{\partial p^\hvarphi}
 \begin{pmatrix}
  p^\hvarphi \wf' \\ p^\hvarphi \wf'' 
 \end{pmatrix} \right] = 
 -\frac{1}{\tau} 
 \begin{pmatrix}
  \wf' - \wf'_{\rm (eq)}\\
  \wf'' - \wf''_{\rm (eq)}
 \end{pmatrix},  \label{eq:boltz_couette_cons}
\end{multline}
where the flow was assumed to be homogeneous with respect to the $\varphi$ and 
$z$ coordinates and the $p^{\hatz}$ degree of freedom was reduced as described 
in Sec.~\ref{sec:LB:red} , while $\wf' = f' R$ and $\wf'' = f'' R$. 
The reduced distributions $f'$ and $f''$ were defined in Eq.~\eqref{eq:fred}.
The above equation can be shown to be equivalent to the 
equations used in Refs.~\cite{sharipov99,an12}.

As pointed out in Ref.~\cite{leveque02}, the numerical implementation of
hyperbolic equations in the $\widetilde{f}$ formulation
(i.e., by computing the numerical fluxes at the level of $\wf = f \sqrt{g}$)
is problematic since the preservation of 
a constant (analytic) solution is not guaranteed numerically.

In the $\chi$ formulation, the variable $\chi^R$ can be introduced 
via Eq.~\eqref{eq:chi_def}, following Ref.~\cite{falle96}:
\begin{equation}
 \chi^R = \frac{R^2}{2}.
\end{equation}
The Boltzmann equation in the $\chi$ formulation
\eqref{eq:boltz_noncons} can thus be written as follows: 
\begin{multline}
 \frac{\partial}{\partial t} 
 \begin{pmatrix}
  f' \\ f''
 \end{pmatrix}
 + \frac{p^\hatR}{m} \frac{\partial}{\partial \chi^R}
 \begin{pmatrix}
  f' R \\ f'' R
 \end{pmatrix}
 + \frac{1}{mR} \left[(p^\hvarphi)^2 \frac{\partial}{\partial p^\hatR} 
 \begin{pmatrix}
  f' \\ f''
 \end{pmatrix}\right.\\
 \left.- p^\hatR \frac{\partial }{\partial p^\hvarphi}
 \begin{pmatrix}
  p^\hvarphi f' \\ p^\hvarphi f'' 
 \end{pmatrix} \right] = 
 -\frac{1}{\tau} 
 \begin{pmatrix}
  f' - f'_{\rm (eq)}\\
  f'' - f''_{\rm (eq)}
 \end{pmatrix}. \label{eq:boltz_couette_kom}
\end{multline}

More details regarding our numerical implementation of the 
above equation and its order of accuracy are provided in 
Subsecs.~\ref{sec:num_sch:n} and 
\ref{sec:num_sch:order}, respectively.

Let us now consider the grid stretching procedure described in Sec.~\ref{sec:num_sch:stretch}
for the case of the radial coordinate. Defining $\eta$ in terms of $R$ via 
Eq.~\eqref{eq:eta_def} changes the line element \eqref{eq:ds2_cyl} to
\begin{equation}
 ds^2 = \left[\frac{A_0(R_{\rm out} - R_{\rm in})}{A \cosh^2\eta}\right]^2 d\eta^2 + 
 R^2(\eta) d\varphi^2 + dz^2.
\end{equation}
The triad corresponding to the above metric is:
\begin{equation}
 e_\heta = \frac{A \cosh^2\eta}{A_0(R_{\rm out} - R_{\rm in})} \partial_\eta, \qquad 
 e_{\hvarphi} = \frac{1}{R(\eta)} \partial_\varphi,\qquad 
 e_\hatz = \partial_z,\label{eq:boltz_couette_str_e}
\end{equation}
while the non-vanishing connection coefficients are:
\begin{equation}
 \Gamma^\hvarphi{}_{\heta\hvarphi} = -\Gamma^{\heta}{}_{\hvarphi\hvarphi} = \frac{1}{R(\eta)}.
\end{equation}
The Boltzmann equation in the $\widetilde{f}$ formulation
\eqref{eq:boltz_couette_cons} becomes:
\begin{multline}
 \frac{\partial}{\partial t} 
 \begin{pmatrix}
  \wf' \\ \wf''
 \end{pmatrix}
 + \frac{p^\heta}{m} \frac{\partial}{\partial \eta}\left[
 \frac{A \cosh^2\eta}{A_0(R_{\rm out} - R_{\rm in})}
 \begin{pmatrix}
  \wf' \\ \wf''
 \end{pmatrix}\right]\\
 + \frac{1}{mR(\eta)} \left[(p^\hvarphi)^2 \frac{\partial}{\partial p^\heta} 
 \begin{pmatrix}
  \wf' \\ \wf''
 \end{pmatrix}
 - p^\heta \frac{\partial }{\partial p^\hvarphi}
 \begin{pmatrix}
  p^\hvarphi \wf' \\ p^\hvarphi \wf'' 
 \end{pmatrix} \right]\\ = 
 -\frac{1}{\tau} 
 \begin{pmatrix}
  \wf' - \wf'_{\rm (eq)}\\
  \wf'' - \wf''_{\rm (eq)}
 \end{pmatrix},  \label{eq:boltz_couette_str_cons}
\end{multline}
while 
\begin{equation}
 \sqrt{g} = \sqrt{g_\eta} = \frac{A_0(R_{\rm out} - R_{\rm in}) R(\eta)}{A \cosh^2\eta}.
\end{equation}
 In the $\chi$ formulation, 
the equivalent of Eq.~\eqref{eq:boltz_couette_str_cons} 
is identical to Eq.~\eqref{eq:boltz_couette_kom}, where 
$p^\hatR$ is replaced by $p^\heta$, $R$ is replaced by $R(\eta)$
and $\chi^R$ is replaced by $\chi^\eta = R^2(\eta) / 2$:
\begin{multline}
 \frac{\partial}{\partial t} 
 \begin{pmatrix}
  f' \\ f''
 \end{pmatrix}
 + \frac{p^\heta}{m} \frac{\partial}{\partial \chi^\eta}
 \begin{pmatrix}
  f' R(\eta) \\ f'' R(\eta)
 \end{pmatrix}\\
 + \frac{1}{mR(\eta)} \left[(p^\hvarphi)^2 \frac{\partial}{\partial p^\heta} 
 \begin{pmatrix}
  f' \\ f''
 \end{pmatrix}
 - p^\heta \frac{\partial }{\partial p^\hvarphi}
 \begin{pmatrix}
  p^\hvarphi f' \\ p^\hvarphi f'' 
 \end{pmatrix} \right]\\ = 
 -\frac{1}{\tau} 
 \begin{pmatrix}
  f' - f'_{\rm (eq)}\\
  f'' - f''_{\rm (eq)}
 \end{pmatrix}. \label{eq:boltz_couette_str_kom}
\end{multline}

\subsection{Macroscopic equations}\label{sec:couette:macro}

In this Subsection, the macroscopic equations~\eqref{eq:macro} 
are presented for the case when the stationary regime is achieved. 
The continuity equation \eqref{eq:macro_n} reduces to:
\begin{equation}
 \nabla_\hata (nu^\hata) = \frac{1}{R} \partial_R (n R u^{\hatR}) = 0.
\end{equation}
Imposing a vanishing mass flux at the boundaries ($R = R_{\rm in}$ and $R = R_{\rm out}$) 
implies $u^\hatR = 0$ throughout 
the channel. This also implies that $\nabla_\hata u^\hata = 0$ and $u^\hata \nabla_\hata \phi = 0$, 
for any scalar function $\phi$ which does not depend on $t$, $\varphi$ or $z$.

Substituting $\hata \in \{\hatR, \hvarphi, \hatz\}$ into the Cauchy equation 
\eqref{eq:macro_u} gives:
\begin{subequations}
\begin{align}
 \rho (u^\hvarphi)^2 =& \partial_R (R T^{\hatR\hatR}) - T^{\hvarphi\hvarphi},\label{eq:cauchy_R}\\
 \partial_R (R^2 T^{\hatR\hvarphi}) =& 0,\label{eq:cauchy_vphi}\\
 \partial_R (R T^{\hatR\hatz}) =& 0.\label{eq:cauchy_z}
\end{align}
\end{subequations}
Considering that the flow is homogeneous along the $z$ direction, $T^{\hatR\hatz} = 0$ is 
an acceptable solution of Eq.~\eqref{eq:cauchy_z}. 
Next, the nondiagonal component $T^{\hatR\hvarphi}$ 
of the stress-tensor can be expressed analytically as:
\begin{equation}
 T^{\hatR\hvarphi} = T^{\hatR\hvarphi}_{\rm in} \frac{R_{\rm in}^2}{R^2},\label{eq:TRph}
\end{equation}
where $T^{\hatR\hvarphi}_{\rm in}$ is the value of $T^{\hatR\hvarphi}$ in the vicinity 
of the inner cylinder.
It is remarkable that Eq.~\eqref{eq:TRph} is valid for all degrees of rarefaction, while 
$T^{\hatR\hvarphi}_{\rm in}$ depends on the flow parameters, such as ${\rm Kn}$ or 
$\Omega_{\rm in}$.

Finally, the energy equation \eqref{eq:macro_T} reduces to:
\begin{equation}
 \partial_R(R q^\hatR) + R^2 T^{\hatR\hvarphi} \partial_R (R^{-1} u^\hvarphi) = 0.
\end{equation}
Using Eq.~\eqref{eq:TRph} for $T^{\hatR\hvarphi}$ yields:
\begin{equation}
 q^\hatR + u^\hvarphi T^{\hatR\hvarphi} = \frac{Q}{R},\label{eq:Q_def}
\end{equation}
where $Q$ is a constant which depends on the flow parameters.

In Subsections \ref{sec:couette:hydro} and \ref{sec:couette:bal}, 
analytic solutions for $n$, $u^\hvarphi$, $T$ and $q^\hatR$ will 
be derived in the Navier-Stokes and ballistic regimes and highlight
that in the $\widetilde{f}$ formulation, the radial heat flux 
presents a strong jump in the vicinity of the boundaries. 
The $\widetilde{f}$ approach is not considered further 
outside Sections~\ref{sec:couette:cons} and 
\ref{sec:couette:hydro}, respectively. 
The ${\rm Kn}$ dependence of $T_{\rm in}^{\hat{R}\hat{\varphi}}$ 
and $Q$ is discussed in Sec.~\ref{sec:couette:trans} and the 
results are summarized in Fig.~\ref{fig:Q_Trphi}.

\subsection{Comparison of $\widetilde{f}$ and $\chi$ formulations}\label{sec:couette:cons}

This Subsection is dedicated to the comparative a\-na\-ly\-sis of the
$\widetilde{f}$ and $\chi$ implementations of the Boltzmann equation. 
These implementations 
are considered with and without the grid stretching procedure described 
in Sec.~\ref{sec:num_sch:stretch}. The implementation 
of the advection part, described in the general case in Sec.~\ref{sec:num_sch:weno},
is given in Subsec.~\ref{sec:couette:cons:num_sch} for the particular 
cases considered herein. Two test cases are further considered.
The first, consisting of the trivial setup when both cylinders are at 
rest and $f' = f'' = {\rm const}$, is presented in 
Subsec.~\ref{sec:couette:cons:st}. 
The second test case, corresponding to 
rigid rotation (i.e. when the two cylinders rotate at the same angular speed),
is considered in Subsec.~\ref{sec:couette:cons:rigid}.
Our conclusions are presented in Subsec.~\ref{sec:couette:cons:conc}.

\subsubsection{Numerical scheme}\label{sec:couette:cons:num_sch}

As described in Sec.~\ref{sec:num_sch}, the flow domain is 
discretized using $N_R$ cells along the $R$ direction, while 
$N_\varphi= 1$ cells are used along the homogeneous $\varphi$ direction.
For the case of an equidistant grid, the radial coordinates of the 
centers of the $N_R$ cells are given as:
\begin{equation}
 R_s = R_{\rm in} + \frac{s-0.5}{N_R}(R_{\rm out} - R_{\rm in}),
\end{equation}
where $1 \le s \le N_R$, while
$R_{\rm in}$ and $R_{\rm out}$ are the radii of the inner and 
outer cylinders, respectively.
When employing the grid stretching procedure described in Sec.~\ref{sec:num_sch:stretch},
the stretching parameter $\eta$ is discretized equidistantly:
\begin{equation}
 \eta_s = \eta_{\rm in} + \frac{s-0.5}{N_R} (\eta_{\rm out} - \eta_{\rm in}),
\end{equation}
where $\eta_{\rm in}$ and $\eta_{\rm out}$ are defined in Eq.~\eqref{eq:eta_inout}
in terms of $R_{\rm in}$ and $R_{\rm out}$, respectively.

In the $\widetilde{f}$ formulation \eqref{eq:boltz_couette_cons}, 
$\chi^{\widetilde{1}} \equiv \chi^R = R$ and $V^R = p^\hatR / m$, 
such that Eq.~\eqref{eq:flux_def} becomes:
\begin{equation}
 \left(\frac{\partial (V^R \wf)}{\partial R}\right)_{s,1} \simeq 
 \frac{p^\hatR}{m}
 \frac{\widetilde{\mathcal{F}}_{R;s+1/2,1} - 
 \widetilde{\mathcal{F}}_{R; s-1/2,1}}{\delta R},
 \label{eq:flux_cons_R}
\end{equation}
where $\delta R$ is the constant grid spacing along the radial direction.
Similarly, $\chi^{\widetilde{1}} \equiv \chi^\eta = \eta$ 
and $V^\eta = p^\heta e_\heta^{\widetilde{\eta}}$ in 
Eq.~\eqref{eq:boltz_couette_str_cons} such that Eq.~\eqref{eq:flux_def} reduces to:
\begin{multline}
 \left(\frac{\partial (V^\eta \wf)}{\partial \eta}\right)_{s,1} \simeq 
 \frac{p^\heta}{m} \\
 \times 
 \frac{e_{\heta;s+1/2}^{\widetilde{\eta}}\widetilde{\mathcal{F}}_{\eta; s+1/2,1} - 
 e_{\heta;s-1/2}^{\widetilde{\eta}} \widetilde{\mathcal{F}}_{\eta; s-1/2,1}}
 {\delta \eta},
 \label{eq:flux_cons_eta}
\end{multline}
where $\delta \eta$ is the constant grid spacing with respect to $\eta$, while
$e_{\heta}$ is given in Eq.~\eqref{eq:boltz_couette_str_e}.

In the $\chi$ formulation \eqref{eq:boltz_couette_kom}, 
$\chi^R = R^2 / 2$ and Eq.~\eqref{eq:flux_def} becomes:
\begin{multline}
 \left(\frac{\partial (R p^\hatR f / m)}{\partial \chi^R}\right)_{s,1} \simeq 
 \frac{p^\hatR}{m}\\
 \times \frac{R_{s+1/2} \mathcal{F}_{\eta; s+1/2,1} - 
 R_{s-1/2} \mathcal{F}_{\eta; s-1/2,1}}{R_s \delta R}.
 \label{eq:flux_kom_R}
\end{multline}
Similarly, in Eq.~\eqref{eq:boltz_couette_str_kom}, $\chi^\eta = R^2(\eta)/2$, such that 
Eq.~\eqref{eq:flux_def} reduces to:
\begin{multline}
 \left(\frac{\partial [R(\eta) p^\heta f / m]}{\partial \chi^\eta}\right)_{s,1} \simeq 
 2\frac{p^\heta}{m}\\
 \times \frac{R(\eta_{s+1/2}) \mathcal{F}_{\eta; s+1/2,1} - 
 R(\eta_{s-1/2}) \mathcal{F}_{\eta; s-1/2,1}}
 {R^2(\eta_{s+1/2}) - R^2(\eta_{s-1/2})}.
 \label{eq:flux_kom_eta}
\end{multline}

\subsubsection{Cylinders at rest}\label{sec:couette:cons:st}

\begin{table}
\begin{tabular}{lrrrr}
Regime & ${\rm Kn}$ & Model & $N_{\rm vel}$ & $\delta t$\\\hline\hline
Low &
Hydro & ${\rm H}(2;3) \times {\rm H}(2;3)$ & $9$ & \ $5 \times 10^{-4}$\\
Mach & $0.01$ & ${\rm H}(3;4) \times {\rm H}(2;3)$ & $12$ & $3 \times 10^{-3}$\\
& $0.1$ & ${\rm HH}(3;4) \times {\rm H}(2;3)$ & $12$ & $3 \times 10^{-3}$ \\
& $0.5$ & ${\rm HH}(4;12) \times {\rm H}(4;5)$ & $120$ & $3 \times 10^{-3}$ \\
& $1$ & ${\rm HH}(4;16) \times {\rm H}(4;5)$ & $160$ & $2 \times 10^{-3}$ \\
& $100$ & ${\rm HH}(4;40) \times {\rm HH}(2;3)$ & $480$ & $10^{-3}$ \\\hline 
Non-& Hydro & ${\rm H}(4;5) \times {\rm H}(4;5)$ & $25$ & $5 \times 10^{-4}$ \\
negligible & $0.02$ & ${\rm HH}(3;4) \times {\rm H}(4;5)$ & $40$ & $10^{-3}$ \\
Mach & $0.1$ & ${\rm HH}(3;4) \times {\rm H}(4;5)$ & $40$ & $10^{-3}$ \\
& $1$ & ${\rm HH}(4;24) \times {\rm H}(4;11)$ & $528$ & $10^{-3}$ \\
& $10$ & ${\rm HH}(4;60) \times {\rm HH}(3;4)$ & $480$ & $5 \times 10^{-4}$ \\
& $\infty$ & \ ${\rm HH}(4;200) \times {\rm HH}(4;10)$ & \ $8000$ & $2 \times 10^{-5}$
\end{tabular}
\caption{ Mixed quadrature LB models, corresponding
total number of velocities $N_{\rm vel}$ and time step $\delta t$ employed for 
the simulations of the circular Couette flow 
presented in Figs.~\ref{fig:hydro:lowm} and \ref{fig:uw001}
(low Mach number), as well as in 
Figs.~\ref{fig:hydro:highm}, \ref{fig:bal:b05a}, \ref{fig:bal:b05b},
and \ref{fig:trans} (Non-negligible Mach number).
In the hydrodynamic regime (Figs.~\ref{fig:hydro:lowm} and 
\ref{fig:hydro:highm}), the relaxation time $\tau$ is given by 
Eq.~\eqref{eq:hydro_tau} with ${\rm Kn} = 0.001$. 
Outside the hydrodynamic regime (Figs.~\ref{fig:trans} and 
\ref{fig:uw001}), $\tau$ is related to ${\rm Kn}$
through Eq.~\eqref{eq:tau_aoki}.
\label{tab:couette_q}}
\end{table}

\begin{figure}
\begin{center}
\includegraphics[width=0.45\textwidth]{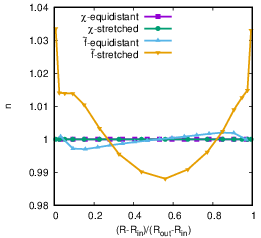} 
\end{center}
\caption{Density profile $n(R)$ for static cylinders having radii
$R_{\rm in} = 1$ and $R_{\rm out} = 2$.
The numerical results were obtained using the ${\rm H}(4;5) \times {\rm H}(4;5)$ model
with $\tau = {\rm Kn} / n$, where ${\rm Kn} = 10^{-3}$.
In all cases, $N_R= 16$ nodes were used and the stretching was performed 
according to $\delta = 0.5$ and $A = 0.95$. 
\label{fig:komissarov}}
\end{figure}

The case when the inner and outer cylinders are at rest ($\Omega_{\rm in} = \Omega_{\rm out} = 0$) 
and at equal temperature ($T_{\rm in} = T_{\rm out} = T_w$) admits the solution 
\begin{equation}
 f' = f'_{\rm (eq)} (n_0, \vu = 0, T_w) 
 = \frac{n_0}{2\pi m T_w} \exp\left(-\frac{p_\hatR^2 + p_\hvarphi^2}{2mT_w}\right),
 \label{eq:f_st}
\end{equation}
while $f'' = T_w f'$.
It can be easily seen that Eq.~\eqref{eq:f_st} satisfies the Boltzmann equation 
\eqref{eq:boltz_couette_cons}, as well as the boundary conditions.

Even though trivial, this simple test case serves as an example which highlights
an important drawback of the $\widetilde{f}$ approaches based on Eqs.~\eqref{eq:boltz_couette_cons} 
and \eqref{eq:boltz_couette_str_cons}.
As seen in Fig.~\ref{fig:komissarov}, the density profile when the $\widetilde{f}$ formulation is 
employed exhibits fluctuations, while the scheme based on the $\chi$ formulations 
\eqref{eq:boltz_couette_kom} and \eqref{eq:boltz_couette_str_kom} recovers Eq.~\eqref{eq:f_st}.
Our conclusion is in agreement with that presented in Ref.~\cite{leveque02}: the numerical fluxes 
associated to $f' \sqrt{g}$ and $f'' \sqrt{g}$ do not vanish, even when $f'$ and 
$f''$ are constant. This leads to a spurious redistribution of $f'$ and $f''$ due 
to which the stationary state does not coincide with the analytic solution.

\subsubsection{Rigid rotation}\label{sec:couette:cons:rigid}

\begin{figure}
\begin{tabular}{c}
\includegraphics[width=0.45\textwidth]{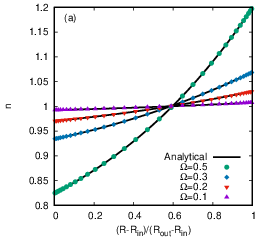} \\
\includegraphics[width=0.45\textwidth]{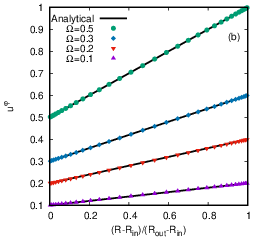}
\end{tabular}
\caption{Numerical results 
obtained using the $\chi$ implementation with the
${\rm H}(4;5) \times {\rm H}(4;5)$ models
for various values of the angular velocity $\Omega$ in the case of
the rigid rotation considered in Sec.~\ref{sec:couette:cons:rigid}.
(a) Density profile compared to Eq.~\eqref{eq:rigid_density}.
(b) Azimuthal velocity compared to the analytic solution $u^\hvarphi = \Omega R$.
}
\label{fig:rigid}
\end{figure}

We now turn our attention to another trivial case in which the two cylinders rotate 
at the same angular speed $\Omega_{\rm in} = \Omega_{\rm out} = \Omega_w$. 
Assuming that the walls have equal temperature $T_{\rm in} = T_{\rm out} = T_w$,
the analytic solution of the Boltzmann equation \eqref{eq:boltz_couette_cons} reads:
\begin{align}
 f'(R) =& f'_{\rm (eq)}[n(R), u^\hvarphi = \Omega R, T_w] \nonumber\\
 =& \frac{n(R)}{2\pi m T_w} \exp\left[-\frac{p_{\hatR}^2 + 
 (p_\hvarphi - m\Omega R)^2}{2m T_w}\right],
 \label{eq:rigid_sol}
\end{align}
and $f''(R) = T_w f'(R)$, while $n(R)$ is given by \cite{cumin02}:
\begin{equation}
 n(R) = N_{\rm tot} \frac{m\Omega^2}{4\pi T_w} 
 \frac{\exp\left[\frac{m \Omega^2}{4T_w}(2R^2 - R_{\rm in}^2 - R_{\rm out}^2)\right]}
 {\sinh\left[\frac{m\Omega^2}{4T_w}(R_{\rm out}^2 - R_{\rm in}^2)\right]},
 \label{eq:rigid_density}
\end{equation}
where $N_{\rm tot}$ represents the total number of particles per unit height 
between the two cylinders. The density normalization is chosen such that 
$N_{\rm tot} = \pi (R_{\rm out}^2 - R_{\rm in}^2)$.

\begin{figure}
 \includegraphics[width=0.45\textwidth]{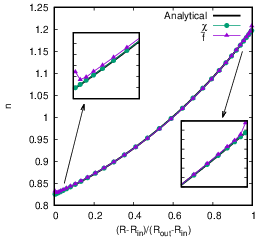}
 \caption{Comparison of the $\widetilde{f}$ and $\chi$ 
 formulations for $\Omega=0.5$, using the ${\rm H}(4;5) \times {\rm H}(4;5)$ model.}
 \label{fig:rigid_consvsnoncons}
\end{figure}

It is worth emphasizing that Eq.~\eqref{eq:rigid_sol} satisfies the Boltzmann equation for all 
values of the relaxation time. Fig.~\ref{fig:rigid} shows that,
in the $\chi$ implementation, our models can successfully 
reproduce both the velocity (top) and the density profile (bottom) for all tested values of 
the angular velocity. 
The models used are ${\rm H}(4;5) \times {\rm H}(4;5)$, the Knudsen number is
${\rm Kn}=0.001$, the time step is set to $\delta t=5 \times 10^{-4}$ 
and $N_R=32$ grid points are employed, stretched according to $\delta = 0.5$ and $A=0.95$. 
In Fig.~\ref{fig:rigid_consvsnoncons}, we highlight
the tendency of the density profile to bend upwards 
in the vicinity of the wall when the $\widetilde{f}$ formulation is employed, 
while the $\chi$ approach matches the analytic solution
with very high accuracy.

\subsubsection{Summary}\label{sec:couette:cons:conc}

The simple tests considered in this Subsection highlight two important drawbacks of 
the $\widetilde{f}$ formulation. First, the trivial solution $f' = f'' = {\rm const}$
cannot be fully recovered in this formulation, as shown in Fig.~\ref{fig:komissarov}.
This is in agreement with the discussion in Ref.~\cite{leveque02}.
Second, spurious terms are induced in the density profile in the 
vicinity of the boundaries. Even though the magnitude of these terms is small,
they are not present in the $\chi$ formulation. 

We thus conclude that the $\chi$ formulation is superior to the $\widetilde{f}$ 
one for the applications 
considered in this paper.
It is worth emphasizing that the conservation of the total number of particles 
is retained in the $\chi$ formulation, as highlighted in Ref.~\cite{falle96}
(see also Sec.~\ref{sec:num_sch:n} for more details).

\subsection{Navier-Stokes regime}\label{sec:couette:hydro}

The hydrodynamic regime is achieved in kinetic theory by taking the limit when the Knudsen 
number satisfies ${\rm Kn} \ll 1$. In the BGK formulation of the collision operator, 
we set the relaxation time in the form:
\begin{equation}
 \tau = \frac{{\rm Kn}}{n T}, \label{eq:hydro_tau}
\end{equation}
where ${\rm Kn}$ is set to $10^{-3}$ in order to achieve 
the hydrodynamic regime. The form \eqref{eq:hydro_tau} for the 
relaxation time ensures that the viscosity $\mu$ and heat conductivity $\kappa$ 
remain constant throughout the simulation, as implied by Eq.~\eqref{eq:CE_tcoeff}.

The analytic solution of the Navier-Stokes equations is obtained in 
Subsec.~\ref{sec:couette:hydro:an}. This solution is used in Subsecs.~\ref{sec:couette:hydro:lowm}
and \ref{sec:couette:hydro:highm} to validate our implementation in the low and 
moderate Mach number regimes. The numerical simulations were performed by 
fixing the inner cylinder radius at $R_{\rm in} = 1$, while the radius of the outer cylinder 
is allowed to vary in order to check the sensitivity of our implementation to curvature
effects \cite{dongari13a,budinsky14,titarev06}. We thus set
$R_{\rm out} \in \{2, 4, 8, 16\}$, resulting in the radii ratios 
$\beta = R_{\rm in} / R_{\rm out} \in \{0.5, 0.25, 0.125, 0.0625\}$.
The number of nodes employed is $64$ and $96$ for the low and non-negligible values of 
the Mach number, respectively, while the time step was set to $\delta t= 5 \times 10^{-4}$.
Since in the hydrodynamic regime, the flow is close to equilibrium, the full-range 
Gauss-Hermite quadrature is employed on all momentum space directions.

\subsubsection{Analytic analysis}\label{sec:couette:hydro:an}

In order to obtain the analytic solution in the Navier-Stokes regime, the constitutive equations 
\eqref{eq:CE_T} and \eqref{eq:CE_q} are employed for the nonequilibrium parts $\delta T^{\hata\hatb}$ 
and $\delta q^\hata$ in Eq.~\eqref{eq:noneq}, where the transport coefficients 
$\mu$ and $\kappa$ are assumed to be constant. The cylinders are assumed to have equal 
temperatures $T_{\rm in} = T_{\rm out} = T_w$, the outer cylinder is kept at rest 
(i.e. $\Omega_{\rm out}=0$), while the angular velocity 
$\Omega_{\rm in}$ of the inner cylinder is left arbitrary.
Noting that $\nabla_\hata u^\hata = 0$, the non-vanishing components of the stress-tensor are:
\begin{subequations}
\begin{align}
 T^{\hatR\hatR} =& T^{\hvarphi\hvarphi} = T^{\hatz\hatz} = P, \label{eq:NS_P_aux}\\
 T^{\hatR\hvarphi} =& -\mu R \frac{\partial}{\partial R} (R^{-1} u^\hvarphi). \label{eq:NS_TRph_aux}
\end{align}
\end{subequations}
Substituting Eq.~\eqref{eq:NS_TRph_aux} into Eq.~\eqref{eq:TRph} gives the Navier-Stokes 
solution for the velocity \cite{kundu15,rieutord15}:
\begin{equation}\label{eq:NS_U}
 u^\hvarphi = R^{-1} \frac{\Omega_{\rm in}}{R_{\rm in}^{-2} - R_{\rm out}^{-2}} - 
 R \frac{\Omega_{\rm in} R_{\rm in}^2}{R_{\rm out}^2 - R_{\rm in}^2},
\end{equation}
where the conditions $u^\hvarphi(R =R_{\rm in}) =\Omega_{\rm in} R_{\rm in}$ and $u^\hvarphi(R = R_{\rm out}) = 0$ 
were imposed on the inner and outer cylinders, respectively. 
The tangential stress $T^{\hatR\hvarphi}$ \eqref{eq:TRph} reads:
\begin{equation}
 T^{\hatR\hvarphi} = T^{\hatR\hvarphi}_{\rm in} \frac{R_{\rm in}^2}{R^2}, \qquad
 T^{\hatR\hvarphi}_{\rm in} = 
 \frac{2\mu \Omega_{\rm in} R_{\rm out}^2}{R_{\rm out}^2 - R_{\rm in}^2}.\label{eq:NS_TRph}
\end{equation}

Next, the temperature can be obtained by substituting Eq.~\eqref{eq:CE_q} into Eq.~\eqref{eq:Q_def}:
\begin{multline}
 T = T_w + \frac{\mu}{\kappa} \frac{\Omega_{\rm in}^2}{R_{\rm in}^{-2} - R_{\rm out}^{-2}} \\\times
 \left[\frac{R_{\rm in}^{-2} - R^{-2}}{R_{\rm in}^{-2} - R_{\rm out}^{-2}} - 
 \frac{\ln(R/R_{\rm in})}{\ln(R_{\rm out} /R_{\rm in})}\right],\label{eq:NS_T}
\end{multline}
where the boundary conditions $T(R=R_{\rm in}) = T(R = R_{\rm out}) = T_w$ were imposed.
The heat flux $q^\hatR = -\kappa \partial_R T$ can be obtained as follows:
\begin{multline}
 q^\hatR = -\frac{\mu}{R}\frac{\Omega_{\rm in}^2}{R_{\rm in}^{-2} - R_{\rm out}^{-2}} \\\times
 \left[\frac{2R^{-2}}{R_{\rm in}^{-2} - R_{\rm out}^{-2}} - \frac{1}{\ln(R_{\rm out}/R_{\rm in})}\right],\label{eq:NS_QR}
\end{multline}
while the constant $Q$ in Eq.~\eqref{eq:Q_def} is given by:
\begin{equation}
 Q = \frac{\mu \Omega_{\rm in}^2}{R_{\rm in}^{-2} - R_{\rm out}^{-2}} 
 \left[\frac{1}{\ln(R_{\rm out}/R_{\rm in})} - 
 \frac{2R_{\rm in}^2}{R_{\rm out}^2 - R_{\rm in}^2}\right].\label{eq:NS_Q}
\end{equation}

Finally, the equation for the pressure can be obtained by substituting Eq.~\eqref{eq:NS_P_aux} into 
Eq.~\eqref{eq:cauchy_R}:
\begin{equation}
 \partial_R \ln P = \frac{m(u^\hvarphi)^2}{RT}.\label{eq:NS_P_eq}
\end{equation}
To the best of our knowledge, the analytic solution of this equation is not known. 
Thus, the density profile $n = P / T$ must be computed using numerical methods, 
with the constraint that 
\begin{equation}
 2\pi \int_{R_{\rm in}}^{R_{\rm out}} n R dR = 
 \pi (R_{\rm out}^2 - R_{\rm in}^2).
 \label{eq:ntot}
\end{equation}

\subsubsection{Low Mach flows}\label{sec:couette:hydro:lowm}

\begin{figure}
\includegraphics[width=0.45\textwidth]{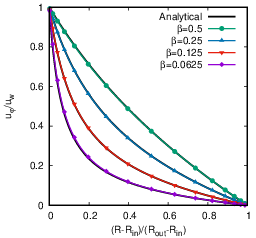} 
\caption{Azimuthal velocity profile $u^{\hvarphi}(R)$ for the 
angular velocity of the inner cylinder of $\Omega_{\rm in}=0.01$.
The curves correspond to various values of $\beta = R_{\rm in} / R_{\rm out}$.
Our numerical results, obtained using the ${\rm H}(2;3) \times {\rm H}(2;3)$ model in the $\chi$ implementation,
are overlapped with the analytic solution \eqref{eq:NS_U}.
}
\label{fig:hydro:lowm}
\end{figure}

The low Mach regime of the circular Couette flow has become a 
preferred benchmark test in the literature for models which deal 
with curved boundaries \cite{budinsky14,watari16,hejranfar17pre}. 
Since in this regime, the flow is essentially incompressible and isothermal, 
we only examine the azimuthal velocity $u^\hvarphi$, which is 
represented in Fig.~\ref{fig:hydro:lowm} for various values of 
$\beta = R_{\rm in} / R_{\rm out}$.
In this regime, the analytic profiles can be recovered using the 
${\rm H}(2;3) \times {\rm H}(2;3)$ model (employing $3\times 3 = 9$ velocities),
which is just the equivalent of the widely-used D2Q9 model
employed in Refs.~\cite{budinsky14,hejranfar17pre}.
However, the vielbein formaism allows only 
one node to be used in the $\varphi$ direction, thus bringing an improvement 
in the computational efficiency of 
several orders of magnitude compared to the implementations
presented in Refs.~\cite{budinsky14,hejranfar17pre}.

\subsubsection{Non-negligible Mach flows} \label{sec:couette:hydro:highm}

\begin{figure*}
\begin{center}
\begin{tabular}{cc}
\includegraphics[width=0.45\textwidth]{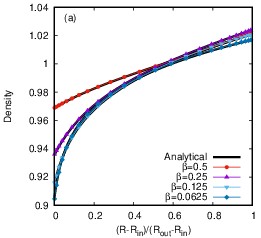} &
\includegraphics[width=0.45\textwidth]{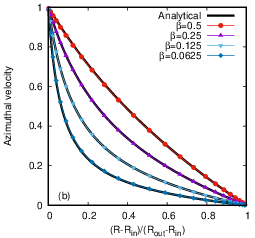}  \\
\includegraphics[width=0.45\textwidth]{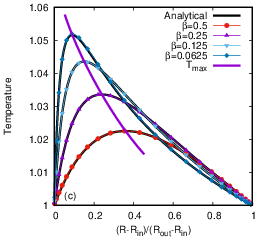} &
\includegraphics[width=0.45\textwidth]{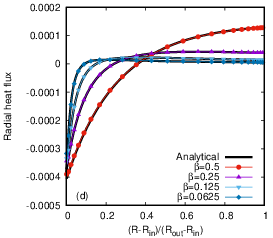}
\end{tabular}
\end{center}
\caption{Comparison between the numerical and analytic results for the profiles of 
(a) $n$ [the analytical curve is obtained numerically by solving Eq.~\eqref{eq:NS_P_eq}];
(b) $u^\hvarphi$ \eqref{eq:NS_U}; 
(c) $T$ \eqref{eq:NS_T}, together with Eq.~\eqref{eq:NS_Rmax} giving the position of the 
maximum in the temperature profile; 
(d) $q^\hatR$ \eqref{eq:NS_QR}.
The curves correspond to various values of $\beta = R_{\rm in} / R_{\rm out}$.
The inner cylinder rotates with $\Omega_{\rm in} = 0.5$, while the outer cylinder is kept at rest.
The numerical results, obtained with the $\chi$ implementation
using the ${\rm H}(4;5) \times {\rm H}(4;5)$ model,
are overlapped with the analytic solutions.}
\label{fig:hydro:highm}
\end{figure*}

We now consider the case when the angular velocity of the inner wall is 
$\Omega_{\rm in} = 0.5$, such that $u^\varphi_{\rm in} = \Omega_{\rm in} R_{\rm in} = 0.5$.

Figure~\ref{fig:hydro:highm} shows a comparison between the results obtained 
with the $\chi$ implementation using the 
${\rm H}(4;5) \times {\rm H}(4;5)$ model
against the analytical solution of the density $n$ 
[computed numerically using Eq.~\eqref{eq:NS_P_eq}], 
tangential velocity $u^{\hvarphi}$ \eqref{eq:NS_U}, 
temperature $T$ \eqref{eq:NS_T} and radial heat flux $q^{\hat R}$ \eqref{eq:NS_QR}.
A very good agreement is observed with the analytic solution for all tested parameters. 
The temperature profile exhibits a maximum when
\begin{equation}
 R= \sqrt{\frac{2\ln(R_{\rm out} / R_{\rm in})}{R_{\rm in}^{-2} - R_{\rm out}^{-2}}}.
 \label{eq:NS_Rmax}
\end{equation}
The above curve is also represented in Fig.~\ref{fig:hydro:highm}(c) and it can 
be seen that the maximum is captured very well.

\begin{figure}[t]
\begin{center}
\includegraphics[width=0.45\textwidth]{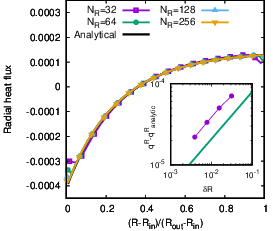}
\end{center}
\caption{Effect of lattice spacing on the behavior of $q^\hatR$ 
near the wall when the $\chi$ implementation on
an equidistant grid is employed. 
The inset shows that the amplitude of the deviation of $q^\hatR$ with respect 
to the expected analytic value \eqref{eq:NS_QR} is proportional to $(\delta R)^{0.58}$.}
\label{fig:hydro:cioc}
\end{figure}

\begin{figure}[b]
\begin{center}
\includegraphics[width=0.45\textwidth]{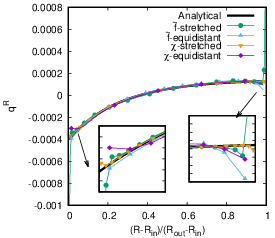}
\end{center}
\caption{Comparison of the $\widetilde{f}$ (stretched and equidistant) and $\chi$
(stretched and equidistant) radial heat flux using $N_R=32$ grid points.}
\label{fig:hydro:cioc_consvsnoncons}
\end{figure}

We note that the radial heat flow profiles are not well recovered near the boundaries, 
where a deviation with respect to the analytic profile can be seen. 
Figure \ref{fig:hydro:cioc} shows the
radial heat flux profile corresponding to equidistant grids having 
$N_R \in \{32,64,128,256\}$ nodes and $\beta=0.5$. 
It can be seen that this deviation occurs in the two points which are nearest to the boundary.
By increasing the resolution in the vicinity of the boundary, the amplitude of the 
deviation of the numerical result compared to the analytic prediction \eqref{eq:NS_QR}
is seen to decrease roughly as $(\delta R)^{0.58}$.
Figure \ref{fig:hydro:cioc_consvsnoncons} shows
the comparison of the $\widetilde{f}$ and $\chi$ formulations with stretched 
and equidistant grids using $N_R=32$ grid nodes. This plot clearly 
shows the advantage of using a stretched grid and the $\chi$ formulation,
which appears to minimize the amplitude of the deviations most efficiently out of 
the previously enumerated approaches.

\subsection{Free molecular flow regime}\label{sec:couette:bal}

In the free molecular flow regime, the collision term in the Boltzmann equation 
vanishes. The analytic solution in this case was derived in Ref.~\cite{aoki03} only 
for the distribution function, density, azimuthal velocity and temperature.
For completeness, we present a similar derivation for the 
distribution function and the macroscopic moments (including the stress tensor and 
heat fluxes which are not derived in Ref.~\cite{aoki03}), which are presented in 
Subsecs.~\ref{sec:couette:bal:f} and \ref{sec:couette:bal:macro}, respectively.
Our numerical scheme is validated by comparison with these results in 
Subsec.~\ref{sec:couette:bal:num}.

\subsubsection{Boltzmann distribution function}\label{sec:couette:bal:f}

\begin{figure}
\centering
%
%
\includegraphics[width=0.97\columnwidth]{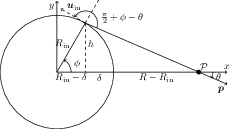}
\caption{Trajectory of a free-streaming particle originating from the inner cylinder,
which passes through point $\mathcal{P}$ at distance $R$ 
from the symmetry axis with momentum $\bm{p}$. Since the particle travels downwards, $\theta < 0$ and 
$\phi > 0$.}
\label{fig:tc-bal-1}
\end{figure}

Since there are no body forces present, the particles in the free molecular flow regime travel
along straight lines between the two bounding cylinders. Due to the symmetry of the flow configuration,
the solution is independent of the azimuth $\varphi$. Let us consider a point $\mathcal{P}$
at a distance $R - R_{\rm in}$ from the first cylinder, as shown in Fig.~\ref{fig:tc-bal-1}. 
The momentum of a particle passing through this point has the components:
\begin{equation}
 p^{\hatR} = p \cos\theta, \qquad p^{\hvarphi} = p \sin\theta,
\end{equation}
where $p = \sqrt{(p^{\hatR})^2 + (p^{\hvarphi})^2}$ and 
$\theta = \arctan (p^{\hvarphi} / p^\hatR)$. 
It is convenient to set the range of $\theta \in (-\pi, \pi)$ 
with $\theta = 0$ corresponding to the radial
direction towards the outer cylinder. With this convention, the particles with 
$\abs{\theta} < \theta_{\rm max} = \arcsin(R_{\rm in}/R)$ originate from the inner cylinder, while 
those with $\theta_{\rm max} < \abs{\theta} < \pi$ are emitted by the outer cylinder.
The coordinate axis $x$ is aligned along the radial direction passing through 
$\mathcal{P}$, such that the radial and azimuthal unit vectors at $\mathcal{P}$ 
are just $\bm{i}$ and $\bm{j}$.

When $\abs{\theta} < \theta_{\rm max}$, the distribution of particles at $\mathcal{P}$ 
having momentum $\bm{p}$ along the direction given by $\theta$ is equal to the distribution 
of particles emitted from the point located at $R_{\rm in}$ and angle 
$-\frac{\pi}{2} < \phi < \frac{\pi}{2}$ with respect to the 
horizontal axis, as shown in Fig.~\ref{fig:tc-bal-1}. We use the convention that 
the angle $\phi$ is positive 
when measured trigonometrically from the horizontal axis and negative otherwise.
From Fig.~\ref{fig:tc-bal-1} it can be seen that:
\begin{multline}
 (\vp - m\vu_{\rm in})^2 = p^2 + p_z^2 + m^2 \Omega_{\rm in}^2 R_{\rm in}^2 \\
 - 2 m p \Omega_{\rm in} R_{\rm in} 
 \cos\left(\frac{\pi}{2} + \phi - \theta\right),\label{eq:bal_sq}
\end{multline}
where $\vu_{\rm in} = \Omega_{\rm in} R_{\rm in} (-\bm{i} \sin\phi + \bm{j} \cos\phi)$.
In the above, it is understood that $\theta$ and $\phi$ have opposite signs, i.e. 
a particle travelling downwards ($\theta < 0$) originates from the upper half of the inner
cylinder ($\phi > 0$), as shown in Fig.~\ref{fig:tc-bal-1}.
The cosine function in Eq.~\eqref{eq:bal_sq} can be evaluated as follows:
\begin{align}
 \cos\left(\frac{\pi}{2} + \phi - \theta\right) =& \sin\theta (\cos\phi - \sin \phi \cot\theta) \nonumber\\
 =& \frac{R}{R_{\rm in}} \sin\theta,
\end{align}
where $\cos\phi = (R_{\rm in} - \delta) / R_{\rm in}$, $\sin \phi = \pm h / R_{\rm in}$ and 
$\cot\theta = \mp (R - R_{\rm in} + \delta) / h$, where the upper sign refers to the case when 
the particle is emitted from above the horizontal axis.

Thus, at radial distance $R$ from the axis of the inner cylinder, the distribution function 
of particles travelling at angle $\theta$ with respect to the radial direction is:

\begin{multline}
 f(R; \theta, p) = \frac{n_{\rm in}}{(2\pi m T_w)^{3/2}} \exp\Big\{-\frac{1}{2mT_w} 
 (p^2 + p_z^2\\ + m^2 \Omega_{\rm in}^2 R_{\rm in}^2 -2mp \Omega_{\rm in} R \sin \theta)\Big\},
 \label{eq:bal_f1}
\end{multline}
where the number density of emitted particles $n_{\rm in}$ 
will be determined in Subsec.~\ref{sec:couette:bal:macro}, 
$T_w$ is the wall temperature and 
$\abs{\theta} < \theta_{\rm max} = \arcsin(R_{\rm in} / R)$. 
Since the outer cylinder is at rest, the distribution function of the emitted particles is 
isotropic, such that, when $\theta_{\rm max} < \abs{\theta} < \pi$, the distribution function is 
given by:
\begin{equation}
 f(R; \theta, p) = \frac{n_{\rm out}}{(2\pi m T_w)^{3/2}} \exp\left[-\frac{p^2+p_z^2}{2mT_w}\right],
 \label{eq:bal_f2}
\end{equation}
where the number density $n_{\rm out}$ 
of the particles emitted by the outer cylinder
will be determined in the next Subsection.

\subsubsection{Macroscopic moments}\label{sec:couette:bal:macro}

Let us introduce the following moments:
\begin{multline}
 M_{s_R, s_\varphi, s_z} \equiv
 \int_{-\infty}^\infty dp_\hatz\int_{-\infty}^\infty dp_{\hat{R}} \int_{-\infty}^\infty dp_{\hat{\varphi}} \, 
 f\, p_{\hat{R}}^{s_R} p_{\hat{\varphi}}^{s_\varphi} p_\hatz^{s_z}\\
 = \int_{-\infty}^\infty dp_\hatz\, p_\hatz^{s_z} \int_0^\infty dp\,p^{s_R + s_\varphi + 1} \\
 \times \int_{-\pi}^\pi d\theta\, f\, (\cos\theta)^{s_R} (\sin\theta)^{s_\varphi}.\label{eq:bal_M_def}
\end{multline}
Using the results \eqref{eq:bal_f1} and \eqref{eq:bal_f2}, the above expression can be written as:
\begin{widetext}
\begin{multline}
 M_{s_R, s_\varphi,s_z} = 
 \frac{1 + (-1)^{s_z}}{2\pi^{3/2}} (2 m T_w)^{\frac{1}{2}(s_R + s_\varphi + s_z)} \Gamma\left(\frac{s_z + 1}{2}\right) \Bigg\{ \\
 \frac{n_{\rm in} e^{-\Rt_{\rm in}^2}}{\Rt^{s_\varphi + 1}} \int_{-\Rt_{\rm in}}^{\Rt_{\rm in}} 
 d\zeta\, \zeta^{s_\varphi} e^{\zeta^2} (1 - \zeta^2/\Rt^2)^{\frac{1}{2}(s_R-1)}
 \int_0^\infty d\xi\, \xi^{s_R+s_\varphi + 1}e^{-(\xi - \zeta)^2}\\
 + n_{\rm out}\, \Gamma\left[1 + \frac{1}{2}(s_R + s_\varphi)\right] 
 \frac{1 + (-1)^{s_\varphi}}{2} \int_{\theta_{\rm max}}^\pi d\theta\, (\cos\theta)^{s_R} (\sin\theta)^{s_\varphi}\Bigg\},
 \label{eq:bal_M}
\end{multline}
\end{widetext}
where the changes of variables $\xi = p / \sqrt{2mT_w}$ and $\zeta = \Rt \sin \theta$ were performed.
The notation 
\begin{equation}
 \Rt = \Omega_{\rm in} R \sqrt{m / 2T_w} \label{eq:Rt}
\end{equation}
represents the square root of the
ratio between the kinetic energy $\frac{m}{2} \Omega_{\rm in}^2 R^2$ induced by the rigid rotation 
at $R$ and the thermal energy $T_w$, while $\Rt_{\rm in} =\Rt R_{\rm in}/R$.

Noting that $\rho u_{\hat{R}} = M_{1,0,0}$, the macroscopic velocity along the radial 
direction can be computed as:
\begin{equation}
 u_{\hatR} = \frac{R_{\rm in}}{\rho R} \sqrt{\frac{m T_w}{2\pi}} (n_{\rm in} - n_{\rm out}),
\end{equation}
where the integration with respect to $\zeta$ in Eq.~\eqref{eq:bal_M} was performed first.
In order to ensure vanishing mass transfer through the bounding cylinders, $u_{\hat{R}}$ must vanish 
at $R = R_{\rm in}$ and at $R = R_{\rm out}$, requiring that:
\begin{equation}
 n_{\rm in} = n_{\rm out} = n_w.\label{eq:n1n2}
\end{equation}
In order to fix $n_w$,
the particle number density $n = M_{0,0,0}$ must be computed:
\begin{multline}
 n(R) = \frac{n_w}{\Rt \pi} e^{-\Rt_{\rm in}^2} 
 \int_{-\Rt_{\rm in}}^{\Rt_{\rm in}} \frac{e^{\zeta^2} d\zeta}{\sqrt{1 - \zeta^2/\Rt^2}}
 \int_0^\infty d\xi\, \xi\, e^{-(\xi - \zeta)^2}\\
 + n_w(1 - \theta_{\rm max} / \pi),
\end{multline}
where $\theta_{\rm max} = \arcsin(R_{\rm in} / R)$. Using the following identity:
\begin{equation}
 \int_0^\infty d\xi\,\xi\,e^{-(\xi-\zeta)^2} = \frac{1}{2} e^{-\zeta^2} + 
 \frac{\sqrt{\pi}}{2} \zeta(1 + \erf\,\zeta),
\end{equation}
the particle number density can be expressed as:
\begin{equation}
 n(R) = n_w\left\{1 - \frac{\theta_{\rm max}}{\pi} + 
 e^{-\Rt_{\rm in}^2} \left[\frac{\theta_{\rm max}}{\pi} + 
 \frac{I_0(\Rt)}{\Rt\sqrt{\pi}} \right]\right\},\label{eq:bal_n}
\end{equation}
where 
\begin{equation}
 I_n(\Rt) = \int_0^{\Rt_{\rm in}} \frac{\zeta^{2n+1} d\zeta}{\sqrt{1 - \zeta^2/\Rt^2}} e^{\zeta^2} \erf\zeta.
\end{equation}
Since the radial integral in Eq.~\eqref{eq:ntot} cannot be performed 
analytically, we resort to numerical methods to find the value of $n_w$.

The macroscopic velocity along the $\varphi$ direction can be computed by noting that 
$\rho u^\hvarphi = M_{0,1,0}$:
\begin{multline}
 \rho u^\hvarphi = \frac{n_w e^{-\Rt_{\rm in}^2}}{\pi \Rt^2} \sqrt{2 m T_w} 
 \int_{-\Rt_{\rm in}}^{\Rt_{\rm in}} \frac{\zeta d\zeta\,e^{\zeta^2}}{\sqrt{1 - \zeta^2/\Rt^2}}\\
 \times \int_0^\infty d\xi\, \xi^2 e^{-(\xi - \zeta)^2}.
\end{multline}
Using the following property:
\begin{multline}
 \int_0^\infty d\xi\,\xi^2 [e^{-(\xi-\zeta)^2} - e^{-(\xi+\zeta)^2}] \\
 = \zeta e^{-\zeta^2} + \frac{\sqrt{\pi}}{2}(1+2\zeta^2) \erf(\zeta),
\end{multline}
the azimuthal velocity can be written as:
\begin{multline}
 u^\hvarphi = \frac{n_w \Omega_{\rm in} R}{2 \pi n(R)} e^{-\Rt_{\rm in}^2} \Bigg\{
 \arcsin\frac{R_{\rm in}}{R} - \frac{R_{\rm in}}{R}\sqrt{1 - \frac{R_{\rm in}^2}{R^2}}\\
 + \frac{\sqrt{\pi}}{\Rt^3} [I_0(\Rt) + 2I_1(\Rt)]\Bigg\}.
 \label{eq:bal_u}
\end{multline}
Noting that
\begin{align}
 I_0 =& \frac{\Rt^3}{\sqrt{\pi}}\left(\theta_{\rm max} - \frac{R_{\rm in}}{R} \sqrt{1 - \frac{R_{\rm in}^2}{R^2}}\right) 
 + O(\Omega_{\rm in}^5),\nonumber\\
 I_1 =& \frac{\Rt^5}{4\sqrt{\pi}}\left[3\theta_{\rm max} - 
 \frac{R_{\rm in}}{R} \sqrt{1 - \frac{R_{\rm in}^2}{R^2}}\left(3+2\frac{R_{\rm in}^2}{R^2}\right)\right] \nonumber\\
 &+ O(\Omega_{\rm in}^7),
\end{align}
it can be seen that, in the small $\Omega_{\rm in}$ limit, Eq.~\eqref{eq:bal_u} reduces to
the expression in Refs.~\cite{willis65,watari16}:
\begin{equation}
 u^\hvarphi = \frac{1}{\pi} \Omega_{\rm in} R \left(\arcsin\frac{R_{\rm in}}{R} - 
 \frac{R_{\rm in}}{R} \sqrt{1 - \frac{R_{\rm in}^2}{R^2}}\right) 
 + O(\Omega_{\rm in}^3).
 \label{eq:bal_u_willis}
\end{equation}
Finally, $u^\hatz = 0$ since $M_{s_R, s_\varphi, 1} = 0$ due to the $[1 + (-1)^{s_z}]/2$ 
prefactor in Eq.~\eqref{eq:bal_M}.

For the computation of the stress tensor, it can be seen that $T^{\hatz\hatz} = \frac{1}{m} M_{0,0,2}$ 
is given by:
\begin{equation}
 T^{\hatz\hatz} = n(R) T_w.
\end{equation}
Since $T^{\hatR\hatz} = \frac{1}{m} M_{1,0,1} = 0$ and $\frac{1}{m} T^{\hvarphi\hatz} = M_{0,1,1} = 0$, 
the only non-vanishing non-diagonal component of the stress tensor is 
$T^{\hatR\hvarphi} = \frac{1}{m} M_{1,1,0}$:
\begin{equation}
 T^{\hatR\hvarphi} = \frac{2 n_w T_w}{\pi \Rt^2} e^{-\Rt_{\rm in}^2}
 \int_0^\infty d\xi\,\xi^3 e^{-\xi^2} 
 \int_{-\Rt_{\rm in}}^{\Rt_{\rm in}} d\zeta\, \zeta e^{2\xi\zeta}.
\end{equation}
The above integrals can be performed analytically, yielding:
\begin{equation}
 T^{\hatR\hvarphi} = T^{\hatR\hvarphi}_{\rm in} \frac{R_{\rm in}^2}{R^2}, \qquad 
 T^{\hatR\hvarphi}_{\rm in} = \frac{n_w T_w}{\sqrt{\pi}} \Rt_{\rm in},
 \label{eq:bal_TRp}
\end{equation}
where, as before, $\Rt_{\rm in} = \Omega_{\rm in} R_{\rm in} \sqrt{m/2T_w}$. The above expression is in agreement 
with the general result \eqref{eq:TRph}.
The last non-vanishing components of the stress-tensor are 
$T^{\hatR\hatR} = \frac{1}{m} M_{2,0,0}$ and $T^{\hvarphi\hvarphi} = \frac{1}{m} M_{0,2,0} - \rho (u^\hvarphi)^2$,
which have the following expressions:
\begin{widetext}
\begin{subequations}
\begin{align}
 T^{\hatR\hatR} =& \frac{n_w T_w}{\pi} e^{-\Rt_{\rm in}^2} \left\{
 \frac{R_{\rm in}}{4R} (4 + 2\Rt_{\rm in}^2 - \Rt^2) \cos\theta_{\rm max} 
 + \frac{1}{4}(4 + \Rt^2) \theta_{\rm max} + \frac{\sqrt{\pi}}{\Rt^3}\left[3 \Rt^2 I_0 + (2\Rt^2 - 3) I_1 - 2I_2\right]\right\}\nonumber\\
 &+ n_w T_w \left(1 - \frac{\theta_{\rm max}}{\pi} - \frac{\sin 2 \theta_{\rm max}}{2\pi}\right),\label{eq:bal_TRR}\\
 T^{\hvarphi\hvarphi} =& \frac{n_w T_w}{\pi} e^{-\Rt_{\rm in}^2}
 \left[ -\frac{R_{\rm in}}{4R} (4 + 2\Rt_{\rm in}^2 + 3\Rt^2) \cos\theta_{\rm max}
 + \frac{1}{4}(4 + 3\Rt^2) \theta_{\rm max} + \frac{\sqrt{\pi}}{\Rt^3}\left(3 I_1 + 2I_2\right)\right]\nonumber\\
 &+ n_w T_w \left(1 - \frac{\theta_{\rm max}}{\pi} + \frac{\sin 2 \theta_{\rm max}}{2\pi}\right)-\rho(u^\hvarphi)^2.\label{eq:bal_Tpp}
\end{align}
\end{subequations}
The temperature can be obtained as follows:
\begin{equation}
 T = T_w - \frac{m}{3} (u^\hvarphi)^2 + 
 \frac{n_w T_w \Rt^2}{3\pi n(R)} e^{-\Rt_{\rm in}^2} \left[
 \theta_{\rm max} -\frac{R_{\rm in}}{R} \cos\theta_{\rm max} + 
 \frac{\sqrt{\pi}}{\Rt} (I_0 + 2I_1)\right],
 \label{eq:bal_T}
\end{equation}
\end{widetext}

Finally, the components of the heat flux can be written as:
\begin{align}
 q^\hatR =& \frac{M_{3,0,0} + M_{1,2,0} + M_{1,0,2}}{2m^2} - u^\hvarphi T^{\hatR\hvarphi},\nonumber\\
 q^{\hvarphi} =& \frac{M_{2,1,0} + M_{0,3,0} + M_{0,1,2}}{2m^2} - u^\hvarphi T^{\hvarphi\hvarphi}\nonumber\\
 &-\frac{1}{2} \rho (u^\hvarphi)^3 - \frac{3}{2} nu^\hvarphi T.
 \label{eq:bal_q}
\end{align}
Noting that $M_{1,0,2} = 0$, $q^\hatR$ can be expressed as in Eq.~\eqref{eq:Q_def}, with
\begin{equation}
 Q = \frac{n_w R_{\rm in} T_{\rm in}^{3/2}}{\sqrt{2\pi m}} \Rt_{\rm in}^2,\label{eq:bal_Q}
\end{equation}
where the notation $\Rt_{\rm in}$ is defined in Eq.~\eqref{eq:Rt}.
The component $q^\hvarphi$ can be obtained from Eq.~\eqref{eq:bal_q} using:
\begin{multline}
 \frac{M_{2,1,0} + M_{0,3,0}}{2m^2} = \frac{n_w T_w^{3/2}\Rt\, e^{-\Rt_{\rm in}^2}}{4\pi\sqrt{2m}}
 \Bigg[\theta_{\rm max}(10 + 3\Rt^2)\\
 - \frac{R_{\rm in}}{R} (10+2\Rt_{\rm in}^2 + 3\Rt^2)\cos\theta_{\rm max} \\
 + \frac{2\sqrt{\pi}}{\Rt^3}(3I_0 + 12I_1 + 4I_2)\Bigg].
\end{multline}
as well as $\frac{1}{2m^2} M_{0,1,2} = \frac{1}{2} n T_w u^\hvarphi$. 

\subsubsection{Numerical results}\label{sec:couette:bal:num}

\begin{figure}[ht]
\begin{tabular}{c}
\includegraphics[width=0.45\textwidth]{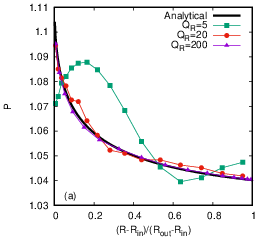}\\
\includegraphics[width=0.45\textwidth]{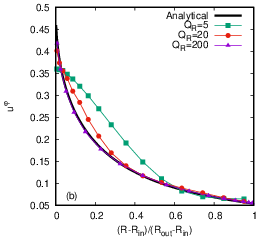}
\end{tabular}
\caption{
Effect of the quadrature order on the oscillations in the 
stationary profile of the pressure and azimuthal velocity in the ballistic regime.
The results were obtained with the models 
${\rm HH}(4;Q_R) \times {\rm HH}(4;10)$ 
and the curves correspond to various values of $Q_R$.
The parameters employed in these simulations are:
$R_{\rm in} = 1$, $R_{\rm out} = 2$, $\Omega_{\rm in} = 1.0$, 
$\Omega_{\rm out} = 0$ and $\delta t = 2\times 10^{-5}$. Our simulations 
reached the stationary state after $500\,000$ iterations.
We used $N_R = 16$ grid points together with the stretching 
given by $A = 0.95$ and $\delta = 0.75$.
The thick continuous curves correspond to the analytic solutions
derived in Subsec.~\ref{sec:couette:bal}
\eqref{eq:bal_n}.
\label{fig:bal:osc}
}
\end{figure}

\begin{figure*}
\begin{center}
\begin{tabular}{cc}
 \includegraphics[width=0.45\textwidth]{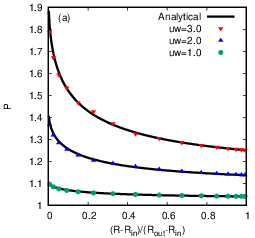} &
 \includegraphics[width=0.45\textwidth]{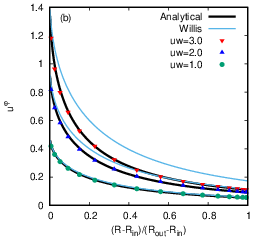} \\

 \includegraphics[width=0.45\textwidth]{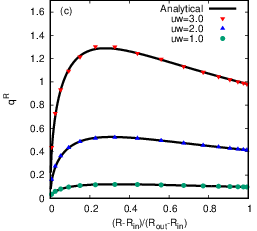}&
 \includegraphics[width=0.45\textwidth]{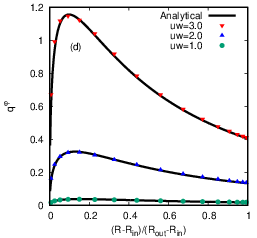}
\end{tabular}
\end{center}
\caption{
Comparison between our numerical results and the analytic predictions
in the ballistic regime. 
(a) $P=nT$ \eqref{eq:bal_n},\eqref{eq:bal_T}; (b) $u^\hvarphi / u_{\rm in}^\hvarphi$ \eqref{eq:bal_u}; (c) 
$q^\hatR$ \eqref{eq:bal_q}; (d) $q^\hvarphi$ \eqref{eq:bal_q}.
The radii of the inner and outer cylinders were $R_{\rm in} = 1$ and 
$R_{\rm out} = 2$, such that $\beta = 0.5$. The quadrature used was 
${\rm HH}(4;200) \times {\rm HH}(4;10)$. 
The curves correspond to various values of $\Omega_{\rm in}$. The 
analytic solution for $u^\hvarphi$ reported by Willis \cite{willis65} 
and reproduced in Eq.~\eqref{eq:bal_u_willis}
is shown in (b) alongside the exact expression \eqref{eq:bal_u}. The time step was 
set to $\delta t = 2\times 10^{-5}$ and the number of nodes was $N_R = 16$.
\label{fig:bal:b05a}}
\end{figure*}

\begin{figure*}
\begin{center}
\begin{tabular}{cc}
 \includegraphics[width=0.45\textwidth]{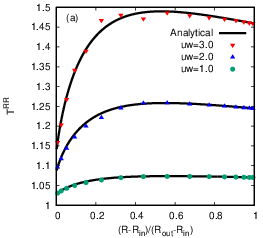} &
 \includegraphics[width=0.45\textwidth]{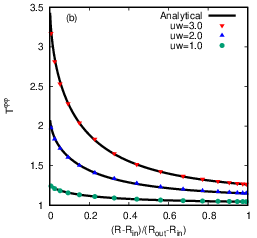}\\

 \includegraphics[width=0.45\textwidth]{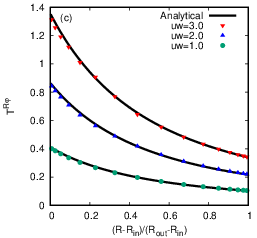}& 
 \includegraphics[width=0.45\textwidth]{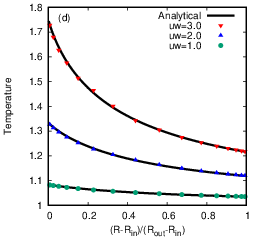}
\end{tabular}
\end{center}
\caption{
Same as Fig.~\ref{fig:bal:b05a} in the case of:
(a) $T^{\hatR\hatR}$ \eqref{eq:bal_TRR}; (b) $T^{\hvarphi\hvarphi}$ \eqref{eq:bal_Tpp};
(c) $T^{\hatR\hvarphi}$ \eqref{eq:bal_TRp}; (d) $T$ \eqref{eq:bal_T}.
The simulation parameters are the same as in Fig.~\ref{fig:bal:b05a}.
\label{fig:bal:b05b}}
\end{figure*}

As also noted in Refs.~\cite{ambrus17arxiv,watari16,ambrus18prc}, a sufficiently high quadrature order 
must be employed at high values of the relaxation time in order to avoid oscillations in the
stationary state. Fig.~\ref{fig:bal:osc} illustrates how increasing the radial quadrature 
order quenches the oscillation amplitude. We note that the quadrature order required to 
reduce the oscillations below a detectable level increases with the number of spatial 
grid points. Since we employ the fifth-order WENO-5 scheme together with an appropriate 
grid stretching, we are able to obtain accurate results with only 16 grid points and a quadrature 
order of $Q_R=200$ \cite{ambrus17arxiv}. 


We tested our models in the high-Mach regime by considering three values of 
the angular velocity of the inner cylinder, namely $\Omega_{\rm in} \in \{1, 2, 3\}$.
In this case, we kept $R_{\rm in} = 1$ and $R_{\rm out} = 2$ fixed, such that 
$\beta = R_{\rm in} / R_{\rm out} = 0.5$.
The time step was set to $\delta t = 2\times10^{-5}$.
For these values of the parameters, we used the models 
${\rm HH}(4;200) \times {\rm HH}(4;10)$ to ensure 
smooth profiles in the stationary state. 
Excellent agreement is found between our simulation results for 
the profiles of $P$, $u^\hvarphi$, $q^\hatR$, $q^\hvarphi$, $T^{\hatR\hatR}$, 
$T^{\hvarphi\hvarphi}$, $T^{\hatR\hvarphi}$ and temperature $T$ and 
the corresponding analytic results derived in Sec.~\ref{sec:couette:bal:macro}, as can be seen 
in Figs.~\ref{fig:bal:b05a} and \ref{fig:bal:b05b}. 

Before ending this Section, it is worth emphasizing that the formula \eqref{eq:bal_u_willis}
derived by Willis \cite{willis65}
is valid only in the limit of low Mach number flows, as shown in Subsec.~\ref{sec:couette:bal:macro}. 
At higher 
values of the Mach number, the ratio $u^\hvarphi / u_w$ no longer coincides with the profile 
predicted by Willis, since the non-linear terms in $u^\hvarphi$ which appear 
in the exact result \eqref{eq:bal_u} and are absent in the result from Willis 
\eqref{eq:bal_u_willis} become important. This discrepancy is highlighted in Fig.~\ref{fig:bal:b05a}(b).

\subsection{Transition flow regime}\label{sec:couette:trans}
 
\begin{figure*}
\begin{tabular}{c}
\begin{tabular}{cc}
\includegraphics[width=0.45\textwidth]{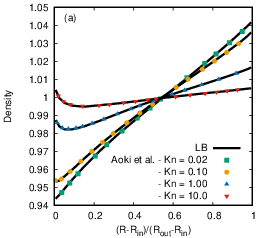}
\quad & \quad
\includegraphics[width=0.45\textwidth]{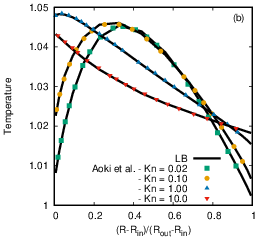}
\end{tabular}
\\
\includegraphics[width=0.45\textwidth]{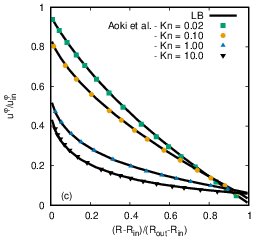}
\end{tabular}
\caption{
Comparison between our simulation results and those reported in 
Ref.~\cite{aoki03}. (a) $n(R)$; (b) $T(R)$; (c) $u^\hvarphi(R) / u^\hvarphi_{\rm in}$.
The angular velocity of the inner cylinder was set to $\Omega_{\rm in}=0.5\sqrt{2}$,
while the Knudsen number is related to the relaxation time through Eq.~\eqref{eq:tau_aoki}.
The models employed in these simulations are summarized in 
the {\it Non-negligible Mach} section of Table~\ref{tab:couette_q}.
\label{fig:trans}}
\end{figure*}

To the best of our knowledge, there is no analytic solution of the Boltzmann-BGK equation 
which is valid for the circular Couette flow in the transition regime. In order to 
validate our models in this regime, we compared our simulation results with those obtained 
by Aoki et al.~\cite{aoki03} using a high-order Discrete Velocity Model (DVM) for 
${\rm Kn} \in \{0.02, 0.1, 1.0, 10.0\}$, where the Knudsen number is related to 
the relaxation time $\tau$ via \cite{aoki03}:
\begin{equation}
 \tau = \frac{{\rm Kn}}{n} \sqrt{\frac{\pi}{8}}.
 \label{eq:tau_aoki}
\end{equation}
The angular velocity of the inner cylinder was set to $\Omega_{\rm in} = 0.5 \sqrt{2}$, 
while $\Omega_{\rm out} = 0$. The radii of the inner and outer cylinders were kept fixed at
$R_{\rm in} = 1$ and $R_{\rm out} = 2$. 
In order to maintain good agreement between our simulation results and those 
reported in Ref.~\cite{aoki03}, the radial and azimuthal quadrature orders 
were increased as ${\rm Kn}$ was increased, as summarized in Table~\ref{tab:couette_q} 
(the time step employed is also shown therein).
The simulation domain comprised $N_R = 16$ nodes stretched 
according to Eq.~\eqref{eq:eta_def} with 
$A = 0.95$ and $\delta = 0.5$.

Figures \ref{fig:trans}(a), \ref{fig:trans}(b) and \ref{fig:trans}(c) 
show comparisons between the profiles of $n$, $T$ and $u^\hvarphi$
obtained using the models summarized in Table~\ref{tab:couette_q}
and those reported by Aoki et al.~\cite{aoki03}.
A very good agreement can be seen.

\begin{figure}
\begin{tabular}{c}
\includegraphics[width=0.45\textwidth]{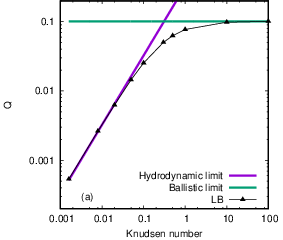}\\

\includegraphics[width=0.45\textwidth]{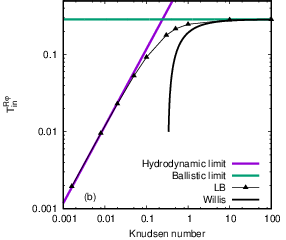} 
\end{tabular}
\caption{
Comparison of our simulation results for the constants $Q$ (a) and $T_{\rm in}^{\hatR\hvarphi}$ (b) 
against the analytic solutions given in Eqs.~\eqref{eq:NS_Q} and \eqref{eq:NS_TRph} in the
hydrodynamic limit, as well as in Eqs.~\eqref{eq:bal_Q} and \eqref{eq:bal_TRp} in the 
free molecular flow limit. The Knudsen number is related to the relaxation time through 
Eq.~\eqref{eq:tau_aoki}. In (b), the analytic result reported by Willis \cite{willis65} 
is represented using a solid black line.}
\label{fig:Q_Trphi}
\end{figure}

In Figures~\ref{fig:Q_Trphi} (a) and \ref{fig:Q_Trphi}(b), 
the variation over ${\rm Kn}$ of the constants 
$Q$ \eqref{eq:Q_def} and $T_{\rm in}^{\hatR\hvarphi}$ \eqref{eq:TRph} is represented. 
The simulation results match the analytic results in the hydrodynamic and ballistic flow regimes.
For $T_{\rm in}^{\hatR\hvarphi}$, our numerical results are compared with the analytic result 
obtained by Willis \cite{willis65} and a good match is observed
at high Knudsen numbers, close to the free molecular flow regime.

\begin{figure}
 \includegraphics[width=0.45\textwidth]{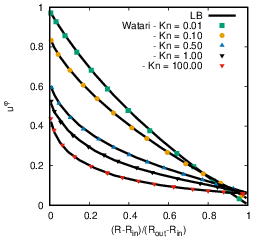}
\caption{
Velocity profile in the low Mach number regime 
$\Omega_{\rm in} = 0.01$. Our simulation results are compared with the 
LB results reported by Watari in Ref.~\cite{watari16}. The models employed in these simulations are summarized in 
the {\it Low Mach} section of Table~\ref{tab:couette_q} alongside 
the corresponding time step. The number of grid points 
was $N_R = 16$, stretched according to Eq.~\eqref{eq:eta_def} 
with $A = 0.95$ and $\delta = 0.5$.
\label{fig:uw001}}
\end{figure}

Finally, we considered the low Mach number case studied in Ref.~\cite{watari16}.
Our simulation results obtained using the models summarized in 
Table~\ref{tab:couette_q} are shown in Fig.~\ref{fig:uw001}. An excellent match 
with the LB results reported by Watari in Ref.~\cite{watari16} can be observed.

\subsection{Performance analysis}\label{sec:couette:msites}

Let us now consider a comparison between the efficiency of our method and that 
of previously published methods.
The lattice Boltzmann implementations employed in Refs.~\cite{budinsky14,hejranfar17pre} 
are validated only in the hydrodynamic regime at 
small Mach numbers and employ the D2Q9 model (employing 9 velocities). 
Our scheme is capable of recovering this regime also with 9 velocities.
However, the implementations of Refs.~\cite{budinsky14,hejranfar17pre} 
do not align the momentum space along the cylindrical coordinate system 
unit vectors, such that the spatial grid employed therein is two-dimensional. Thus, our 
proposed scheme is much more efficient, since our spatial grid is always one-dimensional.

Next, we consider a comparison with the LB implementation proposed by Watari
in Ref.~\cite{watari16}.
This scheme is also restricted to low Mach number flows, however, the whole range 
of the Knudsen number is explored. For ${\rm Kn} \lesssim 0.5$, Watari employed 
models with $40$ velocities in order to obtain accurate results. As Table~\ref{tab:couette_q} shows,
our implementation 
allows us to recover the same results with $12$ and $24$ velocities at ${\rm Kn} = 0.01$ 
and ${\rm Kn} = 0.1$, respectively, while at ${\rm Kn} = 0.5$, we employed 
a model with $120$ velocities. At ${\rm Kn} = 1$, 
Watari employed $4 \times 24 = 96$ velocities, while we required a number of 
$160$ velocities to match the velocity profile. Finally, 
at ${\rm Kn} = 100$, Watari obtained good agreement with the free-streaming solution 
with $4\times 60 = 240$ velocities, while we employed $480$ velocities in this regime.
We note that our implementation requires higher quadrature orders at 
${\rm Kn} \gtrsim 0.5$ due to the inertial forces which act along the 
radial direction, where the distribution function is discontinuous. 
This was also seen in the case of a rarefied gas between parallel plates 
under the effect of gravity \cite{ambrus17arxiv}. Such forces are not present in 
the implementation of Ref.~\cite{watari16}, since there the momentum space is not 
aligned to the cylindrical coordinate system. The gain in efficiency 
at the level of the momentum space compared to our scheme is lost since the 
spatial grid is two dimensional. The number of distribution functions 
required at ${\rm Kn} = 100$ in Ref.~\cite{watari16} is 
$240$ velocities multiplied by $200 \times 100 = 20000$ spatial grid points, 
resulting in $4\,800\,000$ population updates per time step. In our implementation,
we only use $16$ radial points, such that the number of population updates
per time step is just $16\times 480 = 7\,680$, which is significantly more efficient 
than the implementation presented in Ref.~\cite{watari16}.

\begin{figure}
\begin{tabular}{c}
\includegraphics[width=0.9\columnwidth]{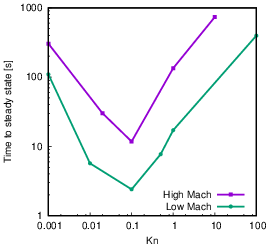}
\end{tabular}
\caption{Time (in seconds) required to achieve steady state
using the models summarized in Table~\ref{tab:couette_q}.
\label{fig:msites_time}
}
\end{figure}

In the transition regime, Aoki et al.~\cite{aoki03} employed a polar decomposition 
of the momentum space using $48$ shells of equal momentum magnitude containing 
$272$ directions, resulting in a velocity set comprising $48 \times 272 = 13\,056$ elements.
As can be seen from Table~\ref{tab:couette_q},
the number of velocities employed by our models is significantly lower at ${\rm Kn} 
\lesssim 10$, with $40$ velocities for ${\rm Kn} \in \{0.02, 0.1\}$,
$528$ velocities at ${\rm Kn} = 1$, and 
$960$ velocities at ${\rm Kn} = 10$.
As ${\rm Kn} \rightarrow \infty$, 
the number of velocities required to obtain accurate results increases to 
$8\,000$, which is still lower than the number of velocities 
employed in Ref.~\cite{aoki03}. Furthermore, the use of the
WENO-5 scheme for the computation of the 
numerical fluxes allows us to recover the analytic solutions in the 
ballistic regime using only $16$ nodes, compared with the $240$ nodes employed 
in Ref.~\cite{aoki03} using the second order numerical scheme introduced in 
Refs.~\cite{sugimoto92,sone93,sone95}.
It can thus be seen that, as the ballistic regime is approached, the efficiency of our 
scheme decreases to that of standard DVM codes. However, in the regime of moderate 
Knudsen numbers, our implementation is significantly more efficient, especially due 
to the use of the half-range Gauss-Hermite quadrature on the radial direction.
This can be seen by looking at Fig.~\ref{fig:msites_time}, where the time required 
to achieve the steady state using the models benchmarked in Figs.~\ref{fig:hydro:lowm},
\ref{fig:hydro:highm}, \ref{fig:trans} and \ref{fig:uw001} and summarized in Table~\ref{tab:couette_q}
is represented with respect to 
${\rm Kn}$, for both the low and the high Mach regimes. It can be seen that 
the lowest runtime is registered around ${\rm Kn} \simeq 0.1$. For completeness, the 
methodology to determine this runtime is presented below.

In each of the simulations presented in Fig.~\ref{fig:msites_time}, the time to 
achieve the steady state is determined by comparing the output of two successive cycles
of duration $\Delta t = 6$ (the number of iterations per cycle is computed 
based on the time step). At the end of cycle $\ell > 1$, the following $L_2$ norms are 
computed:
\begin{align}
 L_{2;u}^{\ell} =& \left[\int_{R_{\rm in}}^{R_{\rm out}} dR \, R
 \left(\frac{u^{\hvarphi}_\ell - u^{\hvarphi}_{\ell-1}}{u_w}\right)^2
 \right]^{1/2}, \nonumber\\
 L_{2;n}^{\ell} =& \left[\int_{R_{\rm in}}^{R_{\rm out}} dR \, R
 \left(\frac{n_{\ell}}{n_{\ell - 1}} -1\right)^2\right]^{1/2}, \nonumber\\
 L_{2;T}^{\ell} =& \left[\int_{R_{\rm in}}^{R_{\rm out}} dR \, R
 \left(\frac{T_{\ell}}{T_{\ell - 1}} -1\right)^2
 \right]^{1/2},\label{eq:msites_steady}
\end{align}
where $u_w$ is the angular velocity of the inner cylinder 
(the outer cylinder is at rest), while 
$R_{\rm in} = 1$ and $R_{\rm out} = 2$ are the radii of the inner and 
outer cylinders.
The integration is performed using the rectangle method by switching to the 
equidistant coordinate $\eta$, as described below for 
an arbitrary function $\mathfrak{f}$:
\begin{multline}
 \int_{R_{\rm in}}^{R_{\rm out}} dR\, R \, \mathfrak{f}(R) = 
 \frac{A_0}{A}(R_{\rm out} - R_{\rm in}) 
 \int_{\eta_{\rm in}}^{\eta_{\rm out}} \frac{R(\eta) d\eta}{\cosh^2 \eta} \mathfrak{f}(\eta) \\
 \simeq \frac{A_0}{A}(R_{\rm out} - R_{\rm in}) \sum_{s = 1}^{N_R} 
 \frac{R_s \delta \eta}{\cosh^2 \eta_s} \mathfrak{f}_s,
\end{multline}
where the quantities bearing the subscript $s$ are evaluated at 
$\eta = \eta_s = \eta_{\rm in} + (s - 0.5) \delta \eta$.
We consider that the steady state is achieved when 
all the $L_2$ norms defined in Eq.~\eqref{eq:msites_steady} 
decrease below the threshold $10^{-5}$.

Let us now discuss the order of algorithmic complexity of the main steps 
of our proposed algorithm, namely:
\begin{enumerate}
 \item Computation of the macroscopic variables;
 \item Relaxation;
 \item Enforcing boundary conditions;
 \item Applying the advection rule;
 \item Applying the forcing terms.
\end{enumerate}
The order of the above steps is arbitrary, since we use a fully explicit algorithm 
and the new populations are stored in a separate memory zone. The complexity 
of steps 1, 2 and 4 is $O(N_{\rm vel} \times N_R)$, where 
$N_{\rm vel} = 2 Q_R Q_\varphi$ is the total number of velocities 
when the half-range and full-range quadratures of orders $Q_R$ and $Q_\varphi$
are employed on the radial and azimuthal directions, while $N_R$ 
is the number of nodes in the radial direction. Step $3$ does not 
depend on the number of nodes (there are only two sites where diffuse reflection 
is applied for the circular Couette problem), so the complexity of this step 
is $O(N_{\rm vel})$. Finally, step 5 involves the computation of the 
momentum space derivatives, which are performed using the kernels introduced 
in Sec.~\ref{sec:LB:force} and in Appendix~\ref{app:force}. It can be seen that 
the complexity for this step is 
$O[(Q_R + Q_\varphi) \times N_{\rm vel} \times N_R]$.
Thus, the time required to perform one iteration can be estimated via:
\begin{multline}
 \Delta T = (a_1 + a_2 + a_4) N_{\rm vel} N_R + 
 a_3 N_{\rm vel} \\
 + a_5 (2Q_R + Q_\varphi) N_{\rm vel} N_R + c,
 \label{eq:msites_DT}
\end{multline}
where $a_i$ ($1 \le i \le 5$) are constants corresponding to the 
steps of the algorithm and the constant $c$ denotes an overhead which 
is due to one-off operations, 
such as memory allocations, input/output operations, etc.

We now consider a series of simulations in order to validate Eq.~\eqref{eq:msites_DT}. 
For simplicity, the number of nodes is kept constant at $N_R = 128$, such that 
Eq.~\eqref{eq:msites_DT} becomes:
\begin{equation}
 \Delta T = a N_{\rm vel} + b(2Q_R + Q_\varphi) N_{\rm vel} + c,\label{eq:msites_DT_glob}
\end{equation}
where $a$, $b$ and $c$ are constants. We now consider three batches of simulations.
In the first batch, the radial and azimuthal quadrature orders are 
varied simultaneously, such that $Q_R = Q_\varphi = Q$, where $4 \le Q \le 30$. 
In this case, $Q = \sqrt{N_{\rm vel} / 2}$ and Eq.~\eqref{eq:msites_DT_glob} 
reduces to:
\begin{equation}
 \Delta T = a N_{\rm vel} + \frac{3 b}{\sqrt{2}} N_{\rm vel}^{3/2} + c.
 \label{eq:msites_equalq}
\end{equation}
The second batch corresponds to keeping $Q_R = 4$ and varying $Q_\varphi$ 
between $4$ and $200$, such that $Q_\varphi = N_{\rm vel} / 8$ and Eq.~\eqref{eq:msites_DT_glob} 
becomes:
\begin{equation}
 \Delta T = a N_{\rm vel} + b N_{\rm vel} \left(8 + \frac{N_{\rm vel}}{8}\right) + c.
 \label{eq:msites_qRfixed}
\end{equation}
Finally, in the third simulation batch, $Q_\varphi = 4$ is kept fixed and 
$Q_R = N_{\rm vel} / 8$ is varied between $4$ and $200$, while $\Delta T$ 
\eqref{eq:msites_DT_glob} is given by:
\begin{equation}
 \Delta T = a N_{\rm vel} + b N_{\rm vel} \left(4 + \frac{N_{\rm vel}}{4}\right) + c.
 \label{eq:msites_qphifixed}
\end{equation}

\begin{figure}
\begin{tabular}{c}
\includegraphics[width=0.9\columnwidth]{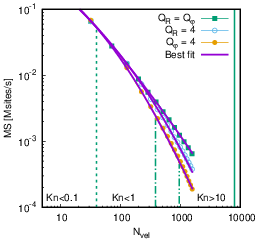}
\end{tabular}
\caption{Number of millions of site updates per second 
in the context of the circular Couette flow for a system with 
$N_R = 128$ nodes in the radial direction when the half-range and 
full-range Gauss-Hermite quadratures 
of orders $Q_R$ and $Q_\varphi$ are employed on the radial and 
azimuthal directions, respectively. The total number of velocities 
is $N_{\rm vel} = 2Q_R Q_\varphi$. The curves correspond to the 
cases when $Q_R = Q_\varphi$; when $Q_R = 4$ is kept fixed and 
$Q_\varphi$ is varied; and when $Q_\varphi = 4$ is kept fixed and 
$Q_R$ is varied. The simulation data are represented using lines 
with points, while the solid lines correspond to Eqs.~\eqref{eq:msites_equalq},
\eqref{eq:msites_qRfixed} and \eqref{eq:msites_qphifixed}, where 
the parameters $a$,$b$ and $c$ are obtained using a fitting routine.
\label{fig:msites_complex}
}
\end{figure}

The time per iteration $\Delta T$ can be used to compute the number of million of sites 
updated per second (Msites/s), which we denote by ${\rm MS}$, being given by:
\begin{equation}
 {\rm MS} = \frac{N_R}{10^6 \Delta T},\label{eq:couette_MS}
\end{equation}
where $\Delta T$ is expressed in seconds.
In order to validate Eq.~\eqref{eq:msites_DT_glob}, ${\rm MS}$ is computed by 
measuring the total simulation time $T$ required to complete 
$32\,000/ N_{\rm vel}$ iterations for a system with $N_R = 128$ nodes 
stretched according to $\delta = 0.5$ and $A = 0.95$,
with $\tau = {\rm Kn} / n$, ${\rm Kn} = 0.001$ and time step 
taken as $\delta t = 10^{-5}$ in order to satisfy the CFL condition 
for all quadrature orders considered in these simulations, by using 
the formula:
\begin{equation}
 {\rm MS} = \frac{0.32 N_R}{N_{\rm vel} T},\label{eq:couette_MS_aux}
\end{equation}
where $T$ is given in seconds.
Figure~\ref{fig:msites_complex} shows the dependence of ${\rm MS}$ with respect to 
$N_{\rm vel}$ for the three batches considered above. For each simulation batch,
the corresponding formula \eqref{eq:msites_equalq}--\eqref{eq:msites_qphifixed} 
is fitted to the numerical values of ${\rm MS}$ in order to determine the coefficients 
$a$, $b$ and $c$. Taking the average between the three sets of values gives
$a \simeq 42.4\ \mu{\rm s}$, $b \simeq 0.944\ \mu{\rm s}$ and 
$c \simeq 198\ \mu{\rm s}$. The curves corresponding 
to Eqs.~\eqref{eq:msites_equalq}--\eqref{eq:msites_qphifixed} with the 
above values for $a$, $b$ and $c$ are represented alongside the numerical 
data and an excellent agreement can be seen. This validates the algorithmic 
complexity proposed in Eq.~\eqref{eq:msites_DT}.

For consistency, all runtime results are calculated for simulations 
performed on a single core of an Intel\copyright Core${}^{\rm TM}$ i7-4790 Processor.

\subsection{Summary}\label{sec:couette:conc}

In this Section, the circular Couette problem was considered at various values of 
the Knudsen number, in the low and moderate Mach number regimes. 
Our numerical results reproduced with high accuracy the analytic solutions 
in the hydrodynamic and ballistic regimes, while at intermediate relaxation times,
we obtained excellent agreement with the discrete velocity model (DVM) results 
reported in Ref.~\cite{aoki03}.

\section{Flow through a gradually expanding channel} \label{sec:canalie}

In this Section, the versatility of the vielbein formalism is demonstrated
in the case of a more complex geometry. The implementation is validated 
for the case of the gradually expanding channel problem initially proposed 
in Ref.~\cite{roache82}.
This type of channel has the advantage that the transition from a 
narrow to a wide channel opening is made gradually, without resorting 
to sharp corners. 

Benchmark results were published in Ref.~\cite{napolitano85} for the 
incompressible Navier-Stokes flow through this channel at 
Reynolds number ${\rm Re} = 100$ in the no-slip regime. In this Section, we 
validate our implementation against these benchmark results and further exploit 
the vielbein formalism in order to study the properties of the flow at non-negligible 
values of the Knudsen number ${\rm Kn}$. In particular, we consider 
flow regimes with Mach numbers of order unity, as well as with 
${\rm Kn} \simeq 0.5$.

In Subsec.~\ref{sec:canalie:gen}, we introduce the vielbein for 
the general case of channels with symmetric walls and show how the 
momentum space can be aligned along them.
The case of the gradually expanding channel is presented in 
Subsec.~\ref{sec:canalie:duct}, where the grid construction 
is discussed. The inlet and outlet boundary conditions, as well as 
specular and diffuse reflection boundary conditions on the 
channel centerline and channel walls, respectively, are discussed in  
Subsec.~\ref{sec:canalie:bcond}. Our implementation is 
validated in the incompressible hydrodynamic regime in 
Subsec.~\ref{sec:canalie:hydro} and simulations in the 
compressible hydrodynamic regime are presented in Subsec.~\ref{sec:canalie:compress}.
To demonstrate the capabilities of the vielbein approach coupled with 
half-range quadratures, the flow through the gradually expanding channel 
is considered for non-negligible values of the Knudsen number 
in Subsec.~\ref{sec:canalie:raref}. 
A comparison with an implementation that does not 
use the vielbein approach and a performance analysis 
are given in Sec.~\ref{sec:canalie:coord}. A brief summary is 
presented in Sec.~\ref{sec:canalie:conc}.

\subsection{General formalism}\label{sec:canalie:gen}

Let us consider the general case of a channel exhibiting a gradual 
symmetric modification of its exterior boundary. Let the top boundary 
be given by the function $x_{\rm top}(y) = \frac{H}{2}[1 + \phi(y)]$, while 
the bottom boundary is located at $x_{\rm bottom}(y) = -x_{\rm top}(y)$.
The normalized tangent vector to the top boundary is:
\begin{equation}
 \bm{t} = \frac{\frac{H}{2} \phi' \bm{i} + \bm{j}}
 {\sqrt{1 + (\frac{H}{2} \phi')^2}},
\end{equation}
while the exterior normal can be obtained as:
\begin{equation}
 \bm{n} = \bm{t} \times \bm{k} = \frac{\bm{i} - \frac{H}{2} \phi' \bm{j}}
 {\sqrt{1 + (\frac{H}{2}\phi')^2}}.
\end{equation}
The incoming flux from the fluid towards the boundary is comprised of the 
particles for which
\begin{equation}
 \bm{p} \cdot \bm{n} = \frac{p^x - \frac{H}{2} \phi' p^y}
 {\sqrt{1 + (\frac{H}{2}\phi')^2}} > 0.\label{eq:canalie_pn}
\end{equation}
The above restriction cuts the momentum space in half along a plane 
given by the equation $p^x = \frac{H}{2} \phi' p^y$, which is point dependent 
due to the presence of $\phi'$. This has the undesirable effect that it does not 
allow the construction of a quadrature rule for the momentum space which is the 
same throughout the fluid domain. In particular, the lattice Boltzmann models
based on half-range quadratures are developed for the case when the 
boundary is orthogonal to one of the momentum space directions (e.g., $p^x$), 
such that the incoming and outgoing fluxes are obtained as momentum space integrals 
of the distribution function restricted to positive or negative values of the momentum 
component along this direction \cite{yang95,li03,li04,li09,lorenzani07,
frezzotti09,frezzotti11,gibelli12,guo13pre,ghiroldi14,ambrus14pre,ambrus14ipht,
ambrus14ijmpc,guo15pre,gibelli15,sader15,ambrus16jcp,ambrus16jocs,ambrus17arxiv,
ambrus18pre}.

In order to make the condition \eqref{eq:canalie_pn} point-independent, the following 
coordinates can be employed:
\begin{equation}
 \lambda = \frac{x}{1 + \phi(y)}, \qquad \xi = y,
\end{equation}
while $z$ remains unchanged.
The boundaries are now located at $\lambda = \pm H/2$.
The line element $ds^2 = dx^2 + dy^2 + dz^2$ becomes:
\begin{equation}
 ds^2 = \left\{[1 + \phi(\xi)] d\lambda + \lambda \phi'(\xi) d\xi\right\}^2 + d\xi^2 + dz^2.
\end{equation}
By writing $ds^2 = g_{\wi\wj} dx^{\wi} dx^{\wj}$, 
it can be seen that the nonvanishing components 
$g_{\wi\wj}$ of the metric tensor are given by:
\begin{gather}
 g_{\widetilde{\lambda} \widetilde{\lambda}} = [1 + \phi(\xi)]^2, \qquad 
 g_{\widetilde{\xi}\widetilde{\xi}} = 1 + \lambda^2 [\phi'(\xi)]^2, \nonumber\\
 g_{\widetilde{\xi}\widetilde{\lambda} } = 
 g_{\widetilde{\lambda}\widetilde{\xi}} = \lambda [1 + \phi(\xi)] \phi'(\xi), \qquad 
 g_{\widetilde{z} \widetilde{z}} = 1,
\end{gather}
while $\sqrt{g} = \sqrt{g_\xi} = 1 + \phi(\xi)$.
Thus, the metric tensor exhibits non-vanishing non-diagonal components.

The components of the momentum vector $\bm{p}$ with respect to the 
coordinates $\lambda$ and $\xi$ are:
\begin{align}
 p^{\widetilde{\lambda}} =& \frac{\partial \lambda}{\partial x} p^x + 
 \frac{\partial \lambda}{\partial y} p^y = \frac{p^x - \lambda \phi' p^y}{1 + \phi},\nonumber\\
 p^{\widetilde{\xi}} =& \frac{\partial \xi}{\partial x} p^x + 
 \frac{\partial \xi}{\partial y} p^y = p^y,
 \label{eq:canalie_pcoord}
\end{align}
while the inverse transformation gives
\begin{equation}
 p^x = (1 + \phi) p^{\widetilde{\lambda}} + \lambda \phi' p^{\widetilde{\xi}}, \qquad
 p^y = p^{\widetilde{\xi}}.
\end{equation}

It can be seen that at $\lambda = H/2$, $p^{\widetilde{\lambda}}$ is proportional to 
$\bm{p} \cdot \bm{n}$, such that Eq.~\eqref{eq:canalie_pn} reduces to:
\begin{equation}
 p^{\widetilde{\lambda}} > 0.
 \label{eq:canalie_pn_tilde}
\end{equation}
The coordinate directions $\partial_\lambda$ and $\partial_\xi$ are not orthogonal, since 
$g_{\widetilde{\lambda}\widetilde{\xi}} \neq 0$. This implies that the momentum vectors 
$\bm{p}_{\lambda}$ and $\bm{p}_\xi$, corresponding to 
$(p^{\widetilde{\lambda}}, p^{\widetilde{\xi}}) = (1,0)$ and $(0,1)$, respectively, 
are not orthogonal:
\begin{equation}
 \bm{p}_\lambda \cdot \bm{p}_\xi = g_{\widetilde{\lambda}\widetilde{\xi}} = 
 \lambda (1 + \phi) \phi'.
\end{equation}

In order to construct an orthogonal momentum space which retains the beauty of 
Eq.~\eqref{eq:canalie_pn_tilde}, it is convenient to work with the 
following triad one-forms:
\begin{gather}
 \omega^{\hat{\xi}} = \frac{\lambda \phi'(1 + \phi)}{\sqrt{1 + \lambda^2 \phi'{}^2}} d\lambda + 
 \sqrt{1 + \lambda^2 \phi'{}^2}\, d\xi, \nonumber\\
 \omega^{\hat{\lambda}} = \frac{1 + \phi}{\sqrt{1 + \lambda^2 \phi'{}^2}} d\lambda, \qquad 
 \omega^{\hat{z}} = dz,
\end{gather}
and the associated triad vectors:
\begin{gather}
 e_{\hat{\lambda}} = \frac{\sqrt{1 + \lambda^2 \phi'{}^2}}{1 + \phi} \partial_\lambda - 
 \frac{\lambda \phi'}{\sqrt{1 + \lambda^2\phi'{}^2}} \partial_\xi,\nonumber\\
 e_{\hat{\xi}} = \frac{1}{\sqrt{1 + \lambda^2\phi'{}^2}} \partial_\xi, \qquad 
 e_{\hat{z}} = \partial_z.
\end{gather}
The connection between the hatted components 
$p^{\hat{\lambda}}$ and $p^{\hat{\xi}}$ and the Cartesian components 
$p^x$ and $p^y$ of 
$\bm{p} = p^{\hat{\lambda}} e_{\hat{\lambda}} 
+ p^{\hat{\xi}} e_{\hat{\xi}} = p^x \partial_x + p^y \partial_y$ 
is given through:
\begin{subequations}\label{eq:canalie_phat}
\begin{align}
 p^{\hat{\lambda}} =& \omega^{\hat{\lambda}}_{\widetilde{\lambda}} p^{\widetilde{\lambda}} + 
 \omega^{\hat{\lambda}}_{\widetilde{\xi}} p^{\widetilde{\xi}} = 
 \frac{p^x - \lambda \phi' p^y}{\sqrt{1 + \lambda^2 (\phi')^2}}, 
 \label{eq:canalie_plambdah}\\
 p^{\hat{\xi}} =& \omega^{\hat{\xi}}_{\widetilde{\lambda}} p^{\widetilde{\lambda}} + 
 \omega^{\hat{\xi}}_{\widetilde{\xi}} p^{\widetilde{\xi}} = 
 \frac{p^y + \lambda \phi' p^x}{\sqrt{1 + \lambda^2 (\phi')^2}}.
 \label{eq:canalie_pxih}
\end{align}
\end{subequations}
The inverse relations are:
\begin{subequations}
\begin{align}
 p^x =& \left(\frac{\partial x}{\partial \lambda} e^{\widetilde{\lambda}}_{\hat{\lambda}} + 
 \frac{\partial x}{\partial \xi} e^{\widetilde{\xi}}_{\hat{\lambda}}\right) p^{\hat{\lambda}} + 
 \left(\frac{\partial x}{\partial \lambda} e^{\widetilde{\lambda}}_{\hat{\xi}} + 
 \frac{\partial x}{\partial \xi} e^{\widetilde{\xi}}_{\hat{\xi}} \right) p^{\hat{\xi}}\nonumber\\
 =& \frac{p^{\hat{\lambda}} + \lambda \phi' p^{\hat{\xi}}}
 {\sqrt{1 + \lambda^2 (\phi')^2}}, 
 \label{eq:canalie_px}\\
 p^y =& \left(\frac{\partial y}{\partial \lambda} e^{\widetilde{\lambda}}_{\hat{\lambda}} + 
 \frac{\partial y}{\partial \xi} e^{\widetilde{\xi}}_{\hat{\lambda}}\right) p^{\hat{\lambda}} + 
 \left(\frac{\partial y}{\partial \lambda} e^{\widetilde{\lambda}}_{\hat{\xi}} + 
 \frac{\partial y}{\partial \xi} 
 e^{\widetilde{\xi}}_{\hat{\xi}}\right) p^{\hat{\xi}}\nonumber\\
 =& \frac{p^{\hat{\xi}} - \lambda \phi' p^{\hat{\lambda}}}
 {\sqrt{1 + \lambda^2 (\phi')^2}}.
 \label{eq:canalie_py}
\end{align}
\end{subequations}
It can be seen that at $\lambda = H/2$, $p^{\hat{\lambda}} = \bm{p} \cdot \bm{n}$
\eqref{eq:canalie_plambdah}. Moreover, 
the triad vectors $e_{\hat{\lambda}}$ and $e_{\hat{\xi}}$ are orthogonal,
thus ensuring that the vectors $\bm{p}_{\hat{\lambda}}$ and $\bm{p}_{\hat{\xi}}$ 
corresponding to $(p^{\hat{\lambda}}, p^{\hat{\xi}}) = (1,0)$ and $(0,1)$ are 
orthogonal:
\begin{equation}
 \bm{p}_{\hat{\lambda}} \cdot \bm{p}_{\hat{\xi}} = 
 g_{\wi\wj} e^{\wi}_{\hat{\lambda}} e^{\wj}_{\hat{\xi}} = 
 \delta_{\hat{\lambda} \hat{\xi}} = 0.
\end{equation}

The only non-vanishing commutator  $[e_{\hat{\lambda}}, e_{\hat{\xi}}]$ 
gives rise to the following connection and Cartan coefficients:
\begin{align}
 \Gamma_{\hat{\lambda}\hat{\xi}\hat{\xi}} =& c_{\hat{\lambda}\hat{\xi}\hat{\xi}} = 
 \frac{\lambda \phi''}{(1 + \lambda^2 \phi'{}^2)^{3/2}},\nonumber\\
 \Gamma_{\hat{\lambda}\hat{\xi}\hat{\lambda}} =& c_{\hat{\lambda}\hat{\xi}\hat{\lambda}} = 
 \frac{\phi'[1 + \lambda^2 \phi'{}^2 - \lambda^2 \phi''(1 + \phi)]}
 {(1 + \phi)(1 + \lambda^2 \phi'{}^2)^{3/2}},\label{eq:canalie_Gamma}
\end{align}
the Boltzmann equation \eqref{eq:boltz_noncons} can be written as:
\begin{multline}
 \frac{\partial f}{\partial t} + 
 \frac{\partial(V^\lambda f)}{\partial \lambda} +
 \frac{\partial(V^\xi f)}{\partial \chi^\xi}\\
 - \frac{1}{m} \Gamma_{\hat{\lambda}\hat{\xi}\hat{\xi}}\left[ (p^{\hat{\xi}})^2 
 \frac{\partial f}{\partial p^{\hat{\lambda}}} - 
 p^{\hat{\lambda}} \frac{\partial(f p^{\hat{\xi}})}{\partial p^{\hat{\xi}}}\right]\\
 - \frac{1}{m} \Gamma_{\hat{\xi}\hat{\lambda}\hat{\lambda}}\left[ (p^{\hat{\lambda}})^2 
 \frac{\partial f}{\partial p^{\hat{\xi}}} - 
 p^{\hat{\xi}} \frac{\partial(f p^{\hat{\lambda}})}{\partial p^{\hat{\lambda}}}\right]
 = -\frac{1}{\tau}(f - \feq),\label{eq:canalie_boltz}
\end{multline}
where homogeneity with respect to the $z$ coordinate was assumed
and the following notation was introduced:
\begin{align}
 V^\lambda =& \frac{\sqrt{1 +\lambda^2 \phi'{}^2}}{1+ \phi} \frac{p^{\hat{\lambda}}}{m}, 
 \nonumber\\
 V^\xi =& (1 + \phi) 
 \frac{p^{\hat{\xi}} - \lambda \phi' p^{\hat{\lambda}}}
 {m \sqrt{1 + \lambda^2 \phi'{}^2}} = (1 + \phi) \frac{p^y}{m},
 \label{eq:canalie_V}
\end{align}
while $\chi^\xi$ is defined through:
\begin{equation}
 d\chi^\xi = (1+\phi) d\xi. \label{eq:chi_canalie}
\end{equation}

Since the channel is symmetric with respect to the central line 
located at $\lambda = 0$, the fluid flow is simulated only in 
the upper half ($0 < \lambda < H/2$).
The fluid domain is thus represented by the rectangle in 
the $(\xi, \lambda)$ space defined by $\xi_{\rm in} < \xi < \xi_{\rm out}$
and $0 < \lambda < H/2$. The non-dimensionalization convention 
is such that $H/2 = 1$.
The $\lambda$ direction is further stretched towards the solid boundary 
according to the coordinate transformation \eqref{eq:eta_def} with 
$\lambda_{\rm left} = 0$, $\lambda_{\rm right} = H/2$ 
and $\delta = 0$, as follows:
\begin{equation}
 \lambda(\eta) = \frac{H}{2 A} \tanh \eta,\label{eq:canalie_stretch}
\end{equation}
where $A = 0.95$ for all simulations presented in this Section.
The resulting grid is shown in Figs.~\ref{fig:canalie_geometry}(a) and 
(b) with respect to the $(x, y)$ and $(\lambda, \xi)$ coordinates, 
respectively, for the wall function $\phi(\xi)$ corresponding to the 
gradually expanding channel, given in Eq.~\eqref{eq:canalie_phi}.

The fluid domain in the $(\eta, \xi)$ variables is divided into 
$N_\eta \times N_\xi$ equally sized cells (where $N_\eta = N_\lambda$) centered on 
coordinates $(\lambda_s, \xi_p)$, where $\lambda_s \equiv \lambda(\eta_s)$ 
and $1 \le s \le N_\lambda$, $1 \le p \le N_\xi$, while
\begin{align}
 \eta_s =& \frac{s - 0.5}{N_\lambda} \arctanh A, \nonumber\\
 \xi_p =& \xi_{\rm in} + \frac{p - 0.5}{N_\xi} (\xi_{\rm out} - \xi_{\rm in}).
\end{align}
Inlet and outlet boundary conditions are imposed at $\xi = \xi_{\rm in}$ 
and $\xi = \xi_{\rm out}$, respectively, while 
specular and diffuse reflection boundary conditions are imposed 
at $\lambda = 0$ and $\lambda = H/2$, respectively, as shown in 
Fig.~\ref{fig:canalie_geometry}(b). 

In order to ensure that $f = {\rm const}$ is accepted as a numerical 
solution, the connection coefficients in Eq.~\eqref{eq:canalie_Gamma} are 
implemented as follows:
\begin{align}
 (\Gamma_{\hat{\lambda}\hat{\xi}\hat{\lambda}})_{s,p} =& \frac{1}{\delta \chi^\xi_{s,p}} \left(
 \frac{1 + \phi_{p + 1/2}}{\sqrt{1 + \lambda_s^2 \phi^{'\,2}_{p+1/2}}} - 
 \frac{1 + \phi_{p - 1/2}}{\sqrt{1 + \lambda_s^2 \phi^{'\,2}_{p-1/2}}}\right),\nonumber\\
 (\Gamma_{\hat{\lambda}\hat{\xi}\hat{\xi}})_{s,p} =& 
 -\frac{\sqrt{1 + \lambda^2_{s+1/2} \phi^{'\,2}_p} - \sqrt{1 + \lambda_{s-1/2}^2 \phi^{'\,2}_p}}
 {\delta \lambda_s (1 + \phi_p)} \nonumber\\
 &\hspace{-40pt} + \frac{\lambda_s}{\delta \chi^\xi_{s,p}} \left(
 \frac{\phi'_{p+1/2}(1 + \phi_{p + 1/2})}{\sqrt{1 + \lambda_s^2 \phi^{'\,2}_{p+1/2}}} - 
 \frac{\phi'_{p-1/2}(1 + \phi_{p - 1/2})}{\sqrt{1 + \lambda_s^2 \phi^{'\,2}_{p-1/2}}}\right)\nonumber\\
\end{align}
where $\delta \chi^\xi_{s,p} = \chi^\xi_{s,p+1/2} - \chi^\xi_{s,p-1/2}$ and $\delta \lambda_{s} = 
\lambda_{s+1/2} - \lambda_{s-1/2}$.

For the remainder of this Section, we consider the reduced form of Eq.~\eqref{eq:canalie_boltz},
obtained by multiplying Eq.~\eqref{eq:canalie_boltz} by $1$ and $(p^{\hatz})^2/m$ followed 
by an integration over the $p^\hatz$ momentum space axis, as described in Sec.~\ref{sec:LB:red}.
The resulting equations for $f'$ ($f''$) are identical with Eq.~\eqref{eq:canalie_boltz},
with $f$ and $\feq$ replaced by $f'$ ($f''$) and $f'_{\rm (eq)}$ ($f''_{\rm (eq)}$), respectively.
In order to ensure constant transport coefficients, the relaxation time is implemented as follows:
\begin{equation}
 \tau = \frac{\rm Kn}{n T}.
 \label{eq:canalie_tau}
\end{equation}

\subsection{Gradually expanding channel} \label{sec:canalie:duct}

\begin{figure}
 \begin{tabular}{c}
 \includegraphics[width=0.45\textwidth]{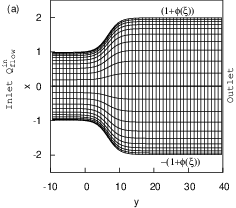} \\
 \includegraphics[width=0.45\textwidth]{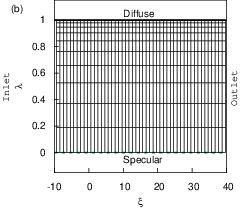}
 \end{tabular}
 \caption{(a) Geometry of the gradually expanding channel for ${\rm Re}_c = 100$.
 The vertical lines correspond to constant values of $\xi$ chosen 
 equidistantly between $-10$ and $40$. The horizontal lines are drawn 
 at constant values of $\lambda$ which are stretched towards the 
 boundaries via Eq.~\eqref{eq:canalie_stretch}.
 (b) Fluid domain in $(\lambda, \xi)$ coordinates, corresponding to the 
 upper half of the channel.
 \label{fig:canalie_geometry}}
\end{figure}

\begin{figure}
 \begin{tabular}{c}
 \includegraphics[width=0.45\textwidth]{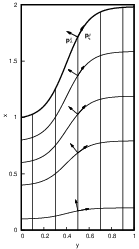}
 \end{tabular}
 \caption{The orientation of the principal axes of the momentum space 
 expressed with respect to the vielbein. The vectors $\bm{p}_{\hat{\lambda}} = e_{\hat{\lambda}}$ and 
 $\bm{p}_{\hat{\xi}} = e_{\hat{\xi}}$ can be regarded as unit vectors along these axes. 
 The vertical lines correspond to constant values of $\xi$, while the horizontal lines 
 represent lines of constant values of $\lambda$. The geometry corresponds to 
 setting ${\rm Re}_c = 6$ in Eq.~\eqref{eq:canalie_phi}.
 \label{fig:canalie_grid}}
\end{figure}

We now turn to the particular case of the 
gradually expanding channel proposed in Refs.~\cite{roache82,napolitano85}, 
for which the function $\phi(\xi)$ defining the position of the wall is 
given as
\begin{equation}
 \phi(\xi) = \frac{1}{2} \left[\tanh(2) - 
 \tanh\left(2 - \frac{30}{{\rm Re}_c} \frac{2\xi}{H}\right) \right].
 \label{eq:canalie_phi}
\end{equation}
The parameter ${\rm Re}_c$ controls the steepness 
of the expanding portion (i.e., its horizontal span). When ${\rm Re}_c$ 
is equal to the Reynolds number ${\rm Re}$ of the flow, the flow features 
become independent of ${\rm Re} = {\rm Re}_c$ in the region $0 < 2\xi/H < 
{\rm Re} / 3$ as ${\rm Re} \rightarrow \infty$. 
In particular, the flow configuration at ${\rm Re} = {\rm Re}_c = 100$ is 
a good approximation for the ${\rm Re} \rightarrow \infty$ case \cite{roache82}. 
In this Section, ${\rm Re}_c = 100$ is employed for all simulations, even when
the Reynolds number of the flow ${\rm Re}$ differs from this value. 
The resulting geometry is shown in Fig.~\ref{fig:canalie_geometry}(a).
Integrating Eq.~\eqref{eq:chi_canalie} gives the following expression for $\chi^\xi$:
\begin{equation}
 \chi^\xi = \left(1 + \frac{\tanh 2}{2}\right) \xi + \frac{H\rm Re_c}{120} 
 \ln \frac{\cosh(2 - 60\xi/{H\rm Re_c})}{\cosh 2}, \label{eq:chi_canalie_exp}
\end{equation}
where the integration constant was fixed such that $\chi^\xi = 0$ when $\xi = 0$.

For all simulations performed in 
the gradually expanding channel, we used a grid comprised of 
$N_\lambda \times N_\xi = 30 \times 200$ nodes.
The relevant flow domain is bounded by $\xi = 0$ and 
$\xi = {\rm Re}_c / 100 \simeq 33.33$. The inlet and outlet 
boundary conditions are imposed at $\xi = \xi_{\rm in} = -10$ and 
$\xi = \xi_{\rm out} = 40$, thus allowing some space for the flow 
to adjust itself before entering the investigated region.

In order to better understand the effect of employing the orthogonal triad, 
Fig.~\ref{fig:canalie_grid} shows the pair of vectors $(\bm{p}_{\hat{\lambda}},
\bm{p}_{\hat{\xi}}) = (e_{\hat{\lambda}}, e_{\hat{\xi}})$ at fixed
$\xi$ and for various values of $\lambda$, represented with respect to the 
$(x,y)$ coordinate frame. In order to maintain the same scale on the horizontal 
and vertical axes, the figure is drawn for a channel with ${\rm Re}_c = 6$, for which 
the horizontal span of the expanding portion of the channel is comparable to its 
vertical span. It can be seen that the two vectors start from being 
parallel to the $x$ and $y$ axes on the horizontal axis ($\lambda = 0$) 
to being aligned perpendicular to, and along, the upper boundary for 
$\lambda = H/2$. 

The momentum space defined with respect to the vielbein is discretized on the $\xi$ and $\lambda$ 
directions separately using $\mathcal{Q}_{\xi} \times \mathcal{Q}_\lambda$ velocities.
On the $\xi$ direction, which is parallel to the walls, the full-range Gauss-Hermite 
quadrature is used, such that $p^{\hat{\xi}} \rightarrow p^{\hat{\xi}}_j$ ($1 \le j \le Q_\xi$),
where $p^{\hat{\xi}}_j$ are the roots of the Hermite polynomial $H_{Q_\xi}(p^{\hat{\xi}})$ of 
order $Q_\xi$. On the $\lambda$ axis, the choice of quadrature depends on the 
value of ${\rm Kn}$. 
The momentum components are indexed as $p^{\hat{\lambda}}_i$, where $1 \le i \le \mathcal{Q}_\lambda$ 
and $\mathcal{Q}_\lambda = Q_\lambda$ when the full-range Gauss-Hermite quadrature of order 
$Q_\lambda$ is employed, while $\mathcal{Q}_\lambda = 2Q_\lambda$ for the case of the 
half-range Gauss-Hermite quadrature of order $Q_\lambda$, 
as discussed in Sec.~\ref{sec:LB:quad}. The expansion orders 
$\mathcal{N}_\xi$ and $\mathcal{N}_\lambda$
are generally constrained by Eq.~\eqref{eq:norder_limit}. We find that 
increasing the expansion orders beyond $4$ does not have a visible effect 
on the simulation results. Thus, the expansion orders 
are computed using 
\begin{equation}
 \mathcal{N}_\lambda = {\rm min}(Q_\lambda - 1, 4), \qquad
 \mathcal{N}_\xi = {\rm min}(Q_\xi -1, 4).
 \label{eq:ntotal_canalie}
\end{equation}

The system at initial time is considered to be in thermal equilibrium 
($f' = f'^{\rm (eq)}$ and $f'' = f''^{\rm (eq)}$) corresponding to the 
temperature $T_0$, density $n_0$ and velocity $\bm{u} = 0$ (the fluid is at rest). 
The non-dimensionalization convention used for the numerical simulations is such that 
$H/2 = 1$, $T_0 = 1$ and $n_0 = 1$, while $\sqrt{K_B T_0 / m} = 1$ is the reference speed.

\subsection{Boundary conditions} \label{sec:canalie:bcond}

\begin{figure}
 \includegraphics[width=0.45\textwidth]{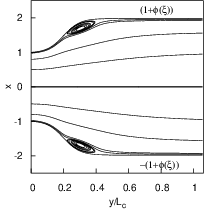}
 \caption{Streamlines for the flow through the gradually expanding channel 
 corresponding to ${\rm Re}_c = 100$, obtained for $Q_0 = 0.1$ and 
 ${\rm Kn} = 0.001$ (${\rm Re} = 100$), corresponding 
 to the incompressible hydrodynamic limit.
 \label{fig:canalie_stream}}
\end{figure}

This Subsection presents our strategy for the implementation 
of the inlet and outlet boundary conditions compatible with the 
approach used in Refs.~\cite{roache82,napolitano85}, as well as 
of the boundary conditions at the wall and channel center. 

\subsubsection{Inlet boundary conditions}\label{sec:canalie:bcond:inlet}

The problem initially proposed in Ref.~\cite{napolitano85} was 
the simulation of the incompressible Navier-Stokes flow through 
the gradually expanding channel introduced in 
Subsec.~\ref{sec:canalie:duct}, subject to an inlet parabolic 
velocity profile at $\xi = 0$ of the following form:
\begin{equation}
 u^y = \frac{3 u_0}{2} \left(1 - \frac{4x^2}{H^2}\right), \qquad 
 u^x = 0,\label{eq:canalie_inlet_def}
\end{equation}
such that the particle flow rate through half of the channel cross section is 
$Q_0 = \frac{H}{2} n_0 u_0$, where $n_0$ is the initial 
fluid particle number density 
throughout the channel. Equation~\eqref{eq:canalie_inlet_def} uses the property 
that $\phi(\xi = 0) = 0$. The Reynolds number is then obtained as follows:
\begin{equation}
 {\rm Re} = \frac{m Q_0}{\mu} = \frac{u_0}{\rm Kn},
 \label{eq:canalie_Re}
\end{equation}
where $H = 2$, $n_0 = 1$ and $u_0 = Q_0$ under the non-dimensionalization employed in 
this Section, while the viscosity 
$\mu = \tau n T = {\rm Kn}$ by virtue of Eq.~\eqref{eq:canalie_tau}. 
As mentioned in Ref.~\cite{napolitano85}, 
this inlet boundary condition immediately raised the concern that 
at $\xi = 0$, the channel already began its expansion, such that the inlet 
condition $u^x = 0$ is not realistic. 

Even though the 
results presented in Ref.~\cite{napolitano85} used Eq.~\eqref{eq:canalie_inlet_def}
as the inlet boundary condition, we instead impose the parabolic profile 
upstream from $\xi = 0$, at a value $\xi_{\rm in}$ where $\phi'(\xi_{\rm in}) \simeq 0$.
Thus, Eq.~\eqref{eq:canalie_inlet_def} can be replaced by:
\begin{align}
 Q^{\rm in}_{\rm flow}(\lambda) =& \frac{3Q_{0}}{H[1 + \phi(\xi_{\rm in})]}
 \left\{1 - \frac{4x^2}{H^2[1 + \phi(\xi_{\rm in})]^2}\right\}\nonumber\\
 =& \frac{3Q_{0}}{H[1 + \phi(\xi_{\rm in})]}
 \left(1 - \frac{4\lambda^2}{H^2}\right),\label{eq:canalie_inlet}
\end{align}
where the inlet particle flow rate $Q^{\rm in}_{\rm flow}(\lambda)$ 
at a given value of $\lambda$ is computed as follows:
\begin{equation}
 Q^{\rm in}_{\rm flow}(\lambda) = \int dp^{\hat{\xi}} dp^{\hat{\lambda}} 
 \, f'\, \frac{p^y}{m}.
\end{equation}
After the discretization of the spatial domain and of the momentum space, 
the above expression can be computed using the numerical flux 
$\mathcal{F}_{\,\widetilde{\xi};s,1/2;i,j}$ corresponding to 
$\bm{p}_{i,j} = (p^{\hat{\lambda}}_i, p^{\hat{\xi}}_j)$, as follows:
\begin{equation}
 Q^{\rm in}_{\rm flow; s} \equiv Q^{\rm in}_{\rm flow}(\lambda_s) = 
 \sum_{i, j} \frac{p^y_{s,1/2;i, j}}{m} 
 \mathcal{F}_{\widetilde{\xi}; s, 1/2; i, j},
 \label{eq:canalie_inlet_s}
\end{equation}
where the labels of $p^y$ \eqref{eq:canalie_py}
indicate its explicit coordinate and momentum space dependence:
\begin{equation}
  p^y_{s,1/2; i,j} = \frac{p^{\hat{\xi}}_j - \lambda_s \phi'(\xi_{1/2}) p^{\hat{\lambda}}_i}
 {\sqrt{1 + \lambda_s^2 \phi'{}^2(\xi_{1/2})}}.
\end{equation}

As also remarked in Ref.~\cite{zhang18}, the inlet and outlet boundary 
conditions can be imposed only at the level of the distribution functions
corresponding to velocities which travel downstream from the inlet 
towards the fluid domain (i.e., $p^y > 0$).
Thus, our strategy for imposing Eq.~\eqref{eq:canalie_inlet} is the 
following. The distributions corresponding to particles travelling upstream 
($p^y < 0$) are extrapolated at zeroth order from the first fluid node:
\begin{equation}
 f'_{s, -1; i,j} = f'_{s, 0; i,j} = f'_{s, 1; i,j},  \qquad p^y_{s,1/2;i,j} < 0,
\end{equation}
A similar boundary condition is imposed for $f''_{s,p; i, j}$. 

The flux for $p^y_{s,1/2;i,j} < 0$
can be computed by noting that $\sigma_3 = 0$ by virtue of Eq.~\eqref{eq:weno_sigma},
such that:
\begin{equation}
 \mathcal{F}_{\widetilde{\xi}; s, 1/2; i, j} = f'_{s,1; i, j}, \qquad 
 (p^y_{s,1/2;i,j} < 0).
\end{equation}

The distribution functions for the particles travelling downstream 
($p^y_{s,1/2;i,j} > 0$) are set using:
\begin{align}
 f'_{s,-2; i, j} =& f'_{s,-1; i,j} = f'_{s,0;i, j} = 
 f'_{{\rm (eq)}; {\rm in}; i,j},\nonumber\\
 f''_{s,-2; i, j} =& f''_{s,-1; i,j} = f''_{s,0;i, j} = 
 T_0 f'_{{\rm (eq)}; {\rm in}; i,j},
\end{align}
where $f'_{{\rm (eq)}; {\rm in}; i,j} \equiv 
f'_{{\rm (eq)}; i, j}(n_s^{\rm in}, \bm{u}_s^{\rm in}, T_0)$ is the reduced 
Maxwell-Boltzmann distribution \eqref{eq:feq_red}, $T_0$ is the initial temperature 
and $(u^x_s, u^y_s) = (0, Q^{\rm in}_{\rm flow; s})$. Since in this case 
$\sigma_1 = 0$ by virtue of Eq.~\eqref{eq:weno_sigma}, the flux is given by:
\begin{equation}
 \mathcal{F}_{\,\widetilde{\xi}; s, 1/2; i, j} = 
 f'_{{\rm (eq)}; {\rm in}; i,j}, \qquad 
 (p^y_{s,1/2;i,j} > 0).
\end{equation}

The density $n_s^{\rm in}$ is then obtained by imposing 
Eq.~\eqref{eq:canalie_inlet_s}:
\begin{equation}
 n_s^{\rm in} = \frac{\displaystyle Q^{\rm in}_{\rm flow;s} - 
 \frac{1}{m}\sum_{p^{y}_{s,1/2;i,j} < 0} f'_{1,p; i, j} p^y_{s,\frac{1}{2};i,j}}
 {\displaystyle \frac{1}{m} \sum_{p^{y}_{s,1/2;i,j} > 0}
 f'_{{\rm (eq)}; i,j}(1, \bm{u}_s^{\rm in}, T_0) p^y_{s,\frac{1}{2};i,j}}.
\end{equation}
Setting the inlet boundary conditions as explained above achieves the desired parabolic velocity 
profile shortly after the simulation is started.

\subsubsection{Outlet boundary conditions}\label{sec:canalie:bcond:outlet}

In order to prevent the build-up of particles inside the flow domain, a similar 
parabolic profile is imposed at the domain outlet (where $\xi = \xi_{\rm out}$). 
The value of $\xi_{\rm out}$ is again chosen sufficiently far downstream such that 
$\phi'(\xi_{\rm out}) \simeq 0$. 
In this case, the equivalent of Eq.~\eqref{eq:canalie_inlet} becomes
\begin{equation}
 Q^{\rm out}_{\rm flow}(\lambda) = \frac{3Q_0}{H[1 + \phi(\xi_{\rm out})]}
 \left(1 - \frac{4\lambda^2}{H^2}\right).\label{eq:canalie_outlet}
\end{equation}
The construction of the outlet boundary conditions is analogous to the 
procedure described for the inlet.

\subsubsection{Specular reflection boundary conditions} \label{sec:canalie:bcond:specular}

Taking advantage of the symmetry of the channel, the simulation 
domain can be restricted to its upper half when specular 
boundary conditions are imposed at the centerline. This amounts to 
populating the nodes with $s \in \{0, -1, -2\}$ as follows:
\begin{align}
 f'_{ 0,p;i, j} =& f'_{1,p; \overline{\imath}, j}, \nonumber\\
 f'_{-1,p;i, j} =& f'_{2,p; \overline{\imath}, j}, \nonumber\\
 f'_{-2,p;i, j} =& f'_{3,p; \overline{\imath}, j},
\end{align}
and similarly for $f''_{s, p; i,j}$, where the notation $\overline{\imath}$ refers to the 
index corresponding to the momentum component $p^{\hat{\lambda}}_{\overline{\imath}}$ which 
satisfies:
\begin{equation}
 p^{\hat{\lambda}}_{\overline{\imath}} = -p^{\hat{\lambda}}_i.
\end{equation}

\subsubsection{Diffuse reflection boundary conditions}\label{sec:canalie:bcond:diffuse}

Diffuse reflection boundary conditions are implemented on the top boundary. Since the vielbein 
is constructed such that the $p^{\hat{\lambda}}_i$ component of the momentum is always 
perpendicular to the top wall, the procedure described in Sec.~\ref{sec:num_sch:bound} 
applies unchanged to this case. In particular, the values of the distributions 
in the ghost nodes are populated 
for the particles travelling back towards the fluid domain 
($p^{\hat{\lambda}}_i < 0$) following Eq.~\eqref{eq:weno_populate}:
\begin{multline}
 f'_{N_\lambda + 1,p;i,j} = f'_{N_\lambda + 2,p;i,j} = f'_{N_\lambda + 3,p;i,j} \\
 = f'_{\rm (eq); i,j}(n_{{\rm w};p}, \vu_{\rm w} = 0, T_{\rm w} = T_0), 
\end{multline}
while for the particles travelling towards the boundary, the 
second-order extrapolation given in Eq.~\eqref{eq:extrapol} is 
employed. Since the wall is at rest, we have $\vu_{\rm w} = 0$, 
while the temperature $T_0 = 1$ is that of
the initial state. The wall density $n_{{\rm w};p}$ is obtained 
using Eq.~\eqref{eq:nw}, as follows:
\begin{equation}
 n_{{\rm w}; p} = - \frac{\sum_{p^{\hat{\lambda}}_i > 0} \sum_j 
 V^{\lambda}_{N_\lambda + 1/2, p; i,j} \mathcal{F}_{\widetilde{\lambda};N_\lambda + 1/2,p; i, j}}
 {\sum_{p^{\hat{\lambda}}_i < 0} \sum_j f'_{\rm (eq)}(n = 1, 0, T_0) 
 V^{\lambda}_{N_\lambda + 1/2, p; i,j}},
\end{equation}
where $\mathcal{F}_{\widetilde{\lambda}; N_\lambda + 1/2,p; i, j}$ is the 
flux along the $\lambda$ direction corresponding to the velocity 
$V^{\lambda}_{N_\lambda+1/2,p;i,j}$ \eqref{eq:canalie_V}.

\subsection{Hydrodynamic regime: validation}\label{sec:canalie:hydro}

\begin{figure}
\begin{tabular}{c}
 \includegraphics[width=0.45\textwidth]{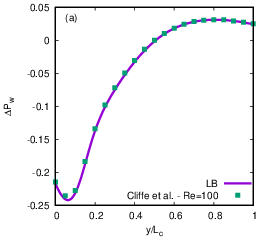}\\

 \includegraphics[width=0.45\textwidth]{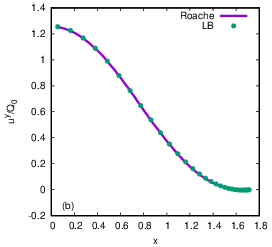} 
\end{tabular}
\caption{Simulation results for the flow through the 
gradually expanding channel with ${\rm Re}_c = 100$, 
$Q_0 = 0.1$ and ${\rm Kn} = 0.001$. 
(a) Normalized wall pressure $\Delta P_{\rm w}$ \eqref{eq:canalie_wall_pressure} 
with respect to the normalized coordinate 
$y / L_c = 3y/{\rm Re}_c$ along the channel, 
validated against the results reported by Cliffe \cite{cliffe82}.
(b) Normalized streamwise velocity $u^y / Q_0$ 
at $\xi = 100 / 12 \simeq 8.33$, validated against the results reported by 
Roache \cite{roache82}.
\label{fig:canalie_hydro}
}
\end{figure}

In this Section, our implementation is validated against 
results obtained in the incompressible limit of the 
Navier-Stokes equations, in the case when 
${\rm Re} = {\rm Re}_c = 100$.
 In order to achieve the incompressible Navier-Stokes regime,
we set $u_0 = Q_0 = 0.1$ and ${\rm Kn} = 10^{-3}$, which 
corresponds to ${\rm Re} = 100$ according to Eq.~\eqref{eq:canalie_Re}.
The results reported in this section are obtained using the 
 ${\rm H}(2;3) \times {\rm H}(2;3)$ model, employing 
 $3\times 3 = 9$ velocities.

In the incompressible (low ${\rm Ma}$) regime,
the continuity equation reduces to $\nabla \cdot \bm{u} = 0$, 
which allows the fluid velocity in planar flows 
to be determined from the vector potential $\bm{\Psi}_{\rm inc} = \bm{k} \psi_{\rm inc}$ 
through $\bm{u} = \nabla \times \bm{\Psi}_{\rm inc}$, such that 
$u_x = \partial_y \psi_{\rm inc}$ and 
$u_y = -\partial_x \psi_{\rm inc}$ \cite{kundu15}. 
However, $\nabla \cdot \bm{u} = 0$
holds only approximately in gas flows. In the kinetic theory
approach, the fluid always presents some degree of compressibility. 
Thus, the correct stream function is computed by noting that in the stationary 
limit, the continuity equation entails:
\begin{equation}
 \nabla \cdot (\rho \bm{u}) = 0.
\end{equation}
The above equation allows the product $\rho \bm{u}$ to be written 
as the curl of the vector potential $\bm{\Psi} = \bm{k} \psi$:
\begin{equation}
 \rho \bm{u} = \nabla \times \bm{\Psi},
\end{equation}
such that \cite{kundu15}:
\begin{equation}
 \rho u_x = \partial_y \psi, \qquad \rho u_y = -\partial_x \psi.
 \label{eq:canalie_psi}
\end{equation}
The stream function $\psi$ can be constructed starting from $\rho u_y = -\partial_x \psi$. 
Setting $\psi= 0$ on the channel centerline 
($s = 1/2$), $\psi$ can be integrated along each line of constant $\xi$ as follows:
\begin{equation}
 \psi_{s+1/2,p} = \psi_{s-1/2,p} - \rho_{s,p} u^y_{s,p} 
 (\lambda_{s+1/2,p} - \lambda_{s-1/2,p}),
 \label{eq:canalie_psi_sol}
\end{equation}
where the Cartesian components $u^x$ and $u^y$ are obtained from the vielbein components 
$u^{\hat{\lambda}}$ and $u^{\hat{\xi}}$ using:
\begin{equation}
 u^x = \frac{u^{\hat{\lambda}} + \lambda \phi'(\xi) u^{\hat{\xi}}}
 {\sqrt{1 + \lambda^2 \phi'{}^2(\xi)}},\qquad 
 u^y = \frac{u^{\hat{\xi}} - \lambda \phi'(\xi) u^{\hat{\lambda}}}
 {\sqrt{1 + \lambda^2 \phi'{}^2(\xi)}}. 
 \label{eq:canalie_u}
\end{equation}

The streamlines corresponding to the gradually expanding channel 
with ${\rm Re}_c = 100$ obtained from a simulation performed with 
the ${\rm H}(2;3) \times {\rm H}(2;3)$ model (employing $9$ 
velocities) are shown in Fig.~\ref{fig:canalie_stream} and 
a good agreement can be seen with the results 
obtained using the D2Q9 LB model 
in Ref.~\cite{hejranfar17cf}. The inlet 
and outlet boundary conditions were imposed at $\xi_{\rm in} = -10$ 
and $\xi_{\rm out} = 40$, respectively, and 
$N_\lambda \times N_\xi= 30 \times 200$ nodes were employed.

We first consider the validation of our numerical results 
by considering the pressure on the channel wall
$P_{{\rm w}; p} \equiv P_{N_\lambda + 1/2; p}$, which 
is obtained via linear extrapolation 
along the $\lambda$ direction from the inner nodes:
\begin{multline}
 P_{{\rm w}; p} 
 = \frac{(x_{\rm w} - x_{N_\lambda-1}) P_{N_\lambda, p}
 - (x_{\rm w} - x_{N_\lambda})P_{N_\lambda - 1, p}}
 {x_{N_\lambda} - x_{N_\lambda - 1}},
\end{multline}
where $x_w \equiv x_{N_\lambda + 1/2}$ is the wall coordinate. 
The value $P_{{\rm w}; c}$ of the wall 
pressure at the center of the channel (where 
$\xi = {\rm Re}_c / 6 \simeq 16.67$) is further subtracted from 
$P_{{\rm w}; p}$ and the result is divided by $\rho_0 u_0^2$ 
in order to conform with the non-dimensionalization conventions 
employed in Ref.~\cite{cliffe82}:
\begin{equation}
 \Delta P_{{\rm w};p} = \frac{P_{{\rm w};p} - P_{{\rm w}; c}}{\rho_0 u_0^2}.
 \label{eq:canalie_wall_pressure}
\end{equation}
It can be seen in Fig.~\ref{fig:canalie_hydro}(a) that our numerical results 
for $\Delta P_{{\rm w};p}$ are in very good agreement with the benchmark data 
reported by Cliffe \cite{cliffe82}.

Figure~\ref{fig:canalie_hydro}(b) validates our results for the 
normalized downstream velocity $u^y/Q_0$ \eqref{eq:canalie_u} at 
$\xi = {\rm Re}_c / 12 \simeq 8.33$, by comparing with the results 
reported by Roache \cite{roache82}. An excellent agreement can be 
seen.

\subsection{Compressibility effects}\label{sec:canalie:compress}

\begin{figure}
 \includegraphics[width=0.45\textwidth]{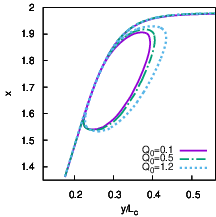}
 \caption{Streamlines for the flow through the gradually expanding channel 
 corresponding to ${\rm Re}_c = 100$, obtained for
 $(Q_0, {\rm Kn}) \in \{(0.1,0.001); (0.5, 0.005); (1.2, 0.012)\}$,
 such that ${\rm Re} = 100$, highlighting the 
 outermost contour of the vortex. Only the region around the vortex in the 
 upper half of the channel is represented.
 \label{fig:canalie_psiq}}
\end{figure}

\begin{figure*}
\begin{tabular}{cc}
  \includegraphics[width=0.45\textwidth]{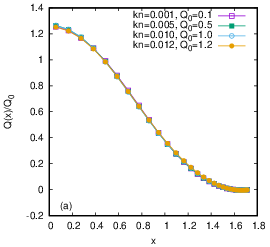}
&
  \includegraphics[width=0.45\textwidth]{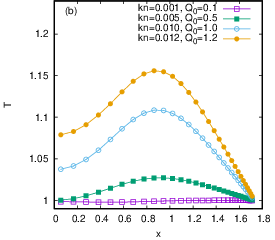}
\end{tabular}
\begin{tabular}{c}
 \includegraphics[width=0.45\textwidth]{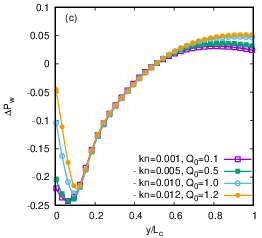}
 
\end{tabular}
\caption{Numerical results for the gradually expanding channel flow 
for (a) normalized local particle flow rate 
$Q(x) / Q_0$ and (b) temperature $T$ across the channel at $\xi = {\rm Re}_c / 12$,
as well as (c) the normalized wall pressure 
difference $\Delta P_{\rm w}$ \eqref{eq:canalie_wall_pressure} 
against the normalized streamwise coordinate $y/L_c = 3y/{\rm Re}_c$, 
at various values of ${\rm Kn}$. The particle flow rate is 
varied according to $Q_0 = 100\ {\rm Kn}$ in order to maintain 
${\rm Re} = 100$ for all simulations.
\label{fig:canalie_Q}}
\end{figure*}

In order to probe the compressible, variable temperature regime of the 
Navier-Stokes equations, we consider four values for the inlet particle 
flow rate, namely $Q_0 = 0.1, 0.5, 1$ and $1.2$.
The value of the Reynolds number is kept at ${\rm Re} = 100$, such that the Knudsen 
number ${\rm Kn}$ is increased, taking the values $0.001$, $0.005$, $0.01$ and 
$0.012$ by virtue of Eq.~\eqref{eq:canalie_Re}. The simulation corresponding to 
${\rm Kn} = 0.001$ was performed 
using the ${\rm H}(2;3) \times {\rm H}(2;3)$ model, while for 
${\rm Kn} = 0.005, 0.01$ and $0.012$, the 
${\rm HH}(3;4) \times {\rm H}(4;5)$ model was employed.

Using Eq.~\eqref{eq:canalie_psi_sol} to compute the stream function $\psi$,
its isocontours corresponding to the outermost closed loops 
of the vortices corresponding to $Q_0 = 0.1$, $0.5$ and $1.2$
are represented in Fig.~\ref{fig:canalie_psiq} 
with purple, green and cyan, respectively.
It can be seen that as $Q_0$ is increased, the vortex is enlarged.

The profile of the normalized local particle flow rate 
$Q(x) / Q_0$ at $\xi = {\rm Re}_c / 12$
is shown in Fig.~\ref{fig:canalie_Q}(a). It can be seen that, for the 
values of ${\rm Kn}$ considered in this Subsection, $Q(x) / Q_0$
is independent of ${\rm Kn}$ and $Q_0$, as long 
as ${\rm Re} = 100$ is kept constant. Thus, the flow remains in the hydrodynamic 
regime even for ${\rm Kn} = 0.012$.
The temperature profile shown in Fig.~\ref{fig:canalie_Q}(b) 
has a non-monotonic behaviour with respect to $x$, exhibiting 
a point of maximum around $x \simeq x_{\rm top} / 2$, where 
$x_{\rm top} \simeq 1.718$. 
Finally, the normalized pressure difference $\Delta P_{\rm w}$
is shown in Fig.~\ref{fig:canalie_Q}(c). It can be seen that $\Delta P_{\rm w}$ 
increases at the onset of the expansion (around $\xi = 0$), as well as towards 
the outlet.

\subsection{Rarefaction effects}\label{sec:canalie:raref}

\begin{figure*}
\begin{tabular}{c}
\begin{tabular}{cc}
 \includegraphics[width=0.45\textwidth]{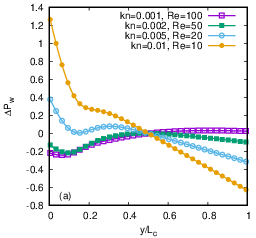}&
 \includegraphics[width=0.45\textwidth]{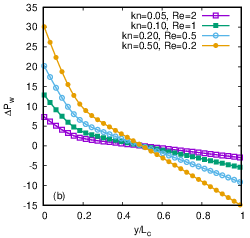}
 \end{tabular}\\
 \includegraphics[width=0.45\textwidth]{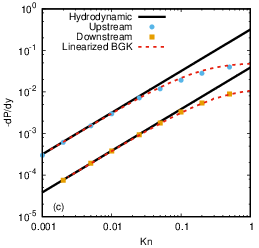} 
\end{tabular}
\caption{
Gradually expanding channel flow results 
for $\Delta P_{\rm w}$ \eqref{eq:canalie_wall_pressure} 
in the hydrodynamic (a) and 
slip flow (b) regimes with respect to the normalized 
downstream coordinate $y/L_c = 3 y / {\rm Re}_c$, as well as (c) for 
$-dP/dy$ in the upstream ($y/L_c < 0$) and downstream ($y/L_c > 0.5$) 
regions. The hydrodynamic limit curves 
$-dP/dy \simeq 0.317\ {\rm Kn}$ (upstream, $\phi \simeq -0.018$) and 
$-dP/dy \simeq 0.0385\ {\rm Kn}$ (downstream, $\phi \simeq 0.982$)
are obtained from Eq.~\eqref{eq:canalie_pois}.
The linearized Boltzmann-BGK results for the pressure-driven Poiseuille 
flow are represented with red dotted lines and are computed using 
Eq.~\eqref{eq:canalie_pois_GP} using the values for $G_{\rm P}^*$ 
reported in Refs.~\cite{cercignani66,lo82}.
The half-channel particle flow rate is taken as $Q_0 = 0.1$.
\label{fig:canalie_kn_press}}
\end{figure*}

\begin{table}
\begin{tabular}{rrrr}
 ${\rm Kn}$ & Model & $N_{\rm vel}$ & $\delta t$\\\hline\hline
 $0.001$ & ${\rm H}(2;3) \times {\rm H}(2;3)$ & $9$ & $10^{-3}$ \\
 $0.002$ & ${\rm H}(4;5) \times {\rm H}(4;5)$ & $25$ & $10^{-3}$ \\
 $0.005$ & ${\rm HH}(3;4) \times {\rm H}(4;5)$ & $40$ & $2 \times 10^{-3}$ \\
 $0.01$ & ${\rm HH}(3;4) \times {\rm H}(4;5)$ & $40$ & $2 \times 10^{-3}$ \\
 $0.05$ & ${\rm HH}(4;8) \times {\rm H}(4;5)$ & $80$ & $ 10^{-3}$ \\
 $0.1$ & ${\rm HH}(4;12) \times {\rm H}(4;5)$ & $120$ & $ 10^{-3}$ \\
 $0.2$ & ${\rm HH}(4;20) \times {\rm H}(4;5)$ & $200$ & $5 \times 10^{-4}$ \\
 $0.5$ & \hspace{10pt} ${\rm HH}(4;40) \times {\rm H}(4;5)$ & 
 \hspace{10pt} $400$ & \hspace{10pt} $5 \times 10^{-4}$ \\\hline
\end{tabular}
\caption{
Mixed quadrature LB models,
total number of velocities $N_{\rm vel}$ 
and time step $\delta t$ employed for the study 
of rarefaction effects in the expanding channel
in Subsec.~\ref{sec:canalie:raref}. The inlet half-channel 
mass flow rate is kept at $Q_0 = 0.1$.
\label{tab:canalie_raref}}
\end{table}

In this Subsection, the capabilities of our models to capture 
non-equilibrium flows are highlighted by performing simulations 
at fixed mass flow rate $Q_0 = 0.1$ for various values of the 
Knudsen number, taken between $0.001 \le {\rm Kn} \le 0.5$.
The models employed in order to conduct these simulations are 
summarized in Table~\ref{tab:canalie_raref}.
The aim of this Subsection is to highlight the transition 
from the hydrodynamic to the rarefied regime 
as the Knudsen layer develops at the diffuse reflective 
boundary. Even though ${\rm Re}$ decreases as ${\rm Kn}$ is increased
according to Eq.~\eqref{eq:canalie_Re}, the simulations are performed 
in the channel corresponding to ${\rm Re}_c = 100$.

We begin this Section with a discussion of the pressure. 
In the limit when the inlet and outlet are positioned 
sufficiently far away, the flow configuration is comprised of
two pressure-driven Poiseuille flow regions separated by 
the expanding portion between them. 

Around the expanding portion and for ${\rm Kn} \lesssim 0.01$, 
the pressure profile exhibits a non-monotonic behaviour, as shown in
Fig.~\ref{fig:canalie_kn_press}(a).
This kind of behaviour was also observed in simulations of 
the micro-orifice flow performed using the Direct Simulation 
Monte Carlo (DSMC) and the Gas-Kinetic Unified Algorithm (GKUA) 
in Refs.~\cite{wang04} and \cite{hou18}, respectively.
As ${\rm Kn}$ is increased, the effect of the expanding portion becomes 
negligible and the pressure profiles decrease monotonically with 
$\xi$, as shown in Fig.~\ref{fig:canalie_kn_press}(b). 

Far from the expanding region, 
the pressure decreases linearly with respect to the streamwise 
coordinate $y$. In the hydrodynamic regime, 
the pressure gradient is given by \cite{rieutord15}:
\begin{equation}
 \frac{dP}{dy} = -\frac{12 \mu Q_{\rm tot}}{n \ell^3} = 
 -\frac{3 Q_0}{(1 + \phi)^3} {\rm Kn},\label{eq:canalie_pois}
\end{equation}
where $Q_{\rm tot} = 2Q_0$ is the particle flow rate through the 
full channel width $\ell = H(1 + \phi)$, while 
$\phi(y \ll 0) \simeq -0.018$ and $\phi(y \gg 0) \simeq 0.982$ 
in the upstream and downstream regions from the expanding portion. 
Outside the hydrodynamic regime, the relation between the 
pressure gradient and the Knudsen number is more complicated. 
Introducing the notation:
\begin{equation}
 \frac{dP}{dy} = -\frac{m Q_{\rm tot} v_0}{\ell^2 G_{\rm P}^*} 
 = -\frac{Q_0}{\sqrt{2}(1 + \phi)^2 G_{\rm P}^*},
 \label{eq:canalie_pois_GP}
\end{equation}
where $v_0 = \sqrt{2 K_B T_0 / m} = \sqrt{2}$ is the most probable 
speed and $\ell = H(1 + \phi) = 2 (1+ \phi)$ is the channel width,
the dependence of the pressure gradient on the Knudsen number 
is contained in the Poiseuille coefficient $G_{\rm P}^*$ 
\cite{sharipov16}. In the linearized limit of the slip regime, 
$G_{\rm P}^*$ can be written as:
\begin{equation}
 G_{\rm P}^* = \frac{\delta}{6} + \sigma_{\rm P},\label{eq:GP}
\end{equation}
where the rarefaction parameter $\delta$ depends on the local channel 
width and Knudsen number ${\rm Kn}$ through:
\begin{equation}
 \delta = \frac{\ell}{{\rm Kn} \sqrt{2}} = \frac{\sqrt{2}}{\rm Kn}[1 + \phi(y)].
 \label{eq:rarefaction_delta}
\end{equation}
The value of $\sigma_{\rm P}$ in Eq.~\eqref{eq:GP} depends on the particle-wall 
interaction, having the value $\sigma_{\rm P} \simeq 1.0162$ for diffuse reflection
\cite{cercignani75,lo82,sharipov99pois,sharipov16}. In the transition and 
free molecular flow regimes, the values of $G_{\rm P}^*$ can be computed 
numerically or semianalytically and are tabulated in a variety of 
papers, of which we recall \cite{cercignani66,lo82,sharipov99pois,sharipov16},
where the linearized limit of the Boltzmann-BGK equation is considered.
The values of $-dP/dy$ obtained from our numerical results 
far upstream and far downstream from the expanding portion are compared 
with the hydrodynamic limit \eqref{eq:canalie_pois} and
the general formula \eqref{eq:canalie_pois_GP} 
in Fig.~\ref{fig:canalie_kn_press}(c), where the values of $G_{\rm P}^*$ 
correspond to the linearized limit of the pressure-driven Poiseuille flow 
and are taken from Refs.~\cite{cercignani66,lo82}. It can be 
seen that the increase of the absolute value of the pressure gradient 
$-dP/dy$ is linear in ${\rm Kn}$ for 
${\rm Kn} \lesssim 0.05$, while for ${\rm Kn} \gtrsim 0.05$, 
$-dP/dy$ increases at a much slower rate, in good agreement with the 
behaviour predicted in Refs.~\cite{cercignani66,lo82,sharipov99pois,sharipov16}. 
This is the first indication 
that at ${\rm Kn} \gtrsim 0.05$, the rarefaction effects become important.

\begin{figure}
\begin{tabular}{c}
 \includegraphics[width=0.45\textwidth]{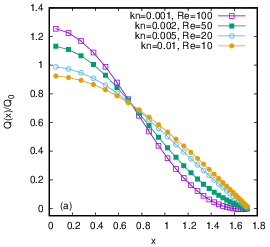}\\
 \includegraphics[width=0.45\textwidth]{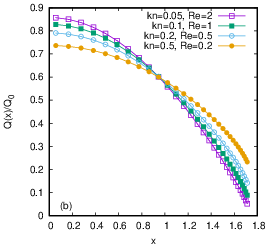}
\end{tabular}
\caption{Numerical results for the normalized local particle flow rate $Q(x) / Q_0$
across the channel at $\xi \simeq 8.33$ in the hydrodynamic (a) and 
slip flow (b) regimes for the gradually expanding channel flow 
corresponding to ${\rm Re}_c = 100$ in Eq.~\eqref{eq:canalie_phi}. 
The half-channel particle flow rate is taken as $Q_0 = 0.1$ for various 
values of ${\rm Kn}$, such that the resulting Reynolds number decreases as 
${\rm Kn}$ is increased.\label{fig:canalie_kn_ux}}
\end{figure}

The normalized local particle flow rate profile at $y = {\rm Re}_c/12$
is shown in Fig.~\ref{fig:canalie_kn_ux} 
for various values of ${\rm Kn}$. The presence of the vortex in 
the ${\rm Kn} = 0.001$ simulation (corresponding to ${\rm Re} = 100$ for 
the flow) is highlighted by the negative values attained by 
$u^y$ close to the boundary. For ${\rm Kn} \gtrsim 0.002$, ${\rm Re}$ 
is significantly decreased, the vortex no longer forms and $u^y$ 
decreases monotonically from the channel centerline towards the boundary. 
In the hydrodynamic flow  regime shown in 
Fig.~\ref{fig:canalie_kn_ux}(a) (${\rm Kn} \lesssim 0.01$),
the particle flow rate regains a parabolic profile as 
${\rm Kn}$ is increased,
while the slip velocity at the wall remains 
negligible. Figure~\ref{fig:canalie_kn_ux}(b) 
shows that the slip velocity becomes 
non-negligible as ${\rm Kn} \gtrsim 0.05$, when the rarefaction effects 
become important, as also noted in the previous paragraph regarding the 
pressure profile.

\begin{figure}
\begin{tabular}{c}
 \includegraphics[width=0.45\textwidth]{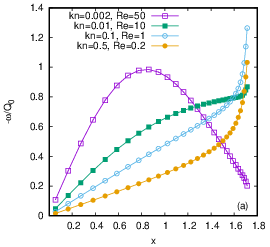}\\
 \includegraphics[width=0.45\textwidth]{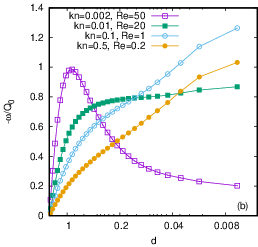}
\end{tabular}
\caption{Numerical results for the normalized vorticity 
$-\omega / Q_0$ as a function of (a) $x$; and 
(b) $d$ \eqref{eq:canalie_delta}, taken at 
$y = {\rm Re}_c / 16 \simeq 8.33$, where 
${\rm Re}_c = 100$ defines the channel geometry through 
Eq.~\eqref{eq:canalie_phi}. The half-channel particle flow rate 
is taken as $Q_0 = 0.1$ for various values of ${\rm Kn}$, such that the 
resulting Reynolds number of the flow decreases as ${\rm Kn}$ is increased.
\label{fig:canalie_vor_xi833_kn}}
\end{figure}

The previous discussion of the particle flow rate profile clearly highlights the 
development of the Knudsen layer as ${\rm Kn}$ is increased 
above $\sim 0.05$. In order to better assess the capability
of our models to capture the physics of the Knudsen layer, we note that 
the velocity receives contributions of the form
$d \ln d$ inside the Knudsen layer, where $d$ measures the distance from the wall 
\cite{gross57,sone64,yap12,wei15,jiang16}. While this 
term is difficult to highlight when discussing the velocity 
profile, it becomes dominant in the profile of the vorticity 
$\omega = \partial_x u_y - \partial_y u_x$, which can be written as:
\begin{equation}
 \omega = -\frac{\partial u^x}{\partial \xi} + \frac{1}{1 + \phi(\xi)} 
 \left[\frac{\partial u^y}{\partial \lambda} + 
 \lambda \phi'(\xi) \frac{\partial u^x}{\partial \lambda}\right].
 \label{eq:canalie_vorti}
\end{equation}
The derivatives with respect to $\xi$ are computed using centered differences 
and the second order forward or backward Euler scheme at the inlet and outlet 
nodes, respectively. For the derivatives with respect to the non-equidistantly distributed 
$\lambda$ coordinate, we used the following scheme for bulk nodes 
($1 < s < N_\lambda$):
\begin{multline}
 \left(\frac{\partial f}{\partial \lambda}\right)_{s,p} =  
 \frac{(\lambda_s - \lambda_{s-1})f_{s+1,p}}
 {(\lambda_{s+1}-\lambda_s)(\lambda_{s+1}-\lambda_{s-1})} \\
 + \frac{(\lambda_{s+1} - 2\lambda_s + \lambda_{s-1})f_{s,p}}
 {(\lambda_{s+1} - \lambda_s)(\lambda_s - \lambda_{s-1})}\\
 -\frac{(\lambda_{s+1}-\lambda_s)f_{s-1,p}}
 {(\lambda_{s+1} - \lambda_{s-1})(\lambda_s - \lambda_{s-1})}.
\end{multline}
In the first node ($s = 1$), the following formula is used:
\begin{multline}
 \left(\frac{\partial f}{\partial \lambda}\right)_{1,p} =  
 -\frac{(\lambda_2 + \lambda_3 - 2\lambda_1)f_{1,p}}
 {(\lambda_2-\lambda_1)(\lambda_3-\lambda_1)} \\
 + \frac{(\lambda_3 - \lambda_1)f_{2,p}}
 {(\lambda_2 - \lambda_1)(\lambda_3 - \lambda_2)}
 -\frac{(\lambda_2-\lambda_1)f_{3,p}}
 {(\lambda_3 - \lambda_1)(\lambda_3 - \lambda_2)}. 
\end{multline}
The derivative in the last node ($s = N_\lambda$) is computed using:
\begin{multline}
 \left(\frac{\partial f}{\partial \lambda}\right)_{N_\lambda,p} =  
 -\frac{(2\lambda_{N_\lambda} - \lambda_{N_\lambda-1}- \lambda_{N_\lambda-2})f_{N_\lambda,p}}
 {(\lambda_{N_\lambda}-\lambda_{N_\lambda-1})(\lambda_{N_\lambda}-\lambda_{N_\lambda-2})} \\
 - \frac{(\lambda_{N_\lambda} - \lambda_{N_\lambda-2})f_{N_{\lambda}-1,p}}
 {(\lambda_{N_\lambda} - \lambda_{N_\lambda-1})(\lambda_{N_\lambda-1} - \lambda_{N_\lambda-2})}\\
 +\frac{(\lambda_{N_\lambda}-\lambda_{N_\lambda-1})f_{N_{\lambda-2},p}}
 {(\lambda_{N_\lambda} - \lambda_{N_\lambda-2})(\lambda_{N_\lambda-1} - \lambda_{N_\lambda-2})}. 
\end{multline}

Due to the logarithmic singularity of the gradient of the velocity, 
the vorticity cannot be defined on the diffuse reflective boundary. 
The logarithmic divergence of the vorticity is highlighted 
in Fig.~\ref{fig:canalie_vor_xi833_kn} with respect to (a) the distance 
$x$ from the channel center and (b) the non-dimensionalized distance $d$ 
to the top wall, defined through:
\begin{equation}
 d = 1 + \phi(y) - \frac{2 x}{H}.\label{eq:canalie_delta}
\end{equation}
At ${\rm Kn} = 0.002$, no evidence of the Knudsen layer can 
be seen. This is due to the fact that the point which is closest to the 
boundary is at a non-dimensionalized distance $d \simeq 0.0055$ from the boundary, 
while at ${\rm Kn} = 0.002$, the Knudsen layer is localized 
closer to the boundary. When ${\rm Kn} \gtrsim 0.01$, the 
Knudsen layer becomes visible especially in 
Fig.~\ref{fig:canalie_vor_xi833_kn}(a), where the rapid increase of 
$-\omega$ in the vicinity of the wall can be clearly seen. At ${\rm Kn} = 0.5$, 
$-\omega$ increases roughly linearly with respect to $-\ln d$, except 
for the last few nodes, which may be affected by numerical effects 
caused by our formulation of the diffuse reflection boundary conditions.

\begin{figure}
\begin{tabular}{c}
\includegraphics[width=0.45\textwidth]{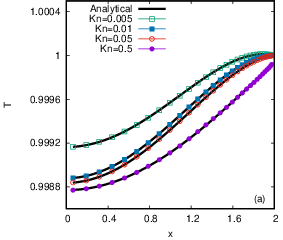}\\
\includegraphics[width=0.45\textwidth]{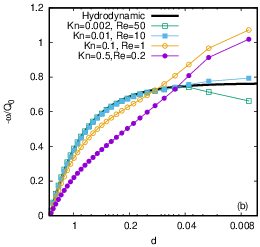}
\end{tabular}
\caption{Numerical results for the temperature $T$(a) 
and normalized vorticity $-\omega/Q_0$(b)
across the channel at $\xi \simeq 33$ 
for the gradually expanding channel flow 
corresponding to ${\rm Re}_c = 100$ in Eq.~\eqref{eq:canalie_phi}. 
The half-channel particle flow rate is taken as $Q_0 = 0.1$ for various 
values of ${\rm Kn}$, such that the resulting Reynolds number decreases as 
${\rm Kn}$ is increased.\label{fig:canalie_kn_outlet}}
\end{figure}

We finally consider the analysis of the flow far downstream from the expanding region.
At $y = {\rm Re} / 3$, the flow enters the regime of the Poiseuille flow. At non-negligible 
values of ${\rm Kn}$, the temperature profile for the Poiseuille flow 
between parallel plates can be written as \cite{ambrus16jocs,mansour97,hess99}:
\begin{equation}
 T(x) = T_0 + \alpha x^2 + \beta x^4,
 \label{eq:canalie_temp}
\end{equation}
where $T_0$ is the temperature on the centerline. 
The bimodal profile for the temperature occurs as a 
rarefaction effect and was shown in Ref.~\cite{xu03pof} to be 
accounted for only at super-Burnett level.
After fitting $T_0$, $\alpha$ and $\beta$ to the numerical data,
it can be seen in Fig.~\ref{fig:canalie_kn_outlet}(a)
that the fluid temperature falls below the temperature of the channel wall.
This effect was also observed in Refs.~\cite{karniadakis05,xu03pof,zheng02jsp,sofonea06epl}
and is due to the fact that the viscous heating is superseded by the gas
expansion \cite{zheng02jsp}.
In the hydrodynamic regime, the streamwise velocity $u^y$ is
approximately given by an expression similar to Eq.~\eqref{eq:canalie_inlet_def},
such that the vorticity becomes:
\begin{equation}
 \omega_{\rm Pois} = -\frac{3u_0 x}{x_{\rm top}^2}.
 \label{eq:canalie_vorti_hydro}
\end{equation}
It can be seen in Fig.~\ref{fig:canalie_kn_outlet}(b) that the 
results corresponding to ${\rm Kn} = 0.002$ and $0.01$ agree very 
well with the hydrodynamic prediction \eqref{eq:canalie_vorti_hydro},
except for the last few nodes which may receive errors from our formulation 
of the boundary conditions. At ${\rm Kn} \gtrsim 0.1$, the 
effects of the Knudsen layer become visible as the magnitude of
the vorticity $-\omega$ increases almost linearly with $-\ln d$.

\subsection{Cartesian decomposition of the momentum space} \label{sec:canalie:coord}

Let us now analyze the case when the momentum space is discretized with respect to its
Cartesian degrees of freedom $(p^x, p^y)$. Making the coordinate change from 
$(x, y)$ to $(\lambda, \xi)$, the Boltzmann equation becomes:
\begin{equation}
 \frac{\partial f}{\partial t} + 
 \frac{p^{\widetilde{\lambda}}}{m} \frac{\partial f}{\partial \lambda} + 
 \frac{p^{\widetilde{\xi}}}{m} \frac{\partial f}{\partial \xi} = 
 -\frac{1}{\tau}[f - \feq],
 \label{eq:canalie_coord_beq}
\end{equation}
where $p^{\widetilde{\lambda}}$ and $p^{\widetilde{\xi}}$ are given in Eq.~\eqref{eq:canalie_pcoord}.
Equation~\eqref{eq:canalie_coord_beq} can be put in conservative form as follows:
\begin{equation}
 \frac{\partial f}{\partial t} + \frac{\partial (V^\lambda f)}{\partial \lambda} + 
 \frac{\partial (V^\xi f)}{\partial \chi^\xi} = -\frac{1}{\tau}[f - \feq],
 \label{eq:canalie_coord_beq_cons}
\end{equation}
where $\chi^\xi$ is defined in Eq.~\eqref{eq:chi_canalie} and
\begin{equation}
 V^\lambda = \frac{p^x - \lambda \phi' p^y}{m(1 + \phi)},\qquad 
 V^\xi = (1 + \phi) \frac{p^y}{m}.
\end{equation}
The advantage of the Boltzmann equation \eqref{eq:canalie_coord_beq_cons} 
written with respect to the original Cartesian components $(p^x, p^y)$ of the momentum 
space is that the force terms appearing in the vielbein equivalent \eqref{eq:canalie_boltz}
are absent. Thus, the coefficient $a_5$ corresponding to the computation of the force 
term can be set to $0$ in the runtime estimate given by Eq.~\eqref{eq:msites_DT}.
However, we anticipate that this apparent improvement of the runtime is 
compensated by increased quadrature orders, as will be discussed below.

The drawback when the vielbein formalism is not employed is that the diffuse 
reflection boundary conditions 
must be implemented judging by the sign of a linear combination of 
$p^x$ and $p^y$. Considering that the momentum space is discretized using Gauss 
quadratures of orders $Q_x$ and $Q_y$ with respect to $p_x$ and $p_y$, respectively,
the density $n_w$ required to construct the wall populations is computed using:
\begin{equation}
 n_w = -\frac{\displaystyle \sum_{V^\lambda_{N_\lambda + 1/2, p; i,j} > 0} 
 \mathcal{F}_{\lambda; N_{\lambda + 1/2}, p; i, j} 
 V^\lambda_{N_\lambda + 1/2, p; i,j}}
 {\displaystyle \sum_{V^\lambda_{ N_\lambda + 1/2, p; i,j} < 0} 
 f'_{\rm (eq)}(n = 1,0, T_0) V^\lambda_{N_\lambda + 1/2, p; i,j}},
 \label{eq:canalie_coord_nw}
\end{equation}
where the discretization of the spatial grid is performed as discussed in 
Subsec.~\ref{sec:canalie:gen}. In the regions where $\phi'$ is non-negligible,
$n_w$ must be computed by integrating over regions of the momentum space 
which are position-dependent. 

We now consider the flow through the 
gradually expanding channel corresponding to ${\rm Re}_c = 100$ in 
Eq.~\eqref{eq:canalie_phi}. As before, the flow region 
of interest is between $\xi = 0$ and $\xi = {\rm Re}_c / 3 \simeq 33.33$.
The inlet and outlet are positioned at $\xi_{\rm in} = -10$ and 
$\xi_{\rm out} = 40$, thus giving enough space for the flow to adjust 
itself before entering the region of interest. For definiteness, we 
consider $Q_0 = 0.1$ and ${\rm Kn} = 0.2$ for the remainder of this Subsection.
The channel is discretized using $N_\xi = 200$ equidistant points along the 
$\xi$ axis and $N_\lambda = 30$ points along the $\lambda$ direction, 
which are stretched according to Eq.~\eqref{eq:canalie_stretch} with
$A = 0.95$.

\begin{figure}
\begin{tabular}{c}
 \includegraphics[width=0.45\textwidth]{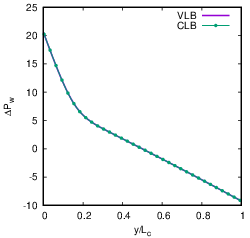}
\end{tabular}
\caption{Comparison of the simulation results 
for the normalized wall pressure $\Delta P_{\rm w}$ \eqref{eq:canalie_wall_pressure} 
obtained using the VLB model ${\rm HH}(4;6) \times {\rm H}(4;5)$ (solid line) 
and the CLB model ${\rm HH}(4;6) \times {\rm HH}(4;6)$ (line and points).
The results are overlapped.
\label{fig:canalie_joscoord_P}}
\end{figure}

\begin{figure}
\begin{tabular}{c}
 \includegraphics[width=0.45\textwidth]{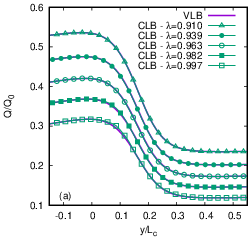}\\
 \includegraphics[width=0.45\textwidth]{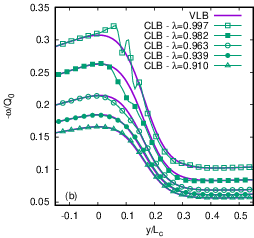}
\end{tabular}
\caption{Comparison of (a) the normalized mass flow rate 
$Q / Q_0$ and (b) the normalized vorticity $-\omega / Q_0$ at 
$\lambda=\{0.997,\,0.982,\,0.963,\,0.939,\,0.910\}$ obtained 
using the VLB model ${\rm HH}(4;6)\times {\rm H}(4;5)$ (solid lines) 
and the CLB model ${\rm HH}(4;6)\times {\rm HH}(4;6)$ (lines and points).
\label{fig:canalie_joscoord}}
\end{figure}

It can be expected that the differences between the vielbein-based lattice Boltzmann (VLB) and 
Cartesian split-based lattice Boltzmann (CLB) implementations will be most significant 
in the expanding region of the channel. Moreover, we expect that the VLB
implementation will be more accurate within the Knudsen layer.
In Fig.~\ref{fig:canalie_joscoord_P}, the normalized wall pressure 
$\Delta P_w$ \eqref{eq:canalie_wall_pressure} obtained using the 
VLB and CLB implementations at similar quadrature orders is shown. 
It can be seen that there are no visible discrepancies at the level of 
the wall pressure. Next, Fig.~\ref{fig:canalie_joscoord} shows a comparison 
of the VLB and CLB results for the normalized flow rate $Q / Q_0$ and 
vorticity $-\omega / Q_0$ around the expansion region, along lines of constant 
$\lambda$. In Fig.~\ref{fig:canalie_joscoord}(a), it can be seen that 
the flow rate results are in general in good agreement, apart from 
along the line which is closest to the wall ($\lambda = 0.997$),
where a small discrepancy can be seen in the expanding region 
(around $y / L_c \simeq 0.1$). Also in the expanding region, 
Fig.~\ref{fig:canalie_joscoord}(b) shows that the 
CLB results for the vor\-ti\-ci\-ty profile present oscillations with respect to $y / L_c$, which 
become more pronounced as the wall is approached. On the other hand, the VLB results 
vary smoothly with respect to $y / L_c$. 

The amplitude of the oscillations observed in the vorticity profile obtained using the 
CLB approach decrease as the quadrature order increases. Similarly, the results 
obtained using the VLB approach exhibit a convergence trend as the quadrature order 
is increased. For the study of the quadrature order dependence of $\omega$, we consider 
the transverse vorticity profile at fixed values of $y / L_c$ inside the expansion
region. 

\begin{figure}
\begin{tabular}{c}
 \includegraphics[width=0.45\textwidth]{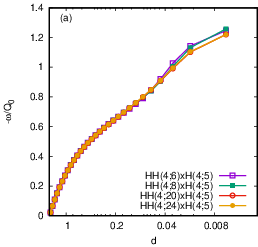}\\
 \includegraphics[width=0.45\textwidth]{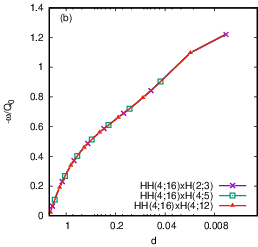}
\end{tabular}
\caption{Convergence study of the normalized vorticity 
with respect to quadrature orders $Q_\lambda$ (a) and 
$Q_\xi$ (b) for the VLB implementation at $y / L_c \simeq 0.25$.
\label{fig:canalie_vor_viel}}
\end{figure}

In Fig.~\ref{fig:canalie_vor_viel}, the typical convergence trend of the 
vorticity profile obtained using the VLB implementation is shown at 
$y/L_c = 0.25$ by varying $Q_\lambda$ at fixed $Q_\xi = 5$ (a) and by 
varying $Q_\xi$ at fixed $Q_\lambda = 16$ (b). The half-range and 
full-range Gauss Hermite quadratures are used on the $\lambda$
and $\xi$ directions, respectively. From Fig.~\ref{fig:canalie_vor_viel}(a),
it can be seen that convergence with respect to $Q_\lambda$ is achieved faster for 
the nodes closer to the channel center than for the nodes in the vicinity of the wall.
Figure~\ref{fig:canalie_vor_viel}(b) demonstrates the remarkable property that the
VLB results for the vorticity corresponding to a fixed value of $Q_\lambda$ are overlapped 
for all values of $Q_\xi \ge 3$. A similar property is also observed in the context 
of the Couette \cite{ambrus16jcp} and Poiseuille \cite{ambrus16jocs} flows 
between parallel plates. It is shared by the VLB implementation because 
the $p^{\hat{\xi}}$ momentum space direction is always parallel to the wall.
We note that $Q_\xi = 3$ is insufficient to capture the temperature 
profile shown in Fig.~\ref{fig:canalie_kn_outlet}(a). For small Mach number flows,
$Q_\xi = 4$ is in general sufficient to obtain accurate results, even for the 
temperature profile. When the Mach number is non-negligible (i.e., 
as considered in Fig.~\ref{fig:canalie_Q}), $Q_\xi = 5$ must be used. 
Our simulations indicate that further increasing the value of $Q_\xi$ 
does not affect the accuracy of the numerical results for all the flow 
parameters considered in this section.

\begin{figure}
\begin{tabular}{c}
 \includegraphics[width=0.45\textwidth]{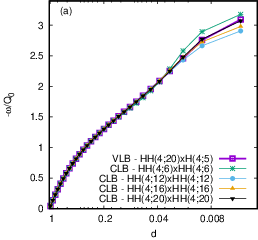}\\
 \includegraphics[width=0.45\textwidth]{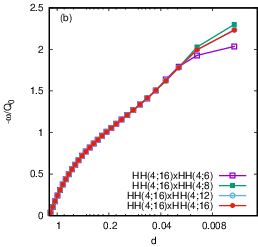}
 \end{tabular}
 \caption{Comparison of the VLB and CLB implementations. 
 Convergence study of the normalized vorticity 
 for the CLB implementation with respect to quadrature order by (a) steadily increasing 
 the quadrature order on both axes at $y / L_c \simeq 0.041 $ and 
 (b) keeping $Q_x$ fixed and varying $Q_y$ at $y / L_c \simeq 0.154$.
 \label{fig:canalie_vor_joscoord}}
 \end{figure}

In order to study the convergence trend of the CLB results, the transverse 
$\omega$ profile is represented in Fig.~\ref{fig:canalie_vor_joscoord}
at selected values of $y / L_c$. According to Eq.~\eqref{eq:canalie_coord_nw},
the computation of the density $n_w$ of the populations emerging from the 
wall back into the fluid requires the recovery of integrals 
over the half of the $(p^x,p^y)$ plane for which 
$p^x - \lambda \phi'(y) p^y$, such that the integration range does 
not cover the full $(-\infty, \infty)$ interval on either $p^x$ or $p^y$. 
Thus, the momentum space is discretized using the half-range Gauss-Hermite 
quadrature for both the $p^x$ and the $p^y$ degrees of freedom.
Figure~\ref{fig:canalie_vor_joscoord}(a)
shows that increasing $Q_x = Q_y$ simultaneously brings the CLB 
results towards the VLB results obtained using $Q_\lambda = 20$ and $Q_\xi = 5$,
confirming that at high quadrature orders, the VLB and CLB implementations 
yield similar results. However, Fig.~\ref{fig:canalie_vor_joscoord}(b) shows 
that, contrary to the VLB implementation, the accuracy 
of the CLB results depends strongly on $Q_y$. 
The results in Figs.~\ref{fig:canalie_vor_joscoord}(a) and 
\ref{fig:canalie_vor_joscoord}(b) are represented at 
$y / L_c \simeq 0.041 $ and $y / L_c \simeq 0.154$, respectively.

It is worth remarking that the profiles of the pressure 
$P$ and flow rate $Q$ can be recovered with much smaller 
quadrature orders compared to the profile of the vorticity $\omega$, even 
at non-negligible values of ${\rm Kn}$. Moreover, 
Figs.~\ref{fig:canalie_joscoord}(b) and \ref{fig:canalie_joscoord_P}
show that the fluctuations in the profiles of $Q$ and $P$ 
are almost negligible, even when the model 
${\rm HH}(4;6) \times {\rm HH}(4;6)$ is employed.

We end this section with a comparative analysis of the performance 
of the CLB and VLB implementations. Since the primary difference of 
these implementations is in the way the momentum space is discretized,
it is reasonable to compare their performance on the same spatial grid,
comprised of $N_\lambda \times N_\xi = 30 \times 200 = 6\,000$ nodes. In the VLB implementation,
the full-range Gauss-Hermite quadrature of order $Q_\xi = 5$ can be
employed along the flow direction, while the half-range Gauss-Hermite 
quadrature of order $Q_\lambda = Q$ is employed along the direction which
is perpendicular to the boundary. In order to ensure the same degree of 
accuracy between the VLB and CLB implementations, 
the half-range Gauss-Hermite quadrature must be employed on both axes
in the CLB implementation, with quadrature orders equal to the one 
employed in the VLB implementation, namely $Q_x = Q_y = Q$.
The total number of velocities in the VLB implementation is 
$N_{\rm vel}^{\rm VLB} = 10Q$, while 
in the CLB implementation, 
$N_{\rm vel}^{\rm CLB} = 4Q^2$ velocities are employed.
The time $\Delta T$ required to perform one iteration can be estimated 
as in Eq.~\eqref{eq:msites_DT} (after minor adjustments to 
account for a two-dimensional grid).
In the case of the VLB implementation, $\Delta T$ can be 
estimated through:
\begin{equation}
 \Delta T_{\rm VLB} = 10 a_{\rm v} Q +
 10 b_{\rm v} (2Q + 5) Q + c_{\rm v}, \label{eq:canalie_msites_DT_VLB}
\end{equation}
while in the case of the CLB implementation, the force term 
is absent ($b_c = 0$):
\begin{equation}
 \Delta T_{\rm CLB} = 4 a_{\rm c} Q^2 + c_{\rm c}. \label{eq:canalie_msites_DT_CLB}
\end{equation}
Formally, the algorithmic complexity of the VLB and CLB implementations 
is similar. At large values of $Q$, 
$\Delta T_{\rm VLB} / \Delta T_{\rm CLB} \simeq 5 b_v / a_c$, 
where $b_v$ and $a_c$ are the values of the coefficients $b$ and $a$ 
corresponding to the VLB and CLB implementations, respectively. 
In the context of the circular Couette flow, the analysis in 
Sec.~\ref{sec:couette:msites} shows that $5b / a \simeq 0.11$, 
thus it can be expected that the VLB implementation is roughly 
one order of magnitude faster than the CLB implementation.

In order to quantitatively assess the computational performance of the VLB and CLB 
implementations, we evaluate the number of million of sites 
updated per second (Msites/s) ${\rm MS}$ \eqref{eq:couette_MS}, which in 
the case of the gradually expanding channel reads:
\begin{equation}
 {\rm MS} = \frac{N_\lambda \times N_\xi}{10^6 \Delta T} = \frac{0.006}{\Delta T},
 \label{eq:canalie_MS}
\end{equation}
where $\Delta T$ is expressed in seconds.
In order to account for runtime fluctuations, we perform 
for each value of $Q$ a series of simulations with total 
number of iterations $N_{\rm iter}$ varying between 
$5 \le N_{\rm iter} \le 15$. For each simulation, the value of 
${\rm MS}$ is computed using the formula:
\begin{equation}
 {\rm MS}(N_{\rm iter}) = \frac{0.006 N_{\rm iter}}{T(N_{\rm iter})}, 
\end{equation}
where $T(N_{\rm iter})$ is the total runtime to complete $N_{\rm iter}$ iterations, 
expressed in seconds. The value of ${\rm MS}$ corresponding to a given 
quadrature order $Q$ is computed by averaging over the values
${\rm MS}(N_{\rm iter})$.
%

\begin{figure}
\includegraphics[width=0.45\textwidth]{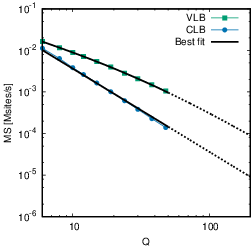}
\caption{ Number of millions of site updates per second 
in the context of the gradually expanding channel flow for a system with 
$N_\lambda \times N_\xi = 30 \times 200$ nodes when the 
VLB ${\rm HHLB}(4; Q) \times {\rm HLB}(4;5)$ (lines and squares)
and CLB ${\rm HHLB}(4; Q) \times {\rm HHLB}(4;Q)$ (lines and circles) 
models are employed.
The solid lines correspond to Eqs.~\eqref{eq:canalie_msites_DT_VLB}
and \eqref{eq:canalie_msites_DT_CLB}, where 
the parameters $a$, $b$ and $c$ are obtained using a fitting routine.
\label{fig:canalie_msites}}
\end{figure}

Figure~\ref{fig:canalie_msites} shows the dependence of ${\rm MS}$ with respect to 
$Q$ for the VLB (lines and squares) and 
CLB (lines and circles) implementations. 
The solid lines correspond to the best fits of Eqs.~\eqref{eq:canalie_msites_DT_VLB}
and \eqref{eq:canalie_msites_DT_CLB} to the numerical data. 
The results of the numerical fits for the particular case of 
a grid comprised of $N_\lambda \times N_\xi = 30 \times 200 = 6000$ nodes 
are $a_{\rm v} \simeq 4.97\ {\rm ms}$, $b_{\rm v} \simeq 0.071\ {\rm ms}$,
$a_{\rm c} \simeq 4.12\ {\rm ms}$, while the free coefficient $c$ 
appears to be negligible in both implementations. 
Thus, at large quadrature orders $Q$, 
it can be expected that the time per iteration ratio 
between the VLB and CLB implementations is $5b_{\rm v} / a_{\rm c} \simeq 0.086$.
For low Mach number flows, $Q_\xi$ can be decreased below the value $Q_\xi = 5$ 
considered above such that the time per iteration ratio becomes 
$Q_\xi b_{\rm v} / a_{\rm c} \simeq 0.0172 Q_{\xi}$.
Thus, it can be expected that the VLB implementation is in general at least 
one order of magnitude faster than the CLB implementation at the same level 
of accuracy.

\subsection{Summary} \label{sec:canalie:conc}
In this Section, the vielbein formalism was employed to study flows 
through channels with non-planar walls. In particular, we considered 
the case of the gradually expanding channel, for which the 
expanding Section is governed by a hyperbolic tangent. Adapting the 
coordinate system to the channel boundary induces a non-diagonal metric.
Our choice for the vielbein field allows the momentum space to be aligned 
along the boundary, such that the diffuse reflection boundary conditions 
can be implemented just like in the case of planar walls. 

Our implementation is validated in the incompressible hydrodynamics
limit, where our results obtained using the ${\rm H}(2;3) \times 
{\rm H}(2;3)$ model (employing $9$ velocities) are 
successfully compared with computational fluid dynamics (CFD) results.
We further presented results for the compressible hydrodynamics 
case, when the temperature is no longer a constant. Our analysis of the 
flow through the gradually expanding channel ends with an analysis of 
rarefaction effects. In particular, 
we highlight the deviations from the hydrodynamic solution of the 
pressure-driven flow in the case when the pressure gradient is 
no longer proportional to ${\rm Kn}$. We further validate the 
results for the temperature profile by successfully fitting a quartic 
function of the distance from the channel center to the numerical data. 
The ability of our implementation to capture rarefaction effects 
was demonstrated by highlighting the logarithmic divergence 
of the vorticity inside the Knudsen layer. 

Finally, we discuss the advantages of using the vielbein 
formalism (VLB) in contrast with the case when the momentum space is 
discretized with respect to its Cartesian degrees of freedom $(p^x, p^y)$ (CLB). 
In the context of the gradually expanding channel, the flow domain cannot be reduced 
to one dimension. However, the VLB formalism allows the momentum space to be 
factorized such that one component is always perpendicular to the wall.
Our analysis shows that this allows a full-range Gauss-Hermite quadrature 
of low order to be employed on the direction which remains parallel to the wall,
while the accuracy of the simulation depends only on the 
quadrature along the direction which is perpendicular to the wall.
In the CLB implementation, the momentum space directions are always parallel to the 
(fixed) $x$ and $y$ axes. Accurate simulation results of the flow inside the 
expanding portion of the channel can be obtained only when the 
half-range Gauss-Hermite quadrature is employed on both axes, at equally 
high order. Moreover, the vorticity profile obtained in the CLB formulation 
exhibits oscillations near the wall (inside the Knudsen layer),
which are not present when the VLB implementation is used. An analysis of the 
runtime of the CLB and VLB implementations at the same level of accuracy 
(same values for the half-range Gauss-Hermite quadratures) shows that, 
at large values of the quadrature order, the VLB implementation is one 
order of magnitude faster than the CLB implementation.

\section{Conclusion} \label{sec:conc}

In this paper, the Boltzmann equation with respect to curvilinear coordinates
was considered, written with respect to orthonormal vielbein fields (triads in $3D$), extending 
the formalism introduced in Ref.~\cite{cardall13} for the relativistic Boltzmann equation 
to the non-relativistic case.
The vielbein can be used to align the momentum space along the coordinate directions, 
while also decoupling the dependence of $(\vp - m\vu)^2$ appearing in the Maxwell-Boltzmann 
equilibrium distribution on the induced metric tensor. 
The vielbein formalism allows the Boltzmann equation to be obtained in conservative form 
for any choice of coordinates using elementary differential geometry.

Choosing a coordinate system adapted to the boundary of the fluid domain allows the 
momentum space to be aligned such that the incoming and outgoing fluxes are 
described by conditions of the form $p^\hata > 0$ and $p^\hata < 0$, respectively.
The separation of incoming and outgoing particles is directly amenable to 
discretizations of the momentum space based on half-range quadratures. 
In the case when the flow shares the symmetries of the curvilinear grid,
aligning the momentum space to the coordinate grid results in a phase space 
which preserves the symmetries of the flow, allowing the spatial
dimensions along which the flow is homogeneous to be suppressed.

To illustrate the advantages of this methodology, we considered 
two applications, namely the circular Couette flow 
between coaxial cylinders and the flow through a gradually expanding 
channel. In the first case, the use of vielbeins in the momentum space
allows a one dimensional spatial grid to be employed. In the second case, 
the vielbeins allow the momentum space degrees of freedom to be aligned along 
the boundary, making the implementation of diffuse reflection using half-range 
Gauss-Hermite quadratures identical to the case of Cartesian geometries.

The validation of our scheme in the context of the circular Couette flow 
was performed by comparing our simulation results with the analytic 
solutions in the hydrodynamic and ballistic regimes and with the 
transition regime results reported in Ref.~\cite{aoki03}, which were obtained
using high-order Discrete Velocity Models. We performed 
simulations in the incompressible (low-Mach number) regime, as well as 
in the non-negligible Mach number regime. In the latter case, we were able to successfully
recover the temperature, stress-tensor and heat flux fields. Thus, we conclude 
that our resulting scheme is applicable for the simulation of the circular 
Couette flow of a compressible gas obeying the Boltzmann-BGK equation 
for all degrees of rarefaction.

In the context of the gradually expanding channel, our numerical results 
were validated in the incompressible limit of the Navier-Stokes regime 
by comparison with the benchmark CFD solutions reported in 
Refs.~\cite{roache82,cliffe82} for the case when the Reynolds number 
is ${\rm Re} = 100$, achieved by setting the inlet debit at $Q_0 = 0.1$ 
and a Knudsen number of ${\rm Kn} = 0.001$. Maintaining ${\rm Re} = 100$ while 
increasing the viscosity $\mu = {\rm Kn}$ brings the flow in the compressible, 
non-isothermal regime, where we highlighted the temperature variation in the 
transverse direction, as well as the enhancement of the vortex dimensions with the 
increase of the debit at the inlet. Finally we explored the rarefaction effects 
by keeping $Q_0 = 0.1$ for increasing values of ${\rm Kn}$. We highlighted 
deviations from the Hagen-Poiseuille law for the pressure gradient, as well as 
the formation of a Knudsen layer where the vorticity diverges logarithmically 
with the distance to the boundary.

Since our quadrature-based lattice Boltzmann models are off-lattice, 
we employed high-order finite-difference methods such as the 
total variation diminishing third-order Runge-Kutta (TVD RK-3) method 
developed in Ref.~\cite{shu88} for the time-stepping procedure, together with 
the fifth-order weighted essentially non-oscillatory (WENO-5) method for the
computation of the numerical fluxes. 
Noting that the non-trivial features of the flow form 
predominantly near the domain boundaries, we employed a grid stretching
method inspired from Refs.~\cite{mei98,guo03}.
We were thus able to obtain accurate simulation results with a comparatively small number 
of grid nodes, ranging from $96$ points to $16$ points in the 
hydrodynamic and ballistic regimes for the circular Couette flow and 
$30 \times 200 = 6000$ nodes for the gradually expanding channel.

During the analysis of the circular Couette flow, we considered
two formulations of the Boltzmann equation,
namely the $\widetilde{f}$ and $\chi$ formulations. 
In the $\widetilde{f}$ formulation, the time evolution 
and advection are performed at the level of 
$\widetilde{f} = f\sqrt{g}$ and the spatial derivative 
is taken with respect to the radial coordinate $R$. 
In the $\chi$ formulation, the time evolution and 
advection are performed at the level of the distribution 
function $f$, while the spatial derivative is taken with 
respect to $\chi^R = R^2 / 2$.
We found that applying the TVD RK-3 and WENO-5 schemes to solve the 
Boltzmann equation in the $\widetilde{f}$ formulation 
could not recover the simple solution 
$f = {\rm constant}$ in the case when both cylinders were kept at rest and at the same 
temperature. We further demonstrated that in the $\widetilde{f}$ 
formulation, the macroscopic 
variables (number density $n$, temperature $T$ and radial and tangential heat fluxes 
$q^\hatR$ and $q^\hvarphi$) develop sharp jumps near the boundaries, 
as well as non-physical 
oscillations when the lattice spacing is coarse.
With our implementation of the $\chi$ formulation of the Boltzmann equation,
we were able to reproduce the exact solution $f = {\rm constant}$ 
in the stationary case, and in the case when the cylinders undergo rotation, 
the resulting stationary profiles of $n$, $T$, $u^\hvarphi$ and $q^\hvarphi$ are smooth. 
However, the radial heat flux still exhibits jumps which are formed 
in the two nodes which are nearest to the boundaries. These jumps were visible only in the 
hydrodynamic regime, while at larger values of the relaxation time 
(i.e.~for $\tau \gtrsim 0.01$), the stationary profile of $q^\hatR$ became smooth.
We found that the effects of these irregularities on the bulk profiles were greatly diminished 
by applying the grid stretching technique to increase the resolution near the boundaries, 
while maintaining a considerably coarser resolution within the bulk of the flow. 
The gain in performance is evident,
since we were able to obtain the same level of accuracy with a stretched grid comprised of 
$32$ points per unit radial length as with the unstretched grid 
employing $128$ points per unit radial length.

We finally draw some conclusions regarding the efficiency of our implementation.
Since the dynamics along the vertical axis in the flows considered in this paper 
is trivial, we integrated out the $p^z$ degree of freedom of the momentum space 
and introduced two sets of reduced distributions.

In the incompressible limit of the Navier-Stokes regime, we recovered 
the analytic solution in the circular Couette flow problem, as well as the 
benchmark solutions of Refs.~\cite{cliffe82,roache82} for the flow through the 
gradually expanding channel using the ${\rm H}(2;3) \times {\rm H}(2;3)$ model (i.e., the 3rd order full-range Gauss-Hermite 
quadrature on both axes) employing $3\times 3 = 9$ velocities. While the number of velocities 
is the same as that employed by the popular D2Q9 lattice Boltzmann model, the efficiency 
of our implementation with respect to, e.g., Refs.~\cite{budinsky14,hejranfar17pre},
is immediately obvious in the context of the circular Couette flow,
since the vielbein approach allows us to employ a one-dimensional discretization of the 
spatial grid (i.e. only along the radial direction). 

In the slip-flow and transition regimes of the circular Couette flow, 
our models employ a number of velocities similar to that used in 
the implementation presented in Ref.~\cite{watari16}, which is based on a Cartesian 
split of the momentum space. Since the latter approach does not preserve the symmetries 
of the geometry, a 2D spatial grid is required, which makes our implementation 
more efficient by at least two orders of magnitude. Furthermore, the number of velocities
employed in Ref.~\cite{aoki03}, where the cylindrical symmetry in the momentum 
space is retained (allowing a one-dimensional spatial grid to be used) is significantly 
larger than the one employed in our models, mainly due to the fact that our models 
employ the half-range Gauss-Hermite quadrature in order to implement the boundary conditions.
Thus, our implementation is at least two orders of magnitude faster than that employed 
in Ref.~\cite{aoki03} for ${\rm Kn} \lesssim 10$. It is worth mentioning that at larger 
values of ${\rm Kn}$, the number of velocities required for our models increases dramatically, 
becoming of the same order of magnitude as the number of velocities employed in Ref.~\cite{aoki03}.

The versatility of our models to probe rarefaction effects in non-Cartesian geometries 
is demonstrated by our simulations performed in the context of the gradually expanding 
channel for values of ${\rm Kn}$ up to $0.5$, highlighting the formation of a Knudsen 
layer where the vorticity presents a logarithmic divergence with respect to the distance 
to the channel wall. To the best of our knowledge, our results represent the first account 
for rarefaction effects in the gradually expanding channel geometry.
Our investigations show that the simulation of rarefied flows 
in the geometry of the gradually expanding channel is around one order of magnitude 
faster in the vielbein approach than when a Cartesian decomposition 
of the momentum space is employed.

\begin{acknowledgments}
This work was supported by a grant of the Romanian National Authority for Scientific 
Research and Innovation, CNCS-UEFISCDI, project number PN-II-RU-TE-2014-4-2910. 
Computer simulations were done using the Portable Extensible Toolkit for Scientific Computation (PETSc) 
developed at Argonne National Laboratory, Argonne, Illinois \cite{petsc-web-page,petsc-efficient}.
The authors are grateful to Professor 
Victor Sofonea (Romanian Academy, Timi\cb{s}oara Branch, Romania) for 
encouragement, as well as for sharing with us the computational infrastructure 
available at the Timi\cb{s}oara Branch of the Romanian Academy.
\end{acknowledgments}

\appendix

\section{Boltzmann equation with respect to general coordinates}\label{app:cov}

It is easy to check that Eq.~\eqref{eq:boltz_cov} is in covariant form, i.e.~that its form remains 
unchanged under a change of coordinate system from $\{x^\wi\}$ to some new coordinates 
$\{x^{\overline{i}}\}$. Also, it can be checked that Eq.~\eqref{eq:boltz_cov} reduces to the 
Boltzmann equation \eqref{eq:boltz} when Cartesian coordinates are employed.

For completeness, this appendix presents a derivation of the form in Eq.~\eqref{eq:boltz_cov} 
without the use of the tools of differential geometry. The first step in writing the Boltzmann 
equation with respect to the new coordinates is to consider the differential of $f$:
\begin{subequations}
\begin{align}
 df =& \frac{\partial f}{\partial t} dt + \left(\frac{\partial f}{\partial x^i}\right)_{p^j} dx^i + 
 \frac{\partial f}{\partial p^i} dp^i\label{eq:df_old}\\
 =& \frac{\partial f}{\partial t} dt + \left(\frac{\partial f}{\partial x^\wi}\right)_{p^\wj} dx^\wi + 
 \frac{\partial f}{\partial p^\wi} dp^\wi\label{eq:df_new},
\end{align}
\end{subequations}
where the notation $(\partial f /\partial x^i)_{p^j}$ refers to the derivative of $f$ with respect 
to $x^i$ while keeping $p^j$ constant. In order to replace the derivatives occurring in 
Eq.~\eqref{eq:df_old} with those occurring in Eq.~\eqref{eq:df_new}, the following results can be used:
\begin{align}
 dx^\wi = \frac{\partial x^\wi}{\partial x^j} dx^j, \qquad 
 dp^\wi = \frac{\partial x^\wi}{\partial x^j} dp^j + 
 p^j \frac{\partial^2 x^\wi}{\partial x^k \partial x^j} dx^k.
\end{align}
Thus, the Boltzmann equation takes the form:
\begin{equation}
 \frac{\partial f}{\partial t} + \frac{p^\wi}{m} \frac{\partial f}{\partial x^\wi} + 
 \left(F^\wi + \frac{1}{m}\frac{\partial^2 x^\wi}{\partial x^j \partial x^k} p^j p^k\right) \frac{\partial f}{\partial p^\wi} = J[f].
 \label{eq:boltz_aux}
\end{equation}
Writing:
\begin{align}
 \frac{\partial^2 x^\wi}{\partial x^j \partial x^k} p^j p^k =& 
 \frac{\partial x^j}{\partial x^\wj} \frac{\partial x^k}{\partial x^\wk} 
 \frac{\partial^2 x^\wi}{\partial x^j \partial x^k} p^\wj p^\wk\nonumber\\
 =& -\frac{\partial x^\wi}{\partial x^\ell} \frac{\partial^2 x^\ell}{\partial x^\wj \partial x^\wk}
 p^\wj p^\wk,
\end{align}
the identification \eqref{eq:christoffel} can be made on the last line above,
such that Eq.~\eqref{eq:boltz_aux} reduces to \eqref{eq:boltz_cov}.

\section{Boltzmann equation with respect to orthonormal triads} \label{app:triad}

The same methodology as in appendix~\ref{app:cov} can be applied in the case when orthonormal 
triads are employed:
\begin{subequations}
\begin{align}
 df =& \frac{\partial f}{\partial t} dt + \left(\frac{\partial f}{\partial x^\wi}\right)_{p^\wj} dx^\wi + 
 \frac{\partial f}{\partial p^\wi} dp^\wi\label{eq:df_old_triad}\\
 =& \frac{\partial f}{\partial t} dt + \left(\frac{\partial f}{\partial x^\wi}\right)_{p^\hata} dx^\wi + 
 \frac{\partial f}{\partial p^\hata} dp^\hata\label{eq:df_new_triad}.
\end{align}
\end{subequations}
In this case, it is possible to express $dp^\hata$ as follows:
\begin{equation}
 dp^\hata = d(\omega^\hata_\wi p^\wi) = \omega^\hata_\wi dp^\wi + 
 p^\wi \frac{\partial \omega^\hata_\wi}{\partial x^\wj} dx^\wj.
\end{equation}
Thus, the Boltzmann equation becomes:
\begin{multline}
 \frac{\partial f}{\partial t} + \frac{p^\hata}{m} e_\hata^\wi \frac{\partial f}{\partial x^\wi}\\
 + \left[F^\hata + \frac{1}{m} p^\wi p^\wj \left(\frac{\partial\omega^\hata_\wi}{\partial x^\wj} 
 - \omega^\hata_\wk \Gamma^\wk{}_{\wi \wj}\right)\right] \frac{\partial f}{\partial p^\hata} = J[f].
 \label{eq:boltz_aux_triad} 
\end{multline}
The connection coefficients $\Gamma^\hata{}_{\hatb\hatc}$ are related to the covariant 
derivative of $\omega^\hata_\wi$ through:
\begin{align}
 \nabla_\wj \omega^\hata_\wi =& \frac{\partial \omega^\hata_\wi}{\partial x^\wj} - 
 \omega^\hata_\wk \Gamma^\wk{}_{\wi\wj} \nonumber\\
 =& \omega_\wj^\hatc \nabla_\hatc \omega^\hata_\wi \nonumber\\
 =& -\Gamma^\hata{}_{\hatb\hatc} \omega^\hatb_\wi \omega^\hatc_\wj.
\end{align}
The above result is sufficient to render Eq.~\eqref{eq:boltz_aux_triad} in the form of Eq.~\eqref{eq:boltz_triad}.

\section{Boltzmann equation in conservative form}\label{app:cons}

Starting from Eq.~\eqref{eq:boltz_triad}, it is possible to arrive at Eq.~\eqref{eq:boltz_cons} 
by forcing a $g^{-1/2}$ factor in front of each term on the left hand side, as follows:
\begin{multline}
 \frac{1}{\sqrt{g}} \frac{\partial (f \sqrt{g})}{\partial t} + 
 \frac{1}{\sqrt{g}} \frac{\partial}{\partial x^{\wi}} \left(\frac{p^\hata}{m} e_\hata^\wi f \sqrt{g}\right)\\
 + \frac{1}{\sqrt{g}} \frac{\partial}{\partial p^\hata} \left[\left(F^\hata - 
 \frac{1}{m} \Gamma^\hata{}_{\hatb\hatc} p^\hatb p^\hatc\right) f \sqrt{g}\right] \\
 - f \left[ \frac{p^\hata}{m} \frac{1}{\sqrt{g}} \frac{\partial}{\partial x^\wi} \left(e^\wi_\hata \sqrt{g}\right) - 
 \left(\Gamma^\hata{}_{\hata\hatb} + \Gamma^\hata{}_{\hatb \hata}\right) \frac{p^\hatb}{m}\right] = J[f].
 \label{eq:boltz_cons_aux}
\end{multline}
The only step required to arrive at Eq.~\eqref{eq:boltz_cons} is to show that the 
last term in the left hand side of Eq.~\eqref{eq:boltz_cons_aux} vanishes. 

First, we use the following property:
\begin{equation}
 \frac{1}{\sqrt{g}} \frac{\partial}{\partial x^\wi} \left(e^\wi_\hata \sqrt{g}\right) = 
 \nabla_{\wi} e^\wi_\hata.\label{eq:nablawi_e}
\end{equation}
We note that in the above, the covariant derivative refers only to the coordinate $\wi$. 
Since the covariant derivative $\nabla_\wi$ transforms as a tensor with respect to changes 
of coordinates, it is possible to express Eq.~\eqref{eq:nablawi_e} in terms of
a covariant derivative in the tetrad index $\hatb$, as follows:
\begin{equation}
 \nabla_{\wi} e^\wi_\hata = \omega_\wi^\hatb \nabla_\hatb e^\wi_\hata.
\end{equation}
The covariant derivative of $e^\wi_\hata$ with respect to $\hatb$ can be written, by definition,
using the connection coefficients $\Gamma^\hatc{}_{\hata\hatb}$, as follows:
\begin{equation}
 \omega_\wi^\hatb \nabla_\hatb e^\wi_\hata = 
 \omega_\wi^\hatb \Gamma^\hatc{}_{\hata\hatb} e^\wi_\hatc.
\end{equation}
Noting that, by construction, $\omega_\wi^\hatb e^\wi_\hatc = \delta^\hatb{}_{\hatc}$, the following 
result is obtained:
\begin{equation}
 \frac{1}{\sqrt{g}} \frac{\partial}{\partial x^\wi} \left(e^\wi_\hata \sqrt{g}\right) = \Gamma^\hatb{}_{\hata\hatb}.
\end{equation}
With the above result, the last term in the left hand side of Eq.~\eqref{eq:boltz_cons_aux} reduces to:
\begin{equation}
 \frac{p^\hata}{m} \frac{1}{\sqrt{g}} \frac{\partial}{\partial x^\wi} \left(e^\wi_\hata \sqrt{g}\right) - 
 \left(\Gamma^\hata{}_{\hata\hatb} + \Gamma^\hata{}_{\hatb \hata}\right) \frac{p^\hatb}{m} = 
 - \Gamma^\hata{}_{\hata\hatb} \frac{p^\hatb}{m}.
\end{equation}
Expression \eqref{eq:boltz_noncons} is obtained after noting that 
$\Gamma^\hata{}_{\hata\hatb} = 0$, due to the antisymmetry
of the connection coefficients in the first pair of indices.

\section{Projection of the force term onto the space of orthogonal polynomials}\label{app:force}

In this Section of the appendix, the implementation of the momentum space derivatives 
$\partial_{p^\hata} f$ and $\partial_{p^\hata} (fp^\hata)$
of the distribution function $f$ (or its reduced versions $f'$ and $f''$ introduced 
in Sec.~\ref{sec:LB:red}) in the LB models employed in this paper 
is reviewed for the cases when the full-range and half-range Gauss-Hermite quadratures 
are employed. We consider that the momentum space is two-dimensional,
since in the applications considered in this paper, the third dimension is reduced
by analytic integration, as described in Sec.~\ref{sec:LB:red}.
It is understood that all instances of $f$ can be replaced directly by the reduced 
distributions $f'$ and $f''$.

\subsection{Projection on the space of full-range Hermite polynomials}

\subsubsection{Projection of $\partial f / \partial p^\hata$}

The projection of $\partial f / \partial p^\hata$ onto the space of full-range 
Hermite polynomials has been discussed in the context of 
the LB models employed in this paper in Refs.~\cite{ambrus16jocs,ambrus17arxiv}. For completeness, 
we include in this Subsection a brief review of the results presented therein.

Let us consider the expansion of the distribution function $f$
with respect to the momentum component $p^\hata$ in terms of full-range Hermite polynomials:
\begin{equation}
 f = \frac{e^{-p_\hata^2 / 2}}{\sqrt{2\pi}} \sum_{\ell = 0}^\infty \frac{1}{\ell!} \mathcal{F}_\ell H_\ell(p^\hata).
 \label{eq:force:HLB:f}
\end{equation}
The expansion coefficients $\mathcal{F}_\ell$ can be obtained 
using:
\begin{equation}
 \mathcal{F}_\ell = \int_{-\infty}^\infty dp^\hata\,f\,H_\ell(p^\hata),
 \label{eq:force:HLB:F}
\end{equation}
where the following orthogonality relation of the Hermite polynomials was used:
\begin{equation}
 \int_{-\infty}^\infty \frac{dx}{\sqrt{2\pi}} e^{-x^2/2} H_\ell(x) H_{\ell'}(x) = \ell! \delta_{\ell,\ell'}.
\end{equation}
The derivative of $f$ with respect to $p^\hata$ is given by:
\begin{equation}
 \frac{\partial f}{\partial p^\hata} = -\frac{e^{-p_\hata^2 / 2}}{\sqrt{2\pi}} 
 \sum_{\ell = 0}^\infty \frac{1}{\ell!} \mathcal{F}_\ell H_{\ell+ 1}(p^\hata),
 \label{eq:force:HLB:df_aux}
\end{equation}
where the relation $\partial_x [e^{-x^2/2} H_\ell(x)] = -e^{-x^2/2} H_{\ell+1}(x)$ was used.

For definiteness, let us consider a full-range Gauss-Hermite quadrature of order $Q_1$
along the first momentum space direction, such that
$p^{\hat{1}}$ takes the discrete values $p_i^{\hat{1}}$ ($i = 1, 2, \dots Q_1$) satisfying
$H_{Q_1}(p_i^{\hat{1}}) = 0$. The other component $p^{\hat{2}} \rightarrow \{p^{\hat{2}}_j\}$ 
($j = 1, 2, \dots \mathcal{Q}_2$) is also discretized according to an arbitrary quadrature, 
such that Eq.~\eqref{eq:force:HLB:f} is replaced by:
\begin{equation}
 f_{ij} = w_i^H(Q_1) \sum_{\ell = 0}^{Q_1 - 1} \frac{1}{\ell!} \mathcal{F}_{\ell; j} H_\ell(p^{\hat{1}}_i),
 \label{eq:f_H}
\end{equation}
where $w_i^H(Q_1)$ is the full-range Gauss-Hermite quadrature weight defined in Eq.~\eqref{eq:w_H}.
The above definition of $f_{ij}$ allows the integral in Eq.~\eqref{eq:force:HLB:F} 
to be exactly recovered using the full-range Gauss-Hermite quadrature formula 
\cite{hildebrand87,shizgal15}:
\begin{equation}
 \mathcal{F}_{\ell;j} = \sum_{i = 1}^{Q_1} f_{ij} H_\ell(p^{\hat{1}}_i).
\end{equation}
Truncating Eq.~\eqref{eq:force:HLB:df_aux} following the above recipe gives:
\begin{equation}
 \left(\frac{\partial f}{\partial p^{\hat{1}}}\right)_{ij} = \sum_{i' = 1}^{Q_1} 
 \mathcal{K}_{i,i'}^{\hat{1},H} f_{i'j},
\end{equation}
where the elements of the 
$Q_1 \times Q_1$ matrix $\mathcal{K}_{i,i'}^{\hat{1},H}$ are given in Eq.~\eqref{eq:df_K_H}.

\subsubsection{Projection of $\partial (f p^{\hat{a}}) / \partial p^{\hat{a}}$}

Starting from the expansion \eqref{eq:force:HLB:f} of $f$ with respect to $p^{\hat{a}}$,
a similar expansion for $\partial (fp^\hata) / \partial p^\hata$ can be assumed:
\begin{equation}
 \frac{\partial (fp^\hata)}{\partial p^\hata} = 
 \frac{e^{-p_\hata^2 / 2}}{\sqrt{2\pi}} \sum_{\ell = 0}^\infty \frac{1}{\ell!} \mathcal{F}_\ell' H_\ell(p^\hata).
 \label{eq:force:HLB:dfp}
\end{equation}
The coefficients $\mathcal{F}_{\ell}'$ can be obtained by multiplying 
Eq.~\eqref{eq:force:HLB:dfp} by $H_\ell(p^\hata)$ 
and integrating with respect to $p^\hata$:
\begin{equation}
 \mathcal{F}_\ell' = -\int_{-\infty}^\infty dp^\hata\, f\,p^\hata
 \frac{\partial H_\ell(p^\hata)}{\partial p^\hata},\label{eq:force:HLB:dfp_aux}
\end{equation}
where integration by parts was used to arrive at the above result. Using the property
$x H_\ell'(x) = \ell H_\ell(x) + \ell(\ell - 1) H_{\ell - 2}(x)$, the integral in Eq.~\eqref{eq:force:HLB:dfp_aux}
can be performed in terms of the coefficients $\mathcal{F}_\ell$:
\begin{equation}
 \mathcal{F}_\ell' = -\ell \mathcal{F}_{\ell} - \ell(\ell - 1) \mathcal{F}_{\ell - 2}.
\end{equation}
We now assume that $a$ represents the first momentum space direction 
and $p^{\hat{1}} \rightarrow p^{\hat{1}}_i$ ($i = 1, 2, \dots Q_1$) 
according to a full-range Gauss-Hermite quadrature 
of order $Q_1$. In this case, $\mathcal{F}_\ell'$ can be written as:
\begin{equation}
 \mathcal{F}_{\ell;j}' = -\sum_{i' = 1}^{Q_1} f_{i'j} [\ell H_\ell(p^{\hat{1}}_{i'}) + 
 \ell(\ell - 1) H_{\ell - 2}(p^{\hat{1}}_{i'})].
\end{equation}
Thus, $\partial (fp^{\hat{1}})/\partial p^{\hat{1}}$ \eqref{eq:force:HLB:dfp} can be written 
as a linear combination of $f_{ij}$:
\begin{equation}
 \left[\frac{\partial (fp^{\hat{1}})}{\partial p^{\hat{1}}}\right]_{ij} = 
 \sum_{i' = 1}^{Q_1} \widetilde{\mathcal{K}}^{\hat{1},H}_{i,i'} f_{i'j},
\end{equation}
where the elements of the 
$Q_1 \times Q_1$ matrix $\widetilde{\mathcal{K}}^{\hat{1},H}_{i,i'}$ are given in 
Eq.~\eqref{eq:dfp_K_H}.

\subsection{Projection on the space of half-range Hermite polynomials}

\subsubsection{Projection of $\partial f / \partial p^\hata$}

The construction of the derivative $\partial f / \partial p^\hata$ in the frame 
of LB models based on the half-range Gauss-Hermite quadrature was presented in 
Ref.~\cite{ambrus17arxiv}. In this Subsection, the construction 
procedure and the main results are briefly reviewed.

The idea behind LB models based on half-range Gauss-Hermite quadratures is to acknowledge that 
the wall interaction induces a discontinuity in the distribution function,
since the distribution of particles emitted by the diffuse reflective boundary has in general 
a different functional form compared to that of
the distribution of the incident particles.
Thus, it is natural to separate the space of incoming and outgoing particles as follows:
\begin{equation}
 f(p^\hata) = \theta(p^\hata) f^+(p^\hata) + \theta(-p^\hata) f^-(p^\hata).
 \label{eq:force:hh:f}
\end{equation}
Taking the derivative of Eq.~\eqref{eq:force:hh:f} with respect to $p^\hata$ gives:
\begin{equation}
 \frac{\partial f}{\partial p^\hata} = \theta(p^\hata) 
 \left(\frac{\partial f}{\partial p^\hata}\right)^+ + 
 \theta(-p^\hata) \left(\frac{\partial f}{\partial p^\hata}\right)^-,
\end{equation}
where
\begin{equation}
 \left(\frac{\partial f}{\partial p^\hata}\right)^\pm = 
 \frac{\partial f^\pm}{\partial p^\hata} + \delta(p^\hata) [f^+(0) - f^-(0)].
 \label{eq:force:hh:dfpm}
\end{equation}
The Dirac delta function is obtained as the derivative of the Heaviside step functions:
\begin{equation}
 \delta(x) = \pm \partial_x \theta(\pm x).
\end{equation}
In obtaining Eq.~\eqref{eq:force:hh:dfpm}, we used $\delta(p^\hata) \rightarrow
\delta(p^\hata)[\theta(p^\hata) + \theta(-p^\hata)]$,
while $f^\pm(0)$ are defined through:
\begin{equation}
 f^+(0) = \lim_{p^\hata \rightarrow 0+} f(p^\hata), \qquad
 f^-(0) = \lim_{p^\hata \rightarrow 0-} f(p^\hata).
\end{equation}
In general, $f^+(0) \neq f^-(0)$ due to the interaction with the boundary.

Let us now consider the expansion of $f^\pm(p^\hata)$ with respect to the 
half-range Hermite polynomials \cite{ambrus16jcp,ambrus17arxiv}:
\begin{equation}
 f^\pm = \frac{e^{-p_\hata^2 / 2}}{\sqrt{2\pi}} \sum_{\ell = 0}^\infty 
 \mathcal{F}_\ell^\pm \hh_{\ell}(|p^\hata|),
\end{equation}
where the expansion coefficients $\mathcal{F}^\pm_\ell$ are given as:
\begin{equation}
 \mathcal{F}^+_\ell = \int_0^\infty dp^\hata\, f\, \hh_{\ell}(p^\hata), \qquad 
 \mathcal{F}^-_\ell = \int_{-\infty}^0 dp^\hata\, f\, \hh_{\ell}(-p^\hata).
 \label{eq:F_hh}
\end{equation}

The expansion of $(\partial f / \partial p^\hata)^\pm$ \eqref{eq:force:hh:dfpm} was obtained in
Ref.~\cite{ambrus17arxiv}:
\begin{multline}
 \left(\frac{\partial f}{\partial p^\hata}\right)^\pm = 
 \pm \frac{e^{-p_\hata^2 / 2}}{\sqrt{2\pi}} \Bigg\{\\
 \sum_{\ell = 0}^\infty \mathcal{F}^\pm_\ell
 \sum_{s= \ell + 1}^\infty \left[\frac{\hh_{s,0} \hh_{\ell,0}}{\sqrt{2\pi}} - 
 \frac{1}{a_s} \delta_{\ell, s+1}\right] \hh_s(|p^\hata|)\\
 -\frac{1}{2\sqrt{2\pi}} \left[\sum_{\ell = 0}^\infty (\mathcal{F}^+_\ell + \mathcal{F}^-_\ell) \hh_{\ell,0}\right]
 \left[\sum_{s = 0}^\infty \hh_{s,0} \hh_s(|p^\hata|)\right]\Bigg\},
 \label{eq:force:hh:df_aux}
\end{multline}
where the notation $\hh_{\ell,s}$ is defined in Eq.~\eqref{eq:hh_exp}.

Let us now consider that $\hata$ refers to the first momentum space direction 
and $p^{\hat{1}}$ is discretized using $p^{\hat{1}}_i (i = 1, 2, \dots 2Q_a)$ 
according to the half-range Gauss-Hermite quadrature, as described in 
Eq.~\eqref{eq:hh_roots}. Considering also that $p^{\hat{2}} \rightarrow p^{\hat{2}}_j$
according to an arbitrary quadrature method, the equivalent of Eq.~\eqref{eq:f_H}
becomes:
\begin{align}
 f_{ij} =& w_i^\hh(Q_1) \sum_{\ell = 0}^{Q_1 - 1} \frac{1}{\ell!} \mathcal{F}_{\ell; j}^+ 
 \hh_\ell(p^{\hat{1}}_i),\nonumber\\
 f_{i+Q_1,j} =& w_i^\hh(Q_1) \sum_{\ell = 0}^{Q_1 - 1} \frac{1}{\ell!} \mathcal{F}_{\ell; j}^-
 \hh_\ell(p^{\hat{1}}_i),
 \label{eq:f_hh}
\end{align}
where $i = 1, 2, \dots Q_1$ and the expansion on the second line 
corresponds to the negative momentum semi-axis.
The expansion coefficients $\mathcal{F}_{\ell;j}^\pm$ can be obtained 
using the following quadrature sums:
\begin{equation}
 \mathcal{F}_{\ell;j}^+ = \sum_{i = 1}^{Q_1} f_{ij} \hh_\ell(p^{\hat{1}}_i), \qquad
 \mathcal{F}_{\ell;j}^- = \sum_{i = 1}^{Q_1} f_{i+Q_1,j} \hh_\ell(p^{\hat{1}}_i).
\end{equation}
Truncating Eq.~\eqref{eq:force:hh:df_aux} following the above recipe gives:
\begin{equation}
 \left(\frac{\partial f}{\partial p^{\hat{1}}}\right)_{ij} = 
 \sum_{i' = 1}^{2Q_1} \mathcal{K}_{i,i'}^{\hat{1},\hh} f_{i'j},
\end{equation}
where the elements of the $2Q_1 \times 2Q_1$ matrix $\mathcal{K}^{{\hat{1}},\hh}_{i,i'}$ 
are given in Eq.~\eqref{eq:df_K_hh}.

\subsubsection{Projection of $\partial (f p^\hata) / \partial p^\hata$}

Since the product $p^\hata f$ vanishes at $p^\hata = 0$, the $\delta$ term 
appearing in the expression of $\partial f / \partial p^\hata$ does not appear 
in this case, such that $\partial (p^\hata f) / \partial p^\hata$ can be expanded 
as:
\begin{equation}
 \frac{\partial (p^\hata f)}{\partial p^\hata} = 
 \frac{e^{-p_\hata^2/2}}{\sqrt{2\pi}} \sum_{\ell = 0}^\infty \mathcal{F}^{\sigma}_\ell{}'
 \hh(|p^\hata|),
 \label{eq:force:dfp_hh}
\end{equation}
where, $\sigma = \pm 1$ when $\pm p^\hata > 0$. The coefficients 
$\mathcal{F}^{\hata,\pm}$ can be obtained by virtue of the orthogonality of 
the half-range Hermite polynomials using integration by
parts:
\begin{align}
 \mathcal{F}^{+}_\ell{}' =& -\int_0^\infty dp^\hata\, f\, p^\hata \hh'_\ell(p^\hata),\nonumber\\
 \mathcal{F}^{-}_\ell{}' =& -\int_{-\infty}^0 dp^\hata\, f\, p^\hata \hh'_\ell(-p^\hata).
 \label{eq:Fphi_hh}
\end{align}
The product $x \hh'_\ell(x)$ appearing above can be written as \cite{ambrus16jcp}:
\begin{multline}
 x \hh'_{\ell}(x) = \ell \hh_\ell(x) + \frac{\hh_{\ell,0}^2 + \hh^2_{\ell-1,0}}{a_{\ell - 1}\sqrt{2\pi}} 
 \hh_{\ell - 1}(x)\\
 + \frac{1}{a_{\ell - 1} a_{\ell -2}} \hh_{\ell -2}(x).
\end{multline}
Substituting the above result into Eq.~\eqref{eq:Fphi_hh} and using Eq.~\eqref{eq:F_hh} yields:
\begin{equation}
 \mathcal{F}^{\pm}_\ell{}' =  -\left[\ell \mathcal{F}_\ell^\pm +
 \frac{\hh_{\ell,0}^2 + \hh_{\ell -1,0}^2}{a_{\ell - 1} \sqrt{2\pi}} \mathcal{F}^\pm_{\ell -1} + 
 \frac{1}{a_{\ell - 1} a_{\ell - 2}}\mathcal{F}^\pm_{\ell - 2}\right].
\end{equation}
Let us now consider that the $\hata$ direction corresponds to the first direction 
of the momentum space and $p^{\hat{1}}$ is discretized according to the 
half-range Gauss-Hermite quadrature of order $Q_1$, such that 
Eq.~\eqref{eq:force:dfp_hh} takes the form:
\begin{equation}
 \left[\frac{\partial (p^{\hat{1}} f)}{\partial p^{\hat{1}}}\right]_{ij} = 
 \sum_{i' = 1}^{2Q_1} \widetilde{\mathcal{K}}^{\hat{1},\hh}_{i,i'} f_{i',j},
\end{equation}
where the elements of the $2Q_1 \times 2Q_1$ matrix $\widetilde{\mathcal{K}}^{\hat{1},\hh}_{i,i'}$
are given in Eq.~\eqref{eq:dfp_K_hh}.

\end{document}